\documentclass[aip, jcp, amsmath,amssymb, footinbib,floatfix,citeautoscript, reprint]{revtex4-1}
\usepackage{color}
\usepackage{braket}
\usepackage{graphicx}
\usepackage{units}
\usepackage{dcolumn}
\usepackage{bm}
\DeclareUnicodeCharacter{0301}{\'{e}}
\newcommand{\adj}[1]{#1 ^{\dagger}}
\DeclareMathSymbol{\mdot}{\mathord}{symbols}{"01}
\newcommand{\pdiv}[2]{\frac{\partial #1}{\partial #2}}
 
\newcommand{\ing}[3]{\int_{#1}^{#2} \text{d}#3 \hspace{.1 cm}}
\newcommand{\lr}[3]{\left#1 #2 \right#3}
\newcommand{\tdpo}[1]{\hat{\boldsymbol{\mu}}_{#1}}
\newcommand{\ev}[1]{\left\langle #1 \right\rangle}
\newcommand{\opr}[1]{\hat{#1}}
\newcommand{\erm}{\text{e}}

\newcommand{\smlsub}[1]{{_{#1}}}
\newcommand{\tmdot}{\mdot \mdot \mdot}
\newcommand{\wn}{\,{\rm cm}^{-1}}
\begin{document}

\title{Beyond the Condon limit: Condensed phase optical spectra from atomistic simulations}

\author{Zachary R. Wiethorn}
\affiliation{Department of Chemistry, University of Colorado Boulder, Boulder, Colorado 80309, USA}

\author{Kye E. Hunter}
\affiliation{Department of Chemistry, Oregon State University, Corvallis, Oregon 97331, USA}

\author{Tim J. Zuehlsdorff}
\email{tim.zuehlsdorff@oregonstate.edu}
\affiliation{Department of Chemistry, Oregon State University, Corvallis, Oregon 97331, USA}

\author{Andr\'{e}s Montoya-Castillo}
\email{Andres.MontoyaCastillo@colorado.edu}
\affiliation{Department of Chemistry, University of Colorado Boulder, Boulder, Colorado 80309, USA}

\date{\today}

\begin{abstract}
While dark transitions made bright by molecular motions determine the optoelectronic properties of many materials, simulating such non-Condon effects in condensed-phase spectroscopy remains a fundamental challenge. We derive a Gaussian theory to predict and analyze condensed phase optical spectra beyond the Condon limit. Our theory introduces novel quantities that encode how nuclear motions modulate the energy gap and transition dipole of electronic transitions in the form of spectral densities. By formulating the theory through a statistical framework of thermal averages and fluctuations, we circumvent the limitations of widely used microscopically harmonic theories, allowing us to tackle systems with generally anharmonic atomistic interactions and non-Condon fluctuations of arbitrary strength. We show how to calculate these spectral densities using first-principles simulations, capturing realistic molecular interactions and incorporating finite-temperature, disorder, and dynamical effects. Our theory accurately predicts the spectra of systems known to exhibit strong non-Condon effects (phenolate in various solvents) and reveals distinct mechanisms for electronic peak splitting: timescale separation of modes that tune non-Condon effects and spectral interference from correlated energy gap and transition dipole fluctuations. We further introduce analysis tools to identify how intramolecular vibrations, solute-solvent interactions, and environmental polarization effects impact dark transitions. Moreover, we prove an upper bound on the strength of cross correlated energy gap and transition dipole fluctuations thereby elucidating a simple condition that a system must follow for our theory to accurately predict its spectrum. 
\end{abstract}

\maketitle

\section{Introduction}
\label{main-section:Introduction-GNCT-JCP}

Linear electronic spectroscopy remains one of the most versatile and standard tools to interrogate the optical properties of chemical systems in complex environments, ranging from nanostructured materials to biological chromophores and synthetic dyes. \cite{Amerongen2001,Rivera-Torrente2020,rogl2002, suntivich2014} However, predicting, assigning, and disentangling the physical origin of the often broad and overlapping features in these absorption and fluorescence spectra is challenging as these can arise from electronic-vibrational coupling,\cite{Santoro2008, Baiardi2013, coker2020nonpert_vibronic, cho2020crosscorr, kouppel1984multimode, spano2010spectral, tempelaar2017vibronic, von2017vibrationally,makriflemmingHTJPCB} environmental and electronic polarization effects,\cite{Marini2010,Provorse2016,Zuehlsdorff2020,Hanna2008,fleming1996chromophore,tanimura1993two,van1985time,lai_geva2022electronic} and charge and energy transfer \cite{Aranda2021,Zuehlsdorff2021,mavros2016condensed, jang2004multichromophoric,jang2007multichromophoric,chen2009optical,ishizaki2009unified,gao_geva2020simulating, Krylov2022charge}. Hence, theoretical and simulation approaches that can accurately predict and offer the means to interpret spectra of complex condensed phase systems are crucial in elucidating the processes that underlie these signals and designing chromophores with desired optical properties. 

Accounting for the influence of dark transitions (non-Condon effects) on the optical spectra of chromophores in the condensed phase is particularly challenging. What is more, non-Condon effects are ubiquitous,\cite{Raab1999,Santoro2008,Baiardi2013,Yugin2020, schmidt2005pronounced, loparo2007variation, Czernuszewicz1993, SPIRO1990541, HE2012ZnP, webertpp2020, MINAEV2006308, makriflemmingTPPHT} arising when molecular motions break the symmetry of a molecule that otherwise disallows an electronic transition. Such effects can manifest as additional peaks and broadening in the spectrum, as is the case for the S$_0 \rightarrow $ S$_1$ transition in benzene \cite{li2010symmetrybenzene, deSouza2018, beguvsic2018fly}, and as strong symmetry breaking between absorption and fluorescence lineshapes \cite{Craig1969,makriflemmingHTJPCB}. A variety of approaches has thus emerged to tackle specific aspects of this thorny problem.

For instance, one can routinely employ the microscopic harmonic approximation to treat non-Condon effects in small and fairly rigid molecules in the gas phase via a low-order expansion of the transition dipole around a minimum energy reference structure \cite{barone2011computational}. However, the high dimensionality of condensed phase potential energy surfaces (PESs) makes finding such reference structures impractical and often impossible. Furthermore, thermally accessible low-frequency anharmonic motions, such as torsions and collective solvent fluctuations, raise questions about the applicability of the harmonic approximation\cite{avila2014lineshape, improta2014quantum,cerezo2019adiabatic}. Alternatively, in the limit of slow nuclei, one can employ the ensemble approach that averages vertical excitations over thermally accessible but frozen configurations.\cite{Isborn2012, Milanese2017, Crespo-Otero2012, Ge2015, Zuehlsdorff2017, marenich2015electronic} This approach, however, misses dynamical effects, including vibronic progressions and spectral diffusion arising from molecules experiencing a changing local environment. In contrast, approaches based on the second-order truncation of the cumulant expansion\cite{Mukamel-book,Zwier2007,Loco2018,Loco2018b, Zuehlsdorff2019,cho2008coherent,zhuang2009coherent} can unambiguously tackle the condensed phase, capture the inhomogeneous (slow nuclei) limit, the quantum dynamics of intramolecular vibrations and spectral diffusion, and the effects of anharmonic molecular motions. Yet, it is limited to the Condon approximation\cite{Condon1926} where the transition dipoles are assumed to be insensitive to thermally accessible nuclear motions. Thus, developing a method that faithfully renders atomistic interactions, accounts for disorder and finite temperature effects, and captures the quantum dynamics of electronic excitations and nuclear fluctuations beyond the Condon limit is of paramount importance.

Here we introduce a statistical dynamical theory to treat non-Condon effects in linear spectroscopy that achieves these goals. It is compatible with both electronic structure and molecular dynamics (MD) toolboxes and enables the simulation of spectra of molecules in the gas and condensed phases. Unlike earlier methods that require a reference configuration on which to build perturbatively small non-Condon corrections,\cite{Small1971,Santoro2008,Baiardi2013,deSouza2018,Toutounji2019,makriflemmingHTJPCB} our treatment leverages the statistics of the energy gap and transition dipole fluctuations to construct a simple and physically interpretable framework that can treat non-Condon fluctuations of arbitrary strength. It further enables us to employ simulation data that incorporates anharmonic atomistic interactions, disorder, and thermal effects to construct the exact quantum dynamical Gaussian optical response of the system. 

Unlike the Gaussian theory of Condon spectroscopy (second-order cumulant approach) \cite{Mukamel-book}, which requires only one spectral density, our Gaussian non-Condon theory employs three. Common to both theories is the energy gap spectral density that quantifies the impact of photoinduced nonequilibrium nuclear rearrangements, or solvation, on the transition energy. Our work introduces the remaining two. These quantify the effects of photo-induced nonequilibrium nuclear rearrangements through the lens of the transition dipole and the cross correlation of the transition dipole and energy gap. While our transition dipole spectral density is analogous to those used to describe the nuclear motion-induced fluctuations of hopping terms in charge transfer theories \cite{liao2002numerical, cook2006exact, mavros2016condensed}, our work introduces these for the first time in the context of the transition dipole for spectroscopy. 

To offer a protocol to calculate these spectral densities directly from atomistic simulations, we employ the Kubo quantum correction factor\cite{bader1994quantum, Egorov1999, Kim2002b, mano2004quantum, ramirez2004quantum} to re-express the quantum time correlation functions in terms of classical counterparts that can be calculated using classical MD simulations. However, our theory can also be used with exact or approximate quantum dynamics schemes \cite{makrictpi, tuckerman2018open, cao1994formulation, jang1999cmd, mano2004quantum, markland2018nuclear}. We illustrate the ability of our theory to accurately capture absorption lineshapes by considering the phenolate anion in polar and nonpolar solvents. 

We conclude with an analysis that leverages model spectral densities to derive physical insight into the various spectral limits of our non-Condon theory. These include cases where both Condon and non-Condon contributions are prominent, and where non-Condon contributions dominate and even appear to split electronic transitions. We further illustrate how negative absorption intensity can arise when the strength of cross correlated fluctuations of the energy gap and transition dipole exceeds that of the respective autocorrelated fluctuations.

\section{Theory}
\label{sec:theory}

Here we introduce our Gaussian theory for optical spectroscopy in the condensed phase beyond the Condon approximation. To do this, we first establish the notation by reviewing the linear response framework for optical spectroscopy and consider the Gaussian treatment of energy gap fluctuations that underlies the celebrated second-order cumulant method. The second-order cumulant offers a way to calculate the optical spectra of chromophores in the gas and condensed phases employing atomistic simulations, although exclusively in the Condon limit. We then generalize the second-order cumulant approach beyond the Condon approximation to account for nuclear motions that cause Gaussian fluctuations of the transition dipole and dynamically render dark transitions bright. Our non-Condon optical response function consists of a sum of four terms that account for statistical Condon, non-Condon, and mixed contributions. To make our theory compatible with atomistic simulation, we introduce two new spectral densities and offer a simple protocol to calculate these using MD simulations. Lastly, we rewrite the absorption spectrum to introduce a closed expression in terms of our non-Condon spectral densities, elucidating how Condon and non-Condon contributions impact the absorption spectrum of a chromophore. 

\subsection{Linear response expressions}

We restrict our attention to the pure dephasing limit where chromophores do not transfer energy and the shape of the spectral response arises from the modulation of the energy gap by nuclear motions.\cite{skinner1988theory, PhysRevB.65.195313, PhysRevB.81.245419, vezvaee2022noise} A minimal Hamiltonian describing this situation contains two electronic states coupled differently to the nuclear components, 
\begin{equation}\label{eq: pure-dephasing-hamiltonian}
    \hat{H} = \ket{g} H_g(\hat{\mathbf{p}}, \hat{\mathbf{q}}) \bra{g} + \ket{e} H_e(\hat{\mathbf{p}}, \hat{\mathbf{q}}) \bra{e}. 
\end{equation}
Here, $g$ and $e$ denote the ground and excited adiabatic electronic states, and $H_i(\hat{\mathbf{p}}, \hat{\mathbf{q}}) = T(\hat{\mathbf{p}}) + V_i(\hat{\mathbf{q}}) + \omega_{ig}^0$, are the nuclear Hamiltonians respectively coupled to the electronic states with reference frequency $\omega_{ig}^0=\omega_i - \omega_g$. We note, however, that our treatment is general and can easily be applied beyond two electronic levels. 
 
The light-matter interaction in an optical experiment, whose matter component is encoded via the transition dipole operator, can induce an electronic transition, 
\begin{equation}\label{eq:transition-dipole-operator}
    \hat{\boldsymbol{\mu}}(\hat{\mathbf{q}}) = \ket{g} \hat{\boldsymbol{\mu}}_{ge}(\hat{\mathbf{q}}) \bra{e} + \ket{e} \hat{\boldsymbol{\mu}}_{eg}(\hat{\mathbf{q}}) \bra{g}.
\end{equation}
The transition dipole operator generally depends on the nuclear configuration of the system, $\hat{\mathbf{q}}$, and $\hat{\boldsymbol{\mu}}_{ij}(\hat{\mathbf{q}}) = \bra{i}\hat{\boldsymbol{\mu}}(\hat{\mathbf{q}})\ket{j}$ is the transition dipole matrix element connecting electronic states $i$ and $j$. 

The linear absorption spectrum of the system may be written as,\cite{Mukamel-book} 
\begin{equation}\label{eq:linear-absorption-general}
    \sigma_{\rm a}(\omega) = s(\omega) \int \text{d}t \, \erm^{i \omega t} \chi^{(1)}(t),
\end{equation}
where $s(\omega) \propto \omega$ is a prefactor with linear-frequency dependence for single-photon absorption spectroscopy\cite{Mukamel-book}. The linear optical response function, $\chi^{(1)}(t)$, encodes the non-equilibrium response of the system to the interaction with light. Within linear response theory, $\chi^{(1)}(t)$ can be written as,\cite{Mukamel-book}
\begin{equation}\label{eq:equilibrium-linear-optical-response}
\begin{split}
    \chi^{(1)}(t) &= \text{Tr}_{\rm nuc} \Big[\rho_g( \hat{\mathbf{p}}, \hat{\mathbf{q}}) \boldsymbol{\mu}_{ge} (\hat{\mathbf{q}} ; t) X(\hat{\mathbf{q}} ; t) \boldsymbol{\mu}_{eg} (\hat{\mathbf{q}}) \Big] \\ &\equiv \Big \langle \boldsymbol{\mu}_{ge} (\hat{\mathbf{q}} ; t) X(\hat{\mathbf{q}} ; t) \boldsymbol{\mu}_{eg} (\hat{\mathbf{q}}) \Big \rangle.
\end{split}
\end{equation}
Here, $\rho_g( \hat{\mathbf{p}}, \hat{\mathbf{q}}) \equiv \erm^{-\beta H_g ( \hat{\mathbf{p}}, \hat{\mathbf{q}})}/\mathrm{Tr}_{\rm nuc}[\erm^{-\beta H_g ( \hat{\mathbf{p}}, \hat{\mathbf{q}})}]$ is the canonical nuclear distribution for the ground electronic state PES, with respect to which thermal averages are evaluated, $\langle \tmdot \rangle$.\footnote{For fluorescence, the initial condition is the equilibrium distribution of molecular nuclei on the excited state PES, $\rho_{e}^\textrm{eq} = \erm^{-\beta \hat{H}_e} / \mathrm{Tr}_{\rm nuc}[\erm^{-\beta \hat{H}_e}]$.} Time evolution is also performed with respect to the ground state nuclear Hamiltonian, $\boldsymbol{\mu}_{ge} (\hat{\mathbf{q}} ; t) = \erm^{i \hat{H}_g t } \boldsymbol{\mu}_{ge} (\hat{\mathbf{q}}) \erm^{-i \hat{H}_g t }$, and $X(\hat{\mathbf{q}} ; t) = \erm^{i H_g( \hat{\mathbf{p}}, \hat{\mathbf{q}}) t } \erm^{-i H_e( \hat{\mathbf{p}}, \hat{\mathbf{q}}) t }$. Defining the energy gap operator, $\hat{U}(\hat{\mathbf{q}}) \equiv H_e( \hat{\mathbf{p}}, \hat{\mathbf{q}}) - H_g( \hat{\mathbf{p}}, \hat{\mathbf{q}})$, rewriting it in terms of its thermal average and fluctuations, $U(\hat{\mathbf{q}}) = \omega_{eg}^{\rm av} + \delta U(\hat{\mathbf{q}})$, and employing the Dyson expansion, one can write, 
\begin{equation}\label{eq:time-ordered-egap-statistics}
    X(\hat{\mathbf{q}};t) = \erm^{-i \omega_{eg}^{\rm av} t} \, \erm_+^{-i \int_0^{t}\text{d}\tau \, \delta U({\hat{\mathbf{q}}; \tau) } } .
\end{equation}
Manipulation of the optical response function into this form traditionally precedes the adoption of the Condon approximation and its subsequent treatment within the second-order cumulant approximation. 

\subsection{Statistical mapping in the Condon limit}
\label{subsection:Condon-statistical-mapping}

Other than the assumption of linear response theory, the treatment of the optical response has been exact. We now turn to a versatile statistical treatment that is applicable regardless of the atomistic interactions and the shape of the underlying PESs. Specifically, for systems in the condensed phase, a macroscopic number of nuclear motions, $\mathbf{q}$, modulate the energy gap fluctuations, $\delta U(\mathbf{q})$. In the limit where these motions \textit{individually} modulate the energy gap fluctuations only weakly, the central limit theorem holds and one can treat these motions as random Gaussian variables.\cite{makri1999linear, mukamel1985fluorescence, georgievskii1999linear, bader1990role, bader1994quantum, jortner1976temperature, valleau2012alternatives, mahan1990many, Zuehlsdorff2019}

Here, it is customary to first make the Condon approximation, which states that the transition dipole is insensitive to the fluctuating nuclear environment,
\begin{equation}
    \boldsymbol{\mu}(\hat{\mathbf{q}}) \approx \hat{\boldsymbol{\mu}},
\end{equation}
followed by expanding the response function in orders of energy gap cumulants \cite{Mukamel-book}. When the fluctuations of the energy gap obey Gaussian statistics, all cumulants beyond second order exactly vanish \cite{Mukamel-book, Cho-book}. This leads to the celebrated form of the optical response in terms of the second-order lineshape function, $g_2 (t)$,
\begin{equation}
    \chi_\smlsub{\rm GCT}^{(1)} (t) =  |\hat{\boldsymbol{\mu}}_{ge}|^2 \erm^{-i \omega_{eg}^{\rm av} t - g_2(t)}, 
\end{equation}
with GCT denoting the ``Gaussian Condon theory" and,
\begin{equation}\label{eq:GNCT-JCP-cumulant-lineshape}
    g_2(t) = \frac{1}{\pi} \int_{0}^{\infty} \text{d}\omega \, \frac{J(\omega)}{\omega^2} \Omega (\beta, \omega, t). 
\end{equation}
\begin{equation}\label{eq:GCT-universal-function-GNCT-JCP}
    \Omega (\beta, \omega, t) = \coth(\beta \omega / 2)[1 - \cos(\omega t)] + i [\sin(\omega t) - \omega t],
\end{equation}
is a universal function that arises from the Gaussian prescription and is not specific to the chemical system under consideration. Equation~\eqref{eq:GNCT-JCP-cumulant-lineshape} introduces the energy gap spectral density, $J(\omega)$, which is the frequency representation of the imaginary part of the quantum equilibrium time correlation function of the energy gap fluctuations,
\begin{equation}\label{eq:energy-spectral-density-in-terms-of-tcf}
    J(\omega) = i \Theta(\omega) \int_{-\infty}^{\infty} \text{d}t \, \erm^{i \omega t} \, \text{Im} \ev{\delta U(\hat{\mathbf{q}};t) \delta U(\hat{\mathbf{q}};0)}.
\end{equation}
This function encodes the coupling between the electronic excitation and nuclear motions, quantifying the photo-induced nonequilibrium nuclear rearrangements upon transition from the electronic ground to the excited state.

The Brownian oscillator model (BOM) is a minimal, exactly solvable model whose energy gap displays Gaussian statistics. It consists of PESs that can be decomposed into a continuum of noninteracting pairs of harmonic oscillators whose equilibrium positions shift upon transition from the ground to the excited electronic state, 
\begin{subequations}\label{eq: BOM-hamiltonian}
\begin{align}
    H_g(\hat{\mathbf{p}}, \hat{\mathbf{q}}) & = \frac{1}{2} \sum_j \left[ \hat{p}_j^2 + \omega_j^2 \hat{q}_j^2 \right], \\
    H_e(\hat{\mathbf{p}},
    \hat{\mathbf{q}}) & = \omega_{eg}^0+
    \frac{1}{2} \sum_j \left[ \hat{p}_j^2 + \omega_j^2 (\hat{q}_j-d_j)^2 \right]. 
\end{align}
\end{subequations}
Here, $\hat{q}_j = \sqrt{m_j} \hat{Q}_j$ and $\hat{p}_j = \hat{P}_j / \sqrt{m_j}$ are the mass-weighted coordinate and momentum operators for the oscillator with frequency $\omega_j$ whose minimum shifts by $d_j$ upon photo-excitation. Adoption of the second-order cumulant in a generally anharmonic system is equivalent to mapping the problem to an effective BOM. This constitutes a statistical rather than a microscopic harmonic map. Under the Gaussian mapping, Eq.~\eqref{eq:energy-spectral-density-in-terms-of-tcf} can be expressed in terms of the components of the BOM,
\begin{equation}\label{eq:egap-sd-discrete-distribution}
    J(\omega) = \frac{\pi}{2} \sum_j \omega_j^3 d^2_j \delta(\omega-\omega_j) .
\end{equation}
For a system that genuinely displays Gaussian statistics, $J(\omega)$ is independent of the temperature and all cumulants beyond the second order are identically zero. This fact may be used as a diagnostic tool \cite{valleau2012alternatives, Zuehlsdorff2019}.

The second-order cumulant approach can unambiguously tackle the condensed phase, capture the inhomogeneous (slow nuclei) limit, the quantum dynamics of intramolecular vibrations and spectral diffusion, and the effects of anharmonic molecular motions, rendering the second-order cumulant a state-of-the-art technique for predicting condensed phase optical spectroscopy\cite{Mukamel-book,Zwier2007,Loco2018,Loco2018b, Zuehlsdorff2019,cho2008coherent,zhuang2009coherent}. However, a direct atomistic implementation of the second-order cumulant can be computationally expensive. This cost originates from the need to perform MD simulations that properly sample the condensed phase PES and electronic structure calculations of the energy gaps along this trajectory for total simulation times that are sufficiently long to yield converged spectral densities. Importantly, advances in machine learning have reduced this cost by two to three orders of magnitude\cite{xue2020machine, chen2020exploiting, cignoni2023machine, chen2023}. Yet, despite this method's success, a straightforward generalization of the second-order cumulant that accounts for statistical non-Condon effects has proven elusive to date. 

\subsection{\label{subsection:nonCondon-statistical-mapping} Statistical mapping beyond the Condon limit}

To treat \textit{statistical} non-Condon effects, here we relax the Condon approximation and introduce a generalization to the second-order cumulant that leverages the simplicity of the BOM. In particular, we consider a chromophore whose transition dipole, like its energy gap, fluctuates under the influence of a macroscopic number of nuclear motions in the system, i.e., $\boldsymbol{\mu}( \hat{ \mathbf{q} } ) = \ev{\boldsymbol{\mu}} + \delta \boldsymbol{\mu}( \hat{ \mathbf{q} } ) $. Invoking the central limit theorem, we posit that these transition dipole fluctuations exhibit Gaussian statistics whose behavior can be captured by an ensemble of Gaussian random variables. In the spirit of the second-order cumulant mapping to the BOM, we take these random variables to be the coordinates of the oscillators into which the PESs have been decomposed, 
\begin{equation}\label{eq:statistical_dipole_GNCT_JCP}
    \boldsymbol{\mu}(\hat{\mathbf{q}}) = \langle \hat{\boldsymbol{\mu}} \rangle + \sum_j \boldsymbol{\alpha}_j \hat{q}_j.
\end{equation}
Here, $\boldsymbol{\alpha}_j$ is an undetermined coefficient corresponding to the strength and direction of the transition dipole fluctuation corresponding to the $j^{\rm th}$ oscillator coordinate. 

Substituting Eq.~\eqref{eq:statistical_dipole_GNCT_JCP} into Eq.~\eqref{eq:equilibrium-linear-optical-response} yields our non-Condon optical response function, 
\begin{equation}\label{eq:nonCondon-optical-response-summations-GNCT-JCP}
\begin{split}
    \chi_\smlsub{\rm GNCT}^{(1)}(t) &= \big|\big\langle \hat{\boldsymbol{\mu}}_{ge} \big\rangle\big|^2 \big\langle X(\hat{\mathbf{q}}, t) \big\rangle
    \\&+
    \big\langle \hat{\boldsymbol{\mu}}_{ge} \big\rangle\cdot \sum_j \boldsymbol{\alpha}_j \big\langle \hat{q}_j (t) X(\hat{\mathbf{q}}, t) \big\rangle
    \\&+ 
    \big\langle \hat{\boldsymbol{\mu}}_{eg} \big\rangle \cdot \sum_j \boldsymbol{\alpha}_j \big\langle X(\hat{\mathbf{q}}, t) \hat{q}_j (0) \big\rangle
    \\&+ 
    \sum_{j,k} \boldsymbol{\alpha}_j \mdot \boldsymbol{\alpha}_k \big\langle \hat{q}_j (t) X(\hat{\mathbf{q}}, t) \hat{q}_k (0) \big\rangle,
\end{split}
\end{equation}
with ``GNCT" denoting Gaussian non-Condon theory. Employing real-time Feynman path integrals (see Supporting Information (SI) Sec.~\ref{SI-section:non-Conondon-optical-response-derivation-via-path-integrals-GNCT-JCP}), we obtain exact expressions for each of these components in terms of $\Omega(\beta, \omega, t)$ and products of the oscillator shifts $\{ d_j \}$ and the undetermined non-Condon coefficients $\{ \boldsymbol{\alpha}_j \}$ that encode the optical response of a specific chemical system. In Eq.~\eqref{eq:nonCondon-optical-response-summations-GNCT-JCP}, the dynamical quantities in the second and third lines are equivalent, 

\begin{equation}\label{eq:A-term-from-path-integral-GNCT-JCP}
\begin{split}
    \sum_j \boldsymbol{\alpha}_j \big\langle \hat{q}_j (t) X(\hat{\mathbf{q}}, t) \big\rangle &= \sum_j \boldsymbol{\alpha}_j \big\langle X(\hat{\mathbf{q}}, t) \hat{q}_j (0) \big\rangle \\&\equiv \boldsymbol{\mathcal{A}}(t),
\end{split}
\end{equation}
whereas the final term in Eq.~\eqref{eq:nonCondon-optical-response-summations-GNCT-JCP} can be decomposed into $\mathcal{A}^2(t)$ and $\mathcal{B}(t)$, 
\begin{equation}\label{eq:B-n-AA-term-from-path-integral-GNCT-JCP}
    \mathcal{B}(t) \equiv \sum_{j,k} \boldsymbol{\alpha}_j \mdot \boldsymbol{\alpha}_k \big\langle \hat{q}_j (t) X(\hat{\mathbf{q}}, t) \hat{q}_k (0) \big\rangle - \mathcal{A}^2(t).
\end{equation}  
Hence, we rewrite the non-Condon optical response as 
\begin{equation}
\begin{split}
    \chi_\smlsub{\rm GNCT}^{(1)}(t) &=\Big[ |\langle \hat{\boldsymbol{\mu}}_{ge} \rangle|^2 + 2 \text{Re}\{ \langle \hat{\boldsymbol{\mu}}_{ge} \rangle \} \mdot \boldsymbol{\mathcal{A}}(t) \\&\hspace{.6cm}+ \mathcal{A}^2(t) + \mathcal{B}(t) \Big] \erm^{-i\omega_{eg}^\textrm{av}t-g_2(t)}. \label{eq:nC-linear-optical-response-closed}
\end{split}
\end{equation}

Our work introduces the two dynamical quantities, \begin{subequations}\label{eq:discrete-auxiliary-nC-dynamical-quantiites}
\begin{align}
    \boldsymbol{\mathcal{A}}(t) &= \frac{i}{2} \sum_j \frac{\boldsymbol{\alpha}_j d_j}{\omega_j} \frac{\partial \Omega(\beta, \omega_j, t)}{\partial t} \label{eq:nonCondon-A-sumation-GNCT-JCP}, 
    \\ \mathcal{B}(t) &= \frac{1}{2} \sum_j \frac{|\boldsymbol{\alpha}_j|^2}{\omega_j^3} \frac{\partial^2 \Omega(\beta, \omega_j, t)}{\partial t^2}\label{eq:nonCondon-B-sumation-GNCT-JCP},
\end{align}
\end{subequations}  
which account for all non-Condon effects in our Gaussian theory. Crucially, our theory simplifies to the Condon limit when $\boldsymbol{\alpha}_j \rightarrow 0$, which implies that $\boldsymbol{\mathcal{A}}(t) \rightarrow 0$ and $\mathcal{B}(t) \rightarrow 0$. To make our theory broadly usable, we provide a straightforward protocol for obtaining the coefficients $\{d_j\}$ and $\{\boldsymbol{\alpha}_j\}$ from molecular simulation. Using this protocol, we parameterize our quantum dynamical theory with atomistic detail, allowing one to investigate a molecule's spectrum as a function of its chemical structure, solvent environment, and the collective nuclear motions that control \textit{specific} spectral features. 

Before concluding this section, we compare our approach to commonly employed microscopically harmonic theories that go beyond the Condon limit. Specifically, in the FCHT approach, one must find the minimum energy configuration of a chromophore in the gas phase on its ground or excited state PES, around which one performs a \textit{microscopic} harmonic expansion to obtain the normal modes of the system, the equilibrium shifts, mode mixing (Duschinsky rotation), and frequency changes upon photoexcitation. One then has to choose a reference nuclear configuration (e.g., the minimum on the ground or excited state PES) from which to expand the transition dipole to first order in the nuclear displacements, offering a protocol to obtain the FCHT analog of $\{ \boldsymbol{\alpha}_j \}$. The mathematical form of the model that can be parameterized using the FCHT approach \cite{Santoro2008, barone2011computational, deSouza2018, makriflemmingHTJPCB}, when taken in the limit of no Duschinsky rotation or frequency changes upon photoexcitation, is equivalent to our non-Condon BOM. This model is based on non-interacting harmonic oscillators and closed-form solutions for its response function have been derived previously \cite{Santoro2008, barone2011computational, deSouza2018, huh2017cumulant, makriflemmingHTJPCB, tanimura1993real, chen2021correlated}. However, to date, the only available protocol to parameterize this model is based on a direct harmonic approximation to the PESs and linear expansions of the transition dipole. Crucially, these expressions cannot be unambiguously applied in the condensed phase due to their reliance on the microscopically harmonic protocol outlined above. An exception arises when one can employ \emph{a priori} knowledge of a system to motivate additional approximations. For example, one might be able to treat fast and nearly harmonic intramolecular vibrational modes at the FCHT level while statically sampling low-frequency anharmonic modes and solvent degrees of freedom that might be expected to behave classically \cite{improta2014quantum,cerezo2019adiabatic, Zuehlsdorff2021, avila2014lineshape}. Yet, such strategies become system-specific. 

What, then, complicates the direct application of microscopically harmonic methods to condensed phase spectra beyond the Condon limit? For one, the microscopic harmonic approximation of the PESs tends to fail for flexible molecules with torsional modes and becomes inapplicable in the presence of low-frequency anharmonic modes and collective solvent motions. Further, non-Condon corrections in these theories require identifying a reference nuclear configuration from which to perform the Taylor expansion of the transition dipole. Since finding such minima in high dimensional PESs is inconvenient at best and at worst ill-defined, the choice of reference becomes non-unique. Even when one can identify a good reference configuration, the use of a low-order Taylor expansion implies that non-Condon fluctuations must be weak, keeping the strong non-Condon limit beyond the theory's reach. Moreover, going beyond a linear expansion of the transition dipole would greatly increase the computational expense, even for small molecules. In contrast, our statistical decomposition in Eq.~\eqref{eq:statistical_dipole_GNCT_JCP} does not necessitate minimizations on the PESs or a reference nuclear configuration. Instead, our theory requires \textit{only} that the transition dipole and energy gap fluctuations display Gaussian statistics for its validity, regardless of the anharmonicity of the underlying PESs or the strength of non-Condon fluctuations.

\subsection{Obtaining $\{ \boldsymbol{\alpha}_j\}$ from atomistic simulations}
\label{sec:non-Condon-coefficients-from-simulation}

While Eqs.~\eqref{eq:nonCondon-A-sumation-GNCT-JCP} and \eqref{eq:nonCondon-B-sumation-GNCT-JCP} reveal how the energy gap displacements, $d_j$, and the transition dipole fluctuation coefficients, $\boldsymbol{\alpha}_j$, exactly determine the optical response beyond the Condon limit, these do not provide a protocol for obtaining $d_j$ and $\boldsymbol{\alpha}_j$ directly from molecular simulations. Recalling that in the limit of the GCT, $J(\omega)$ allows one to obtain the coefficients $\{d_j^2\}$ that parameterize the optical response, here we take a similar approach to obtain $\{ \boldsymbol{\alpha}_j d_j \}$ and $\{ |\boldsymbol{\alpha}_j|^2 \}$ and construct Eqs.~\eqref{eq:nonCondon-A-sumation-GNCT-JCP} and \eqref{eq:nonCondon-B-sumation-GNCT-JCP} from molecular simulations. Specifically, we introduce two new spectral densities arising from the Fourier transform of the equilibrium time cross correlation function of the transition dipole and energy gap and the autocorrelation function of the transition dipole, respectively (See SI Sec.~\ref{SI-section:spectral-densities-from-molecular-simulation-GNCT-JCP}),
\begin{equation}
\label{eq:discrete_L}
\begin{split}
    \mathbf{L}(\omega) &\equiv i \Theta(\omega) \int_{-\infty}^{\infty} \text{d}t \, \erm^{i \omega t} \, \text{Im} \ev{\delta U(\hat{\mathbf{q}};t) \delta \boldsymbol{\mu}(\hat{\mathbf{q}};0)} \\
    &= \frac{\pi}{2} \sum_{j} \boldsymbol{\alpha}_j d_j \omega_j \delta(\omega - \omega_j),
\end{split}
\end{equation}
and 
\begin{equation}\label{eq:discrete_K}
\begin{split}
    K(\omega) &\equiv i \Theta(\omega) \int_{-\infty}^{\infty} \text{d}t \, \erm^{i \omega t} \, \text{Im} \ev{\delta \boldsymbol{\mu}(\hat{\mathbf{q}};t) \mdot \delta \boldsymbol{\mu}(\hat{\mathbf{q}};0)}. \\
    &= \frac{\pi}{2} \sum_{j} \frac{|\boldsymbol{\alpha}_j|^2}{\omega_j} \delta(\omega - \omega_j).
\end{split}
\end{equation}

Upon substituting Eqs.~\eqref{eq:discrete_L} and \eqref{eq:discrete_K} into
\begin{subequations}\label{eq:continuous-auxiliary-nC-dynamical-quantiites}
\begin{align}
    \boldsymbol{\mathcal{A}}(t) &= \frac{i}{\pi} \int_{0}^{\infty} \text{d} \omega \, \frac{\mathbf{L}(\omega)}{\omega^2} \frac{\partial \Omega(\beta, \omega, t)}{\partial t},\label{eq:non-Condon-A-spectral-density-GNCT-JCP} \\ \mathcal{B}(t) &= \frac{1}{\pi} \int_{0}^{\infty} \text{d} \omega \, \frac{K(\omega)}{\omega^2} \frac{\partial^2 \Omega(\beta, \omega, t)}{\partial t^2}\label{eq:non-Condon-B-spectral-density-GNCT-JCP},
\end{align}
\end{subequations}
one recovers Eqs.~\eqref{eq:nonCondon-A-sumation-GNCT-JCP} and \eqref{eq:nonCondon-B-sumation-GNCT-JCP}. 
Further, just as $J(\omega)$ tracks the non-equilibrium nuclear reorganization seen by the energy gap upon photo-excitation in the GCT, the new spectral densities in our GNCT have a simple physical interpretation: $K(\omega)$ quantifies the non-equilibrium nuclear rearrangement on the transition dipoles and $\mathbf{L}(\omega)$ captures the correlated effect on both the transition dipole and energy gap fluctuations.

What remains to be done is to provide a simple protocol to obtain $\mathbf{L}(\omega)$ and $K(\omega)$ from atomistic simulations. Although one can calculate the equilibrium time correlation functions in Eqs.~\eqref{eq:discrete_L} and \eqref{eq:discrete_K} using exact or approximate quantum dynamics schemes, such as quantum-classical,\cite{makrictpi} open-chain,\cite{tuckerman2018open} centroid,\cite{cao1994formulation, jang1999cmd} and ring-polymer \cite{mano2004quantum, markland2018nuclear} path integrals, here, we choose to employ the rich toolbox of \textit{classical} MD simulations to capture Gaussian non-Condon spectra for complex chemical systems. To achieve this, we approximate the quantum Kubo-transformed correlation functions with their classical analogs and employ the Fourier space detailed balance conditions \cite{bader1994quantum, Egorov1999, Kim2002b, mano2004quantum, ramirez2004quantum} to provide access to the target spectral densities encoding $\boldsymbol{\alpha}_j$ and $d_j$ (See SI Sec.~\ref{SI-section:spectral-densities-from-molecular-simulation-GNCT-JCP}),
\begin{subequations}\label{eq:spectral-densities-in-terms-of-classical-MD}
\begin{align}
    \mathbf{L}(\omega) &\approx  \theta(\omega) \frac{\beta \omega}{2} \int_{-\infty}^{\infty} \text{d}t \, \erm^{i \omega t} \langle \delta U(\hat{\mathbf{q}}; t) \delta \boldsymbol{\mu}(\hat{\mathbf{q}}; 0) \rangle_{\rm cl}, \\ K(\omega) &\approx \theta(\omega) \frac{\beta \omega}{2} \int_{-\infty}^{\infty} \text{d}t \, \erm^{i \omega t} \langle \delta \boldsymbol{\mu}(\hat{\mathbf{q}}; t) \mdot \delta \boldsymbol{\mu}(\hat{\mathbf{q}}; 0) \rangle_{\rm cl}.
\end{align}
\end{subequations}
These expressions for $\boldsymbol{\mathcal{A}}(t)$, and $\mathcal{B}( t)$, in addition to an analogous protocol to obtain $J(\omega)$\cite{mukamel1985fluorescence}, fully connect our Gaussian theory of a non-Condon BOM to MD simulations and electronic structure calculations. 

Since the transition dipole moment $\boldsymbol{\mu}(\hat{\mathbf{q}})$ is a vector quantity and only its absolute value squared (proportional to the oscillator strength of the transition) is a physical observable, one must take care to construct valid classical correlation functions to evaluate Eq.~\eqref{eq:spectral-densities-in-terms-of-classical-MD} directly from MD \cite{Eckart1935, schmidt2005pronounced, Krasnoshchekov2014}. We outline an efficient computational strategy to obtain well-defined spectral densities $\mathbf{L}(\omega)$ and $K(\omega)$ in SI Sec.~\ref{SI-section:computational-details-GNCT-JCP}.

\subsection{Impact of $\mathbf{L}(\omega)$ and $K(\omega)$ on the spectrum}
\label{subsection:GNCT-spectrum-from-GCT-spectrum-KL-convolution-GNCT-JCP}
\begin{figure}
    \centering
    \includegraphics[width=.95\columnwidth]{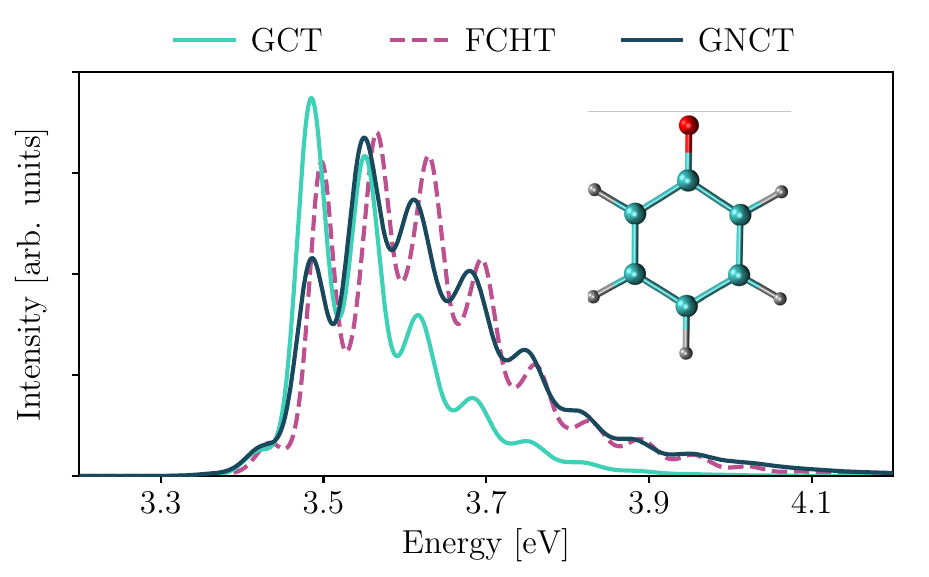}
    \caption{Comparison of the linear absorption spectrum of the phenolate anion in vacuum computed using our Gaussian non-Condon theory (GNCT; dark green) with those calculated using the Gaussian Condon Theory (GCT; teal), and the Franck-Condon Herzberg-Teller (FCHT; fuchsia) method.}
    \label{fig:phenolate_methods}
\end{figure}
While Eqs.~\eqref{eq:linear-absorption-general}, \eqref{eq:nC-linear-optical-response-closed},  \eqref{eq:continuous-auxiliary-nC-dynamical-quantiites} and \eqref{eq:spectral-densities-in-terms-of-classical-MD} provide the full prescription for calculating non-Condon spectra, the mathematical formulation in terms of $\boldsymbol{\mathcal{A}}(t)$ and $\mathcal{B}(t)$ obfuscates how $\mathbf{L}(\omega)$ and $K(\omega)$ impact spectral lineshapes. To clarify this dependence, we rewrite the total optical response as the convolution of the Condon (GCT) lineshape and the non-Condon contributions,
\begin{equation}\label{eq:nC-convolution-linear-spectrum}
\begin{split}
    \sigma_\smlsub{\rm GNCT} (\omega) &= \Big[ |\langle \hat{\boldsymbol{\mu}}^{s}_{ge} \rangle|^2 \delta(\omega) - 2 \text{Re}\{ \langle \hat{\boldsymbol{\mu}}^{s}_{ge} \rangle \} \mdot \frac{\boldsymbol{\rho}_{\mathbf{L}}(\omega)}{\omega} \\
    &+ \frac{\boldsymbol{\rho}_{\mathbf{L}}(\omega)}{\omega} \ast \frac{\boldsymbol{\rho}_{\mathbf{L}}(\omega)}{\omega} + \rho_{K}(\omega) \Big] \ast \sigma_\smlsub{\rm GCT}(\omega).
\end{split}
\end{equation}
Here, $X(\omega) \ast Y(\omega) = \int_0^{\infty} \text{d}z\ X(\omega)Y(\omega-z)$ denotes the convolution operation, and 
\begin{equation}
    \rho_{F}(\omega)= \frac{1}{\pi} \frac{F(\omega)}{1 - \erm^{-\beta\omega}},
\end{equation}
for $ F(\omega) \in \{K(\omega), \mathbf{L}(\omega)\}$. This way of rewriting the response function reveals that a shifted equilibrium average transition dipole, 
\begin{equation}\label{eq:renormalized-thermal-transition-dipole}
    \langle \hat{\boldsymbol{\mu}}^{s}_{ge} \rangle \equiv \langle \hat{\boldsymbol{\mu}}_{ge} \rangle + \boldsymbol{\lambda}_{\mathbf{L}},
\end{equation}
determines the strength of the Condon lineshape. Here, $\boldsymbol{\lambda}_{\mathbf{L}} = \frac{1}{\pi}\ing{0}{\infty}{\omega} \frac{\mathbf{L}(\omega)}{\omega}$ and can be interpreted as the nuclear motion-induced shift in the average transition dipole. Although reminiscent of the positive valued energy gap reorganization energy ($\lambda =\frac{1}{\pi}\int_{0}^{\infty} \text{d} \omega \, \frac{J(\omega)}{\omega}$), $\boldsymbol{\lambda}_{\mathbf{L}}$ is vector-valued and can therefore contain both positive and negative components. $\boldsymbol{\lambda}_{\mathbf{L}}$ describes the extent to which specific nuclear motions simultaneously modulate the energy gap and transition dipole. Additionally, $\boldsymbol{\lambda}_{\mathbf{L}}$ has the ability to either increase ($\boldsymbol{\lambda}_{\mathbf{L}}>0$) or decrease ($\boldsymbol{\lambda}_{\mathbf{L}}<0$) the average transition dipole.

Equation~\eqref{eq:nC-convolution-linear-spectrum} reveals important features of our theory and helps interpret spectral lineshapes. For example, it shows that non-Condon spectral features arise from interfering contributions that convolve the Condon lineshape. Indeed, one can observe that $\langle \hat{\boldsymbol{\mu}}^{s}_{ge} \rangle$ in Eq.~\eqref{eq:renormalized-thermal-transition-dipole} renormalizes the prominence (spectral weight) of the Condon lineshape in Eq.~\eqref{eq:nC-convolution-linear-spectrum}. Because $\mathbf{L}(\omega)$ can have both positive and negative regions, the renormalization term, $\boldsymbol{\lambda}_{\mathbf{L}}$, in Eq.~\eqref{eq:renormalized-thermal-transition-dipole} can enhance or diminish the Condon contribution to the lineshape. Also, since $\boldsymbol{\rho}_{\mathbf{L}}(\omega)$ depends on $\mathbf{L}(\omega)$, the second and third terms in Eq.~\eqref{eq:nC-convolution-linear-spectrum} can lead to both positive and negative contributions to the spectrum that, when added with the contribution from $\boldsymbol{\lambda}_{\mathbf{L}}$, manifest as spectral interference (see Secs.~\ref{section:tight-tdp} and \ref{section:broad-tdp}). Finally, since $\rho_{K}(\omega)$ and $K(\omega)$ are both non-negative functions, this term can only lead to additional shifting and broadening of the Condon lineshape in the non-Condon limit. Below, we illustrate the wealth of non-Condon effects that our theory captures by focusing on chemically transparent model systems and analyze the impact of spectral densities on the resulting optical spectra. 

\section{Solvent effects on the phenolate absorption spectrum}\label{section:phenolate-solvent-effects}

We now rigorously assess our theory's ability to account for non-Condon fluctuations in the spectrum of the phenolate anion in vacuum, cyclohexane, and water. We focus on phenolate as a case study to show how symmetry-breaking fluctuations can significantly influence the resulting spectra. SI Sec.~\ref{SI-section:computational-details-GNCT-JCP} outlines the computational details of the MD simulations and static electronic structure calculations used to construct all spectra presented in this section. 

We begin by analyzing the spectrum of phenolate in vacuum. Because the number of nuclear motions that modulate energy gap and transition dipole fluctuations is far from macroscopic, this example offers a hard test for our Gaussian theory. For a small, rigid molecule in vacuum, one can expect to accurately capture the spectrum, including non-Condon effects, at the level of the Franck-Condon-Herzberg-Teller (FCHT) approach, which we take to give a good approximation to the exact spectrum. We compare the conventional GCT and our GNCT spectra against the FCHT spectrum of phenolate in vacuum in Fig.~\ref{fig:phenolate_methods}. 

Figure~\ref{fig:phenolate_methods} shows that GCT (teal) fails to recapitulate the FCHT spectrum (fuchsia-dashed). Compared to the GCT spectrum, the FCHT spectrum shows a more significant broadening toward the blue, a more pronounced vibronic progression, and a reversal in the maximum intensity peak, which shifts from the peak at $\sim 3.50$ eV to that at $\sim 3.56$ eV. Since the FCHT spectrum contains both non-Condon and non-Gaussian effects (arising from Duschinsky rotation and mismatch between the curvatures of the ground and excited state PESs), any deviation between the two spectra can be expected to arise from failures of the GCT at capturing these effects. However, as we show in SI Fig.~\ref{SI-fig:egap-statistics}, the energy gaps of phenolate in vacuum are nearly Gaussian, with a skewness parameter of only -0.03. Previous work \cite{Zuehlsdorff2019} has shown that such skewness values are sufficiently small for the Gaussian treatment to be valid, suggesting that the inability of the GCT to account for non-Condon effects is what causes the discrepancy between the GCT and FCHT spectra. Indeed, our GNCT (dark green), shows good agreement with the FCHT spectrum, correctly capturing the spectral width and its extension to the blue region. While our GNCT does not agree perfectly with the FCHT intensities for the vibronic progressions, both methods show very similar ordering in the intensities and the reversal in maximum peak intensity between the second and third vibronic peaks. In addition, the fact that the transition dipoles display close to perfect Gaussian statistics with $\mu_x(\hat{\mathbf{q}})$, $\mu_y(\hat{\mathbf{q}})$ and $\mu_z(\hat{\mathbf{q}})$ skewness factors of -0.01, 0.06 and -0.06, respectively, further supports the applicability of our GNCT (see SI Fig.~\ref{SI-fig:transition-dipole-statistics}). The observed differences in the computed spectra can thus be potentially ascribed to the GNCT accounting for PES anharmonicity that appears even in a small and rigid molecule like phenolate, as can be observed in the computed spectral densities $J(\omega)$ and $K(\omega)$ (see SI Sec.~\ref{SI-sec:spectral-density-normal-mode-analysis-GNCT-JCP}). In general, this result demonstrates that our GNCT can successfully capture significant non-Condon effects in the linear spectra of molecules, even in the challenging limit of only a few nuclear degrees of freedom.
\begin{figure}[t]
    \centering
    \includegraphics[width=1.0\columnwidth]{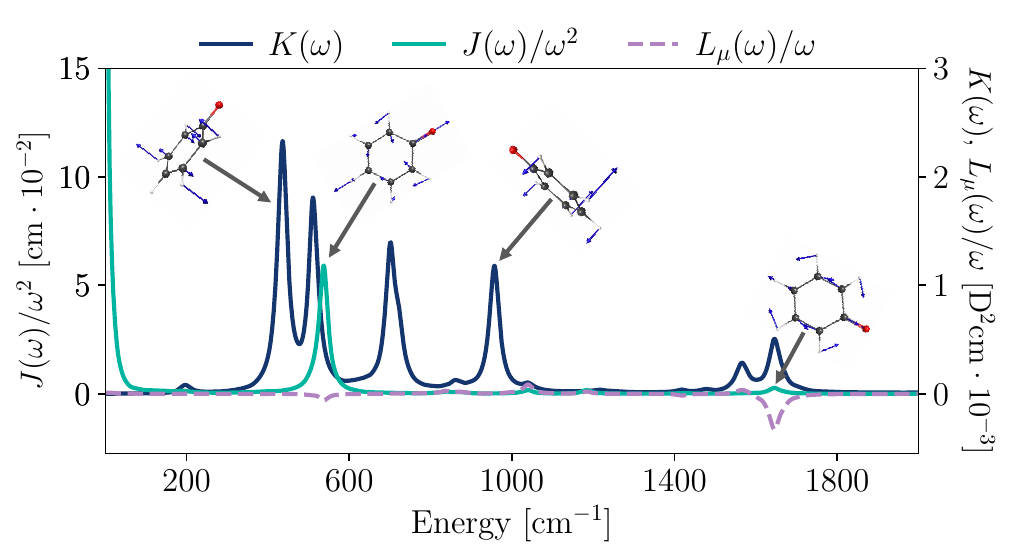}
    \caption{Effective representations of the spectral densities based on their integrand forms in Eq.~\eqref{SIeq:non-Condon-response-discrete-limit}. Here, $L_\mu (\omega) \equiv  \langle \hat{ \boldsymbol{\mu} }_{ge} \rangle \mdot \mathbf{L} (\omega)$. Note the scaled vertical axes for  $J(\omega)/\omega^2$ (displayed on the left) and $K(\omega)$ and $L_{\mu}(\omega)/\omega$ (displayed on the right). We label select normal modes of phenolate present in $J(\omega)$ and $K(\omega)$.}
    \label{fig:phenolate_sds}
\end{figure}

Having shown that our GNCT correctly captures non-Condon effects in the optical spectrum of phenolate, we now interrogate the physical implications of the spectral densities, $K(\omega)$ and $\mathbf{L}(\omega)$, that quantify these effects. Figure~\ref{fig:phenolate_sds} shows all three spectral densities for phenolate in vacuum in terms of their $\omega$-weighted forms as they appear in the integrands that define the response function: $J(\omega)/ \omega^2$, $L_{\mu} (\omega)/ \omega$ with $L_{\mu} (\omega) \equiv \langle \hat{\boldsymbol{\mu}}_{ge} \rangle \mdot \mathbf{L} (\omega)$, and $K(\omega)$ in Eqs.~\eqref{eq:GNCT-JCP-cumulant-lineshape} and \eqref{eq:nC-convolution-linear-spectrum}. As we show in SI Sec.~\ref{SI-sec:spectral-density-normal-mode-analysis-GNCT-JCP}, a normal mode analysis of phenolate reveals that the peaks in $J(\omega)$, which determines the Condon response, primarily correspond to in-plane stretching modes, whereas $K(\omega)$ displays a number of peaks driven by symmetry-breaking vibrational modes that are absent in $J(\omega)$. While $J(\omega)$ and $K(\omega)$ are exclusively positive functions, $\mathbf{L}(\omega)$ can display both positive and negative regions corresponding to energy gap and transition dipole fluctuations that are correlated or anti-correlated, respectively. The peaks in $\mathbf{L}(\omega)$ occur near modes present in both $J(\omega)$ and $K(\omega)$. The low intensities of these peaks indicate that few modes contribute to both Condon and non-Condon optical transitions. The greatest contribution to $\mathbf{L}(\omega)$ is an in-plane C-O stretch in the direction of the average transition dipole moment. The induced fluctuation in the energy gap due to this motion is anti-correlated with the induced fluctuation in the transition dipole moment, resulting in a peak with negative intensity. 

\begin{figure}[t]
    \centering
    \includegraphics[width=1.0\columnwidth]{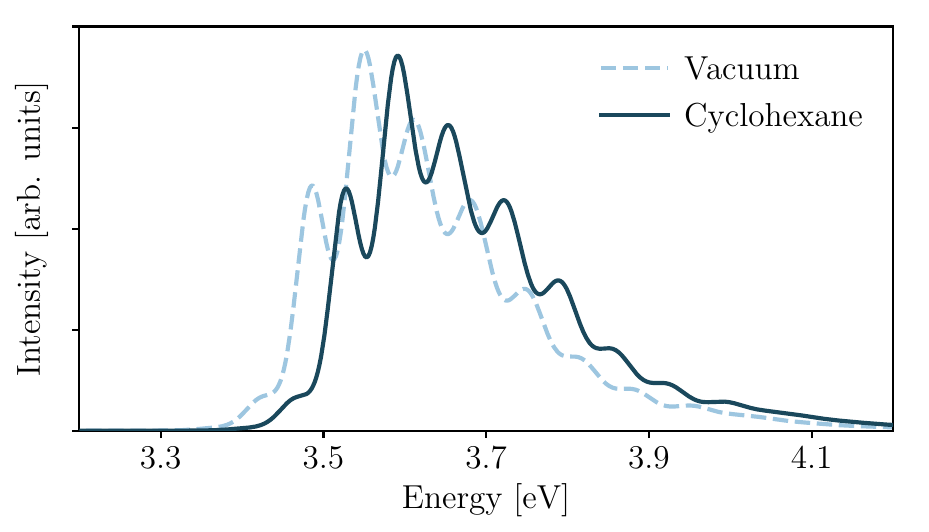}
    \caption{Linear absorption spectra of phenolate in vacuum and cyclohexane.}
    \label{fig:phenolate_cyclo}
\end{figure}
\begin{figure*}
    \centering
\includegraphics[width=0.9\textwidth]{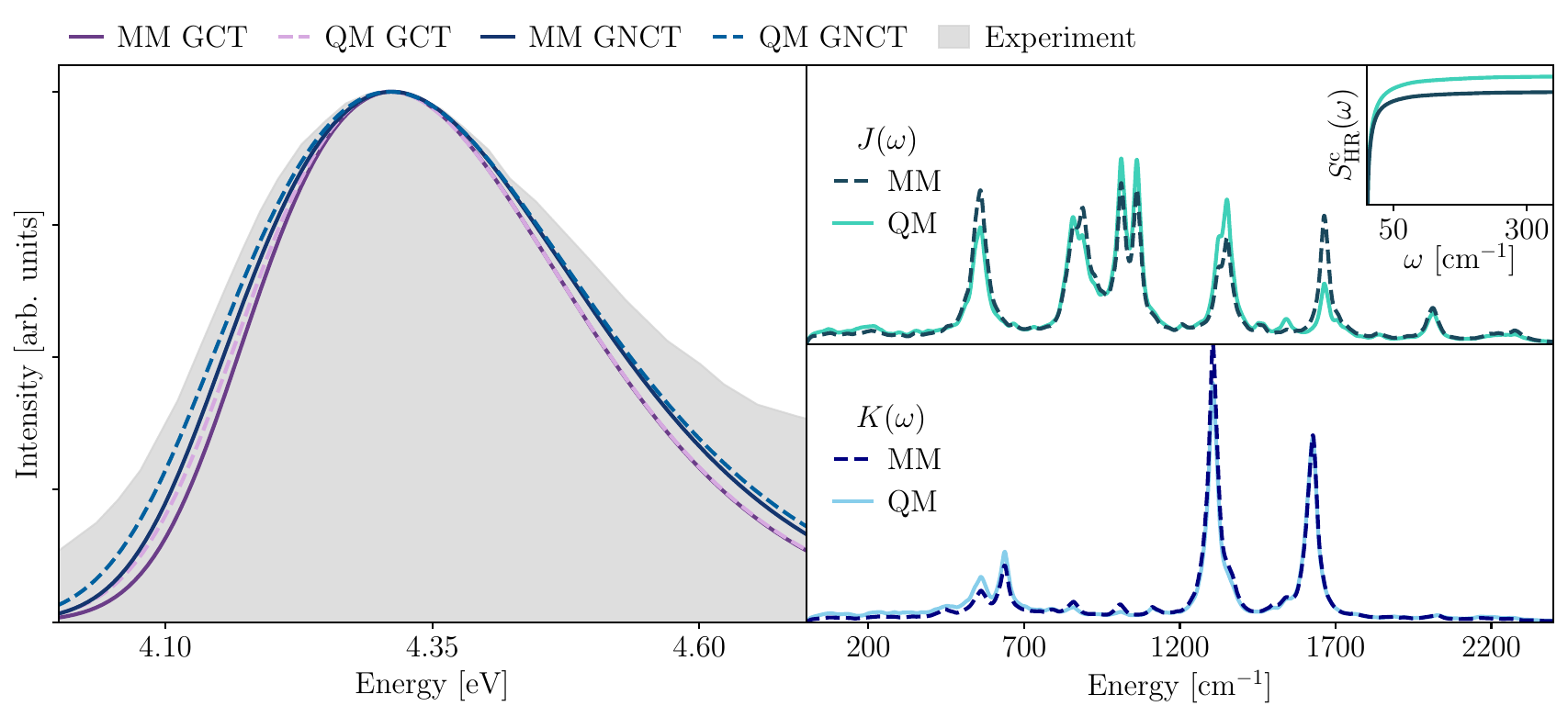}
\caption{Spectra and spectral densities for phenolate in water. (a) Simulated GCT and GNCT spectra compared to the experimental absorption spectrum. (b) Energy gap spectral densities, $J(\omega)$. Inset shows the cumulative Huang-Rhys factor given by $S^{\rm c}_{\rm HR} (\omega) = \frac{1}{\pi}\int_0^{\omega}\text{d} \Omega \, J(\Omega)/\Omega^2$. (c) Transition dipole spectral densities, $K(\omega)$. We show GCT and GNCT results with the solvent environment treated either as classical point charges (MM) or fully quantum mechanically up to a distance of 6~\AA\,(QM). All spectra in water have been scaled and shifted such that the absorption maxima of the computed results align with the experimental spectrum from Ref.~\onlinecite{PhenolateExp}.}
\label{fig:phenolate_water}
\end{figure*}

Spectral densities quantify the sensitivity of chromophore spectra to nuclear motions and, therefore, to the chromophore's solvation environment. Because our present example takes vacuum as the environment, all spectral densities exhibit thin, well-defined peaks associated with specific frequencies of molecular vibrations that can be easily assigned via normal mode analysis in the gas phase (see SI Sec.~\ref{SI-sec:spectral-density-normal-mode-analysis-GNCT-JCP}). For large molecules and chromophores in the condensed phase, where the harmonic approximation in the FCHT method becomes inapplicable, our MD-derived spectral densities become broader and report on more complex motions. We expect our GNCT to excel in such cases. 

Before turning to the condensed phase spectra of phenolate in cyclohexane and water, we first introduce a simple measure, the \textit{non-Condon factor}, to quantify the importance of including transition dipole fluctuations in calculated spectra, 
\begin{equation}
    \varphi = \frac{\gamma_\mu  - \lvert \langle \hat{\boldsymbol{\mu}}_{ge} \rangle \rvert^2}{ \gamma_\mu + \lvert \langle \hat{\boldsymbol{\mu}}_{ge} \rangle \rvert^2 }.
\end{equation}
Here $\gamma_\mu = \int_{0}^{\infty}\text{d}\omega \, K(\omega) $ quantifies the sensitivity of the transition dipole moment to nuclear motions that can shift intensity and render dark states bright. We note that $\varphi$ neglects non-Condon spectral contributions from $\mathbf{L}(\omega)$, but as we show in Sec.~\ref{section:beyond-physics},  $|\mathbf{L}(\omega)|< K(\omega)$, implying that $\gamma_{\mu}$ offers an upper-bound for the strength of correlation encoded by $\boldsymbol{\lambda}_{\mathbf{L}}$. Thus, $\varphi$ measures the relative importance of the average transition dipole to its nuclear motion-induced fluctuations. $\varphi$ ranges between -1 at the Condon limit where assuming a constant transition dipole is no longer an approximation, and 1 where non-Condon effects dominate spectra. For phenolate in vacuum, $\varphi = 0.66$, reiterating that non-Condon effects are important as the differences between the GCT and our GNCT spectra reveal.

We are now equipped to show our GNCT's ability to predict the spectra of solvated chromophores in general, and phenolate in cyclohexane and water in particular.  We begin by moderately increasing the solvent-chromophore interactions by considering the spectrum of phenolate in cyclohexane in Fig.~\ref{fig:phenolate_cyclo}. As one might expect, this non-polar solvent does not introduce appreciable broadening compared to the spectrum calculated in vacuum. In fact, the only notable difference between the two spectra is a minor blue shift of the spectrum in cyclohexane relative to the spectrum in vacuum. Similarly, the importance of non-Condon effects remains largely unchanged with $\varphi = 0.67$ for phenolate in cyclohexane.  

In water, a polar solvent expected to interact strongly with the phenolate anion, one can expect significant changes to the spectrum. To assess the effect of electronic polarization in the closest solvation shell, Fig.~\ref{fig:phenolate_water}a compares the experimental spectrum to the GCT and GNCT spectra obtained when treating the water up to a 6 \AA~radius around the chromophore environment either classically (MM) or quantum mechanically (QM) when constructing the required spectral densities (see SI Sec.~\ref{SI-section:computational-details-GNCT-JCP}). Here, the stronger solute-solvent interactions result in a broad and featureless spectrum that is qualitatively captured by both the GCT and GNCT at both levels of theory for the solvent. 

We find that the quantum mechanical treatment of the solvent leads to greater broadening. This is consistent with previous studies that show that electronic polarization in the form of electron density delocalization can reduce the energy gap and induce a self-screening of the transition dipole during photoexcitation, thus offering an additional spectral broadening mechanism~\cite{Zuehlsdorff2020b} not captured by the MM-treatment~\cite{Zuehlsdorff2016}. Compared to cyclohexane, the polarity of water increases the average transition dipole from $|\langle \hat{\boldsymbol{\mu}}_{ge}\rangle|=0.28$ D in cyclohexane to $|\langle \hat{\boldsymbol{\mu}}_{ge}\rangle|=1.90$ D in MM water and $|\langle \hat{\boldsymbol{\mu}}_{ge} \rangle|=1.60$ D in QM water. However, while the average transition dipole moment increases, the size of the fluctuations, quantified by $\gamma_\mu$, remains largely unchanged, increasing only by 8\%. Together, these result in an attenuation of non-Condon effects: $\varphi$ decreases from 0.67 in cyclohexane to -0.56 in MM water and -0.39 in QM water. Finally, whether the solvent is treated classically or quantum mechanically, the GNCT predicts broader spectra, bringing the results closer to the experimental spectrum. 

While strong phenolate-water interactions subsume detailed structure present in phenolate in cyclohexane/vacuum into broad spectral features, our GNCT bridges atomistic simulations with spectral densities and disentangles how such interactions tune a chromophore's spectrum. Our MM and QM water simulations yield $J(\omega)$ and $K(\omega)$ (see Figs.~\ref{fig:phenolate_water}b and c) that show qualitatively similar structures across these treatments, respectively. Although the most visible difference lies in the high-frequency regions of $J_{\rm MM} (\omega)$ and $J_{\rm QM} (\omega)$, their cumulative Huang-Rhys factors up to frequency $\omega$, $S^{\rm c}_{\rm HR} (\omega) = \frac{1}{\pi}\int_0^{\omega}\text{d} \Omega \, J(\Omega)/\Omega^2$ (inset of Fig.~\ref{fig:phenolate_water}), reveal that the most significant change impacting the optical response lies in the enhanced low-frequency region of the $\omega$-effective ${J_{\rm QM}(\omega)}/\omega^2$ over its MM counterpart. Since this region of the spectral density corresponds to the collective motions of the solvent, this enhancement signals that the QM treatment responds more sensitively than the MM treatment to motions that induce greater spectral broadening. For phenolate in QM water, self-screening decreases the average value of the transition dipole by 20\% while increasing its fluctuations by 12\%. This increases $\varphi$ from -0.56 to -0.38 and the energy gap reorganization energy, $\lambda_{\rm QM}/\lambda_{\rm MM} \sim 3$. The increase $\lambda$ leads to additional broadening from both Condon and non-Condon optical transitions. These results show that spectra can depend sensitively on solvent polarity and quantum mechanical polarizability and ultimately demonstrate that our GNCT has the ability to disentangle and assign the source of specific broadening mechanisms through the atomistic detail encoded by the spectral densities.

\section{Spectral interference in phenolate}
\label{section:phenolate-cross correlation-effects-GNCT-JCP}

An important feature that non-Condon fluctuations introduce is spectral interference, where individual components of the spectrum can redistribute spectral weight and even suppress otherwise connected regions in the observed spectrum. Non-Condon interference arises from $\langle \hat{\boldsymbol{\mu}}_{ge} \rangle \mdot \boldsymbol{\sigma_{\mathcal{A}}}(\omega)$, where $\boldsymbol{\sigma_{\mathcal{A}}} (\omega) \equiv [\boldsymbol{\lambda}_{\mathbf{L}} \delta(\omega) - \frac{\boldsymbol{\rho}_{\mathbf{L}}(\omega)}{\omega}]\ast \sigma_\smlsub{\rm GCT}(\omega)$, redistributing Condon-allowed intensities about the thermally averaged transition energy, $\omega_{eg}^{\rm av}$, whereas $\sigma_{\mathcal{AA}}(\omega) \equiv  [\boldsymbol{\lambda}_{\mathbf{L}} \delta(\omega) -  \frac{\boldsymbol{\rho}_{\mathbf{L}}(\omega)}{\omega} ] \ast \boldsymbol{\sigma_{\mathcal{A}}} (\omega)$ introduces destructive interference near $\omega_{eg}^{\rm av}$ and constructive interference to the red and blue (see Secs.~\ref{section:tight-tdp} and \ref{section:broad-tdp}, and SI~Sec.~\ref{SI-section:symmetries-of-spectral-spectral-interference-GNCT-JCP}). We now examine how the choice of solvent tunes these effects in the phenolate spectrum. 

\begin{figure}[t]
    \centering
    \includegraphics[width=1.0\columnwidth]{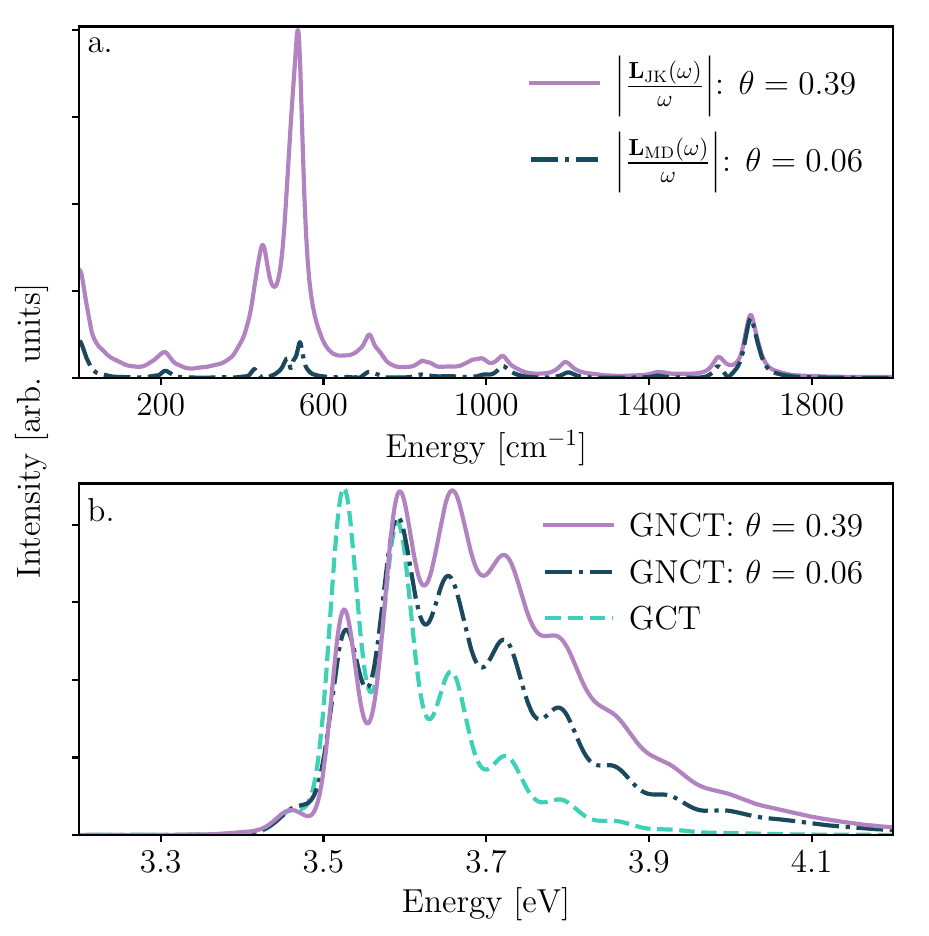}
    \caption{Spectral interference, quantified by $\theta$, and its effects on GNCT spectra. (a) MD-derived $|\mathbf{L}_{\rm MD}(\omega)|$ compared to the highly-correlated limit of $|\mathbf{L}_{\rm JK}(\omega)|=\sqrt{J_{\rm MD}(\omega)K_{\rm MD}(\omega)}$. Note that $\theta_{JK} < 1$ indicates that $J(\omega)$ and $K(\omega)$ are not perfectly correlated, and $\theta < \theta_{JK}$ indicates that the MD data predicts suboptimal correlation. (b) Effects of spectral interference on the GNCT spectra compared to the GCT for phenolate in cyclohexane. Here, $\theta=0.06$ arises from the MD-based $\mathbf{L}(\omega)$and $\theta=0.39$ arises from the highly correlated $|L_{\rm JK}(\omega)|$. The resulting GNCT spectrum was calculated using the purely anticorrelated $-|L_{\rm JK}(\omega)|$. }
    \label{fig:phenolate_gauss_corr}
\end{figure}

Since $\langle \hat{\boldsymbol{\mu}}_{ge} \rangle \mdot \boldsymbol{\sigma_{\mathcal{A}}}(\omega)$ and $\sigma_{\mathcal{AA}}(\omega)$ depend on $\mathbf{L}(\omega)$, the latter's magnitude tracks the importance of non-Condon spectral interference, including the likelihood of spectral peak splitting. Hence, we invoke the Cauchy-Schwartz inequality to introduce a measure, the \textit{correlated-interference factor}, to track the relative importance of $\mathbf{L}(\omega)$, 
\begin{equation}
    \theta = \frac{\lVert \mathbf{L}(\omega)\rVert^2}{\lVert J(\omega) \rVert \lVert K(\omega) \rVert},
\end{equation}
where $\lVert X(\omega) \rVert = \left( \int_0^{\infty}\text{d}\, \omega\ |X(\omega)|^2 \right)^{1/2}$ is the norm of the square-integrable spectral density, $X(\omega)$.
Within our Gaussian theory, $\theta$ is bounded between 0, in which the $J(\omega)$ and $K(\omega)$ do not overlap and lead to $|\mathbf{L}(\omega)|=0$, and 1, where the fluctuations of the energy gap and the transition dipole are perfectly correlated or anticorrelated. This latter case occurs when the nuclear motions causing cross correlated fluctuations of the energy gap and transition dipole follow Gaussian statistics exactly and results in, $J(\omega)$, $K(\omega)$, and $|\mathbf{L}(\omega)|$ sharing the same shape. Thus, the form of $|\mathbf{L}(\omega)|$ that would lead to the greatest possible correlation given two arbitrary shapes for $J(\omega)$ and $K(\omega)$ is $|\mathbf{L}_{\rm JK}(\omega)| = |\sqrt{J(\omega)K(\omega)}|$. Phenolate displays only minor energy gap-transition dipole correlations, with $\theta = 0.06$ in both vacuum and cyclohexane and $0.06$ in both MM and QM water. This implies that for phenolate, non-Condon fluctuations can be expected to only induce additional broadening even at the low values of $\varphi$ in water, as can be confirmed in Figs.~\ref{fig:phenolate_methods}, \ref{fig:phenolate_cyclo}, and \ref{fig:phenolate_water}.

While our MD simulations of phenolate in different solvents reveal weak cross correlation between energy gap and transition dipole fluctuations in $\mathbf{L}(\omega)$, one may ask how enhanced cross correlation could affect spectra. To interrogate this question, we employ the optimally correlated form for the cross correlation spectral density, $\mathbf{L}_{\rm JK}(\omega)$, which yields $\theta = 0.39$. The fact that $\theta<1$ implies that the shapes of $J(\omega)$ and $K(\omega)$ are different, as is evident in Fig.~SI \ref{SIfig:phenol_cyclohex_sds}. Figure~\ref{fig:phenolate_gauss_corr}a shows $|\mathbf{L}_{\rm MD}(\omega)/\omega|$ for phenolate in cyclohexane (black, dashed) with $\theta=0.06$ and a scaled $|\mathbf{L}_{\rm JK}(\omega)/\omega|$ (purple) setting the maximum possible value of $\theta = 0.39$. This modification leads to significant weight in the low-frequency region associated with strong overlap between peaks in $J(\omega)$ and $K(\omega)$. Figure~\ref{fig:phenolate_gauss_corr}b compares GCT and GNCT spectra calculated with $\theta = 0.06$ and $0.39$, respectively, with lineshapes that differ dramatically. Interestingly, the scaled but perfectly anti-correlated GNCT-spectrum (purple) shows a \textit{enhanced linewidth} compared to both the GCT (teal) and MD-derived GNCT (black, dashed) spectra, with most spectral intensity centering at $\sim 3.7$ eV, albeit with a large and slow decaying high energy tail with reduced traces of vibronic peaks. Hence, these results illustrate how energy gap and transition dipole-correlated non-Condon fluctuations can qualitatively change spectra through $\boldsymbol{\sigma_{\mathcal{A}}}(\omega)$ and $\sigma_{\mathcal{AA}}(\omega)$.

\section{Spectral regimes at parameter extrema}\label{sec:parameter-regimes-at-extrema}

In Secs.~\ref{section:phenolate-solvent-effects} and \ref{section:phenolate-cross correlation-effects-GNCT-JCP}, we leveraged our diagnostic parameters, the non-Condon factor $\varphi$ and the correlated-interference factor $\theta$, which characterize spectral lineshapes beyond the Condon limit, to quantify the relative importance of non-Condon fluctuations and spectral interference in phenolate in various solvents. We now employ physically inspired spectral densities to interrogate the wealth of spectral behaviors that emerge at the physical limits of $\varphi$ and $\theta$. Specifically, we use a Debye spectral density to capture collective motions in the condensed phase and incorporate intramolecular vibronic couplings through high-frequency Gaussian profiles. We base all model spectral densities on the spectral densities of phenolate, where we enhance or attenuate contributions from specific motions. For simplicity and ease in tuning the correlated interference, we employ a scalar-valued $L_{\rm JK} (\omega) = \pm \sqrt{J(\omega) K(\omega)}$, implying that energy gap and transition dipole fluctuations are perfectly correlated or anti-correlated. We then analyze the limits where the variance of the fluctuations of the transition dipole dominate or become negligible and where interference effects play a large role, as these are the limits that deviate most markedly from the conventional Condon limit. 

\subsection{Spectral shifts in the low variance limit}
\label{section:tight-tdp}

This limit arises when the average transition dipole moment dominates over the variance of the fluctuations, $\lvert \langle \hat{\boldsymbol{\mu}}_{ge} \rangle \rvert^2 \gg \langle \delta \mu^2 \rangle$, leading $\varphi$ to approach -1. We investigate how spectra change as a function of the energy gap-transition dipole correlation from $\theta \approx 0$ to $\theta \approx 0.5$. In such cases, a balance of Condon and non-Condon effects controls the spectral response, and spectral interference shifts intensity to the red or blue of $\omega_{eg}^{\rm av}$, and exactly at $\omega_{eg}^{\rm av}$ when $\theta = 1$. In the limit where $\theta = 1$, all spectral densities have the same structure (See SI Sec.~\ref{SI-sec:model-spectral-density-build-GNCT-JCP}). Setting $\varphi = -0.35$ for a simpler analysis, the absorption spectrum becomes, 
\begin{equation}\label{eq:tight-transition-dipoles}
\begin{split}
    \sigma_\smlsub{\rm GNCT}(\omega) &\approx  |\langle \hat{\boldsymbol{\mu}}_{ge} \rangle|^2 \sigma_\smlsub{\rm GCT}(\omega) \\&\quad +
    \begin{cases}
      0  &  \text{for} \, \theta=0, \\ 2 \text{Re}\{\ev{\hat{\boldsymbol{\mu}}_{ge}}\}\mdot \boldsymbol{\sigma_{\mathcal{A}}} (\omega)  & \text{for} \, \theta>0,
    \end{cases}
\end{split}
\end{equation}
with $\boldsymbol{\sigma_{\mathcal{A}}} (\omega) \equiv [\boldsymbol{\lambda}_{\mathbf{L}} \delta(\omega) - \frac{\boldsymbol{\rho}_{\mathbf{L}}(\omega)}{\omega}]\ast \sigma_\smlsub{\rm GCT}(\omega)$ being the spectral contribution from $\boldsymbol{\mathcal{A}}(t)$ in Eq.~\eqref{eq:nC-linear-optical-response-closed}.

\begin{figure}[t]
    \centering
\includegraphics[width=1\columnwidth]{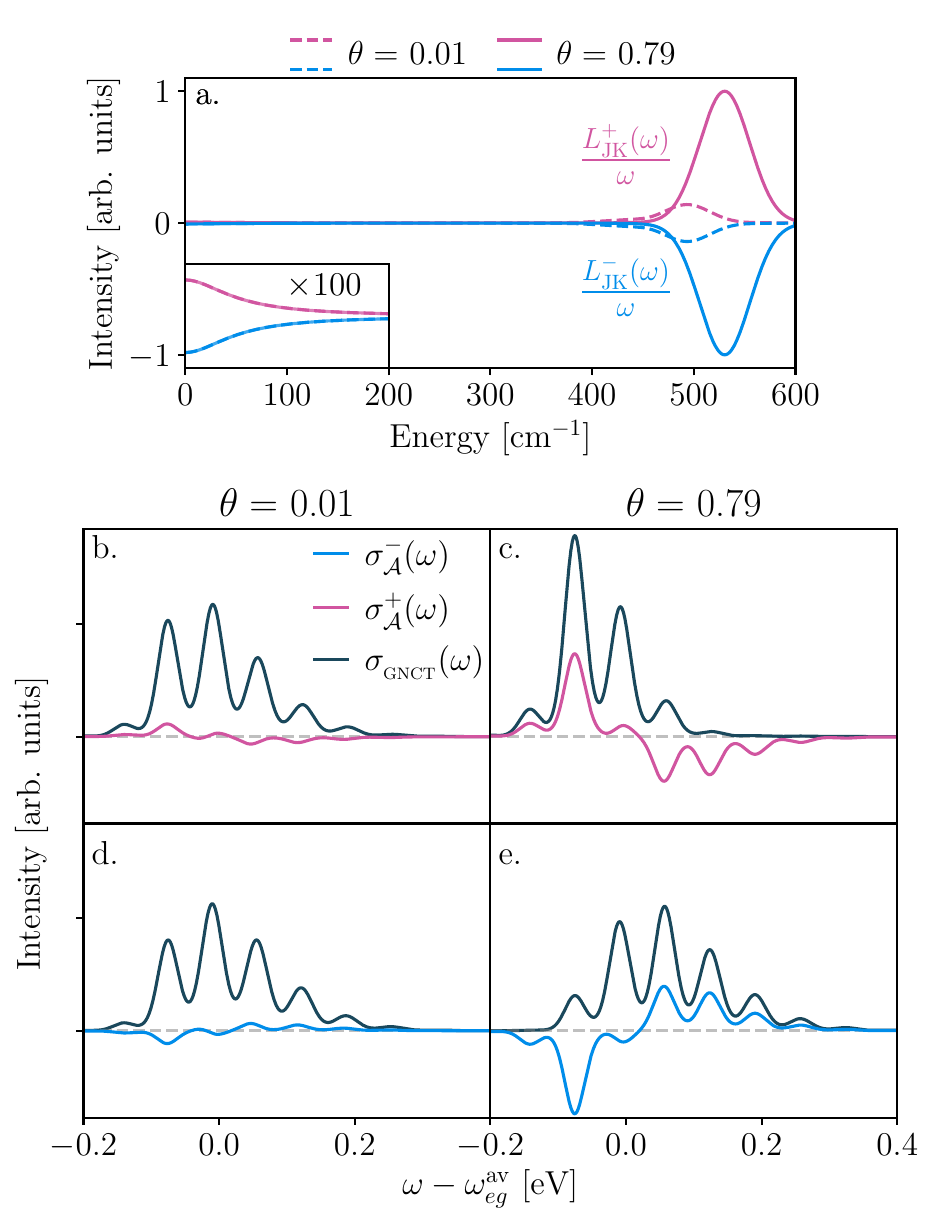}
\caption{The impact of spectral interference with increasing $\theta$ (left to right). Here, interference causes a red-shift ($\sigma^{+}_{\mathcal{A}}(\omega)$) or blue-shift ($\sigma^{-}_{\mathcal{A}}(\omega)$) of spectral intensity depending on whether $\langle \hat{\boldsymbol{\mu}}_{ge} \rangle \cdot \mathbf{L}(\omega)$ is positive or negative, respectively.}
\label{fig:linear-interference-GNCT}
\end{figure}

We find that $\boldsymbol{\sigma}_{\mathcal{A}}(\omega)$ causes intensity to either red- or blue-shift, changing the overall intensity distribution when compared to the GCT spectrum. To illustrate how $\sigma_{\mathcal{A}}(\omega)$ can cause these opposing effects in the full GNCT spectrum, we analyze the structures of model spectral densities that are responsible for red and blue shifts. 

\begin{figure*}
    \centering
\includegraphics[width=.8\textwidth]{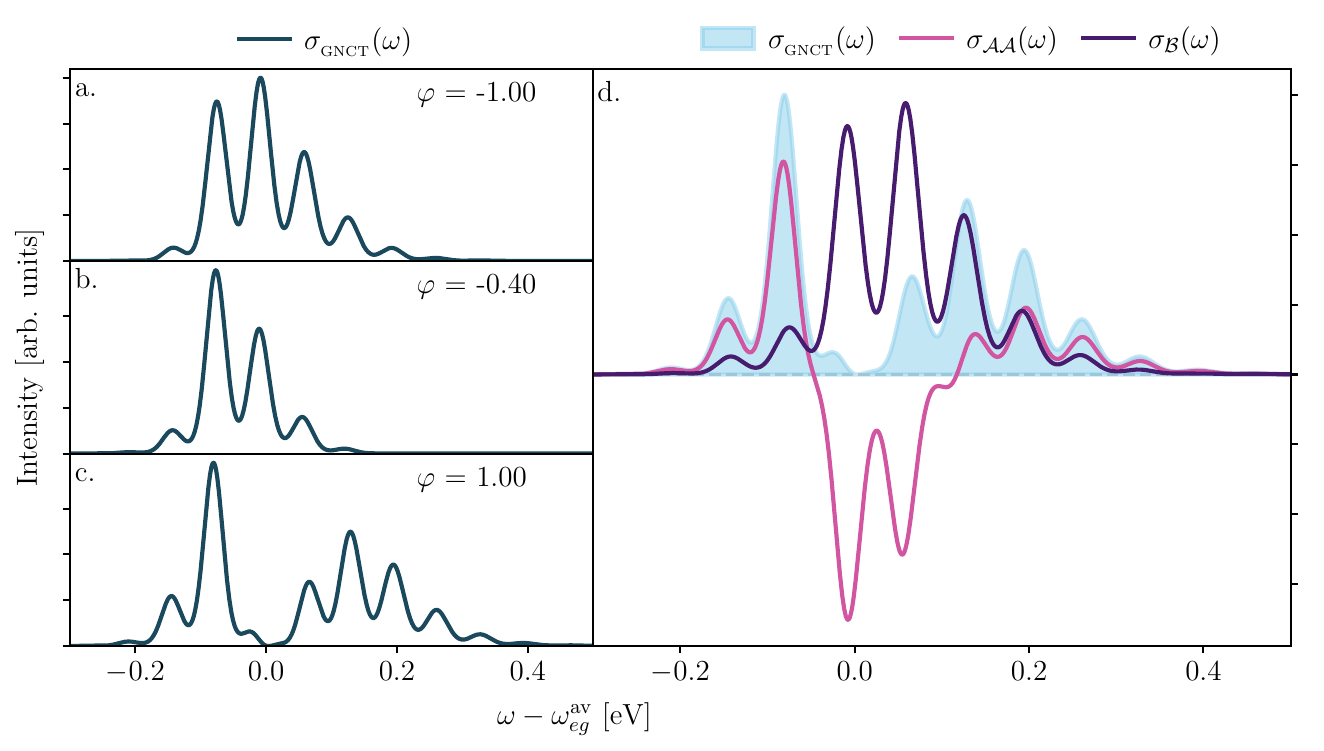}
\caption{Peak splitting can arise as a result of spectral interference as non-Condon effects become more prominent. Here, $\theta=1$ in all cases. (a)-(c) Increasing $\gamma_\mu$ causes non-Condon contributions from $\sigma_{\mathcal{B}}(\omega)$ and $\sigma_{\mathcal{AA}}(\omega)$ dominate the spectrum. (d) Interplay between $\sigma_{\mathcal{B}}(\omega)$ and $\sigma_{\mathcal{AA}}(\omega)$ cause $\sigma_\smlsub{\rm GNCT}(\omega)$ to display two distinct vibronic progressions from the same electronic transition. }
\label{fig:peak_splitting_interference_GNCT}
\end{figure*}

Since the spectral contribution from $\boldsymbol{\sigma}_{\mathcal{A}}(\omega)$ also depends on the vector alignment (dot product) between $\langle \hat{\boldsymbol{\mu}}_{ge} \rangle$ and $\mathbf{L}(\omega)$, in Fig.~\ref{fig:linear-interference-GNCT}a, we consider two types of model spectral densities: $L^+_{\rm JK} (\omega)$ (pink) which aligns parallel to $\langle \hat{\boldsymbol{\mu}}_{ge} \rangle$ and $L^-_{\rm JK} (\omega) = - L^+_{\rm JK} (\omega)$ (blue) which aligns anti-parallel to $\langle \hat{\boldsymbol{\mu}}_{ge} \rangle$. We construct these spectral densities inspired by the peak structures of $J(\omega)$ and $K(\omega)$ for phenolate in cyclohexane. While we assign low-frequency regions to Debye spectral densities that remain the same for both $J(\omega)$ and $K(\omega)$ models, we tune the magnitude of high-frequency peaks in $K(\omega)$ to interrogate the novel spectral features that our GNCT predicts. 

Before turning to the resulting spectra, we comment on how we employ our phenolate data to construct these spectral densities. For $J(\omega)$, we reproduce the peak at $\sim 500 \wn$ corresponding to a symmetric in-plane ring deformation using a Gaussian profile. We model the two peaks in $K(\omega)$ using an identical Gaussian profile to that for $J(\omega)$ and scaled these to obtain $\varphi = -0.35$. We  then tune the correlated interference encoded in $L^{\pm}_{\rm JK} (\omega)$, obtaining values of $\theta = 0.01$ and $0.79$, by using the peaks at $\sim 450 \wn$ and $\sim 500 \wn$ in $K(\omega)$, which correspond to symmetric out-of-plane ring deformations. We then re-center the peak in $K(\omega)$ from $\sim 450 \wn$ to $\sim 500 \wn$ to align with the peak in $J(\omega)$. This causes the center frequency of the high-frequency peak in $L^{\pm}_{\rm JK} (\omega)/\omega$ to shift and its amplitude to increase. This results in an increase in $\theta$ from $0.06$ to $\sim 0.56$ with the increasing overlap between $J(\omega)$ and $K(\omega)$. 

Figure~\ref{fig:linear-interference-GNCT}b-e reveals how spectral interference from our model $L^{\pm}_{\rm JK}(\omega)$ controls lineshapes. As one scans across each row with increasing $\theta$, one observes the effects of interference in $\sigma_\smlsub{\rm GNCT}(\omega)$ (dark green) becoming more prominent. This spectral interference manifests through underlying contributions from $\sigma^+_{\mathcal{A}}(\omega)$ (pink) that cause a red shift in going from panels a to b with increased magnitude of $L^{+}_{\rm JK}(\omega)$ and from $\sigma^-_{\mathcal{A}}(\omega)$ (blue) that cause a blue shift in going from panels c to d with increased magnitude of $L^{-}_{\rm JK} (\omega)$ (see SI Sec.~\ref{SI-section:spectrum-convolution-GNCT-JCP} for more details). Panels c and d uncover how the sign of $\langle \hat{\boldsymbol{\mu}}_{ge} \rangle \cdot \mathbf{L}(\omega)$ can reduce or contribute spectral broadening. Since $\mathbf{L}(\omega)$ generally contains regions of positive or negative intensity, our results demonstrate that regions where $\langle \hat{\boldsymbol{\mu}}_{ge} \rangle\cdot\mathbf{L}(\omega)>0$ \textit{red-shift spectral intensity and reduce linewidths}, whereas regions where $\langle \hat{\boldsymbol{\mu}}_{ge} \rangle\cdot\mathbf{L}(\omega)<0$ \textit{blue-shift and broaden spectra}. 
Thus, these results illustrate how non-Condon effects arising from correlated nuclear modulations of the energy gap and transition dipole can control spectral linewidths through spectral interference based on the relative phases between $\mathbf{L}(\omega)$ and $\langle \hat{\boldsymbol{\mu}}_{ge} \rangle$.
\subsection{Spectral splitting in the high variance limit}
\label{section:broad-tdp}

This regime corresponds to cases where the ratio of the thermal average and variance of the transition dipole becomes small, $|\langle \hat{\boldsymbol{\mu}}_{ge} \rangle|^2/\langle \delta \hat{\mu}_{ge}^2 \rangle \ll 1$ and $\varphi$ approaches unity as $|\langle \hat{\boldsymbol{\mu}}_{ge} \rangle|$ approaches 0. Here, spectra arise from optical transitions that are, on average, transition dipole forbidden. In this regime, we identify two separate mechanisms that cause single electronic excitation signals to split into apparently distinct transitions based on the strength of spectral interference. In this case, the spectrum becomes, 
\begin{equation}
    \sigma_\smlsub{\rm GNCT}(\omega) \approx 
    \begin{cases}
      \sigma_{\mathcal{B}}(\omega) \ &\textrm{for}\ \theta=0, \\
      \sigma_{\mathcal{B}}(\omega) + \sigma_{\mathcal{A} \mathcal{A}} (\omega)\ &\textrm{for}\ \theta>0,
    \end{cases}
\end{equation}
with $\sigma_{\mathcal{B}}(\omega) \equiv \rho_{K}(\omega) \ast \sigma_\smlsub{\rm GCT}(\omega)$ being the spectral contributions from $\mathcal{A}^2(t)$ and $\mathcal{B}(t)$ in Eq.~\eqref{eq:nC-linear-optical-response-closed}, respectively.

We begin in the limit of weak energy gap-transition dipole correlation, $\theta = 0$ and $\varphi=1$, where the GNCT spectrum further reduces to, $\sigma(\omega) = \rho_K(\omega) \ast \sigma_{\rm GCT}(\omega)$. Although a simple expression, this limit can exhibit rich spectral behavior. Specifically, because this limit convolves the Condon (GCT) lineshape with non-Condon fluctuation term, $\rho_K(\omega)$, the detailed shape of the latter can broaden and even split the total spectrum. 

To see how and when such spectral splitting can arise from the structure of $\rho_K(\omega)$, we turn to the transition dipole spectral density, $K(\omega)$. For example, $K(\omega)$ can arise from timescale-separated nuclear motions that modulate transition dipole fluctuations, such as slow collective motions versus fast intramolecular vibrations. In such cases, one can decompose $K(\omega)$ into (largely) non-overlapping spectral densities, $K_n(\omega)$, centered at the characteristic frequencies, $\omega_n$, of such nuclear motions. One can then construct the individual spectra that arise from each contribution of $K_n(\omega)$, revealing that the total spectrum is the sum of individual components, $\sigma_\smlsub{\rm GNCT}(\omega) = [ \sum_n \rho_{K_n} (\omega)] \ast \sigma_{\rm GCT} (\omega)$. The total spectrum arising in this parameter limit is thus a sum over purely positive GCT spectra with amplitudes and widths determined by their convolution with $\rho_{K_n} (\omega)$ that are displaced from their original center-frequency by $\langle \omega \rangle_{\rho_{K_n}}$ (See SI Sec.~\ref{SI-sec:spectral-interference-GNCT-JCP}). A remarkable consequence occurs when a chromophore displays timescale-separated transition dipole fluctuations --- or regions of near-zero spectral weight in $K(\omega)$ separating peaks --- that exceed the characteristic width of the GCT-lineshape (i.e., $\langle \omega \rangle_{\rho_{K_{n+1}}} - \langle \omega \rangle_{\rho_{K_n}} > \langle \delta U^2 \rangle^{\nicefrac{1}{2}}$): peaks arising from a single electronic transition appear to split. Indeed, as we have shown in Ref.~\onlinecite{Wiethorn2023}, this type of timescale separation in the nuclear motions that modulate the transition dipole is the cause of the splitting of the Q-band region in free-base porphyrins.  

In Fig.~\ref{fig:peak_splitting_interference_GNCT}, we consider the limit of strong energy gap-transition dipole correlation, $\theta = 1$ and $\varphi=1$, where spectral splitting arises from spectral interference. As one moves down panels a-c, $\varphi$ increases from -1 to 1, introducing a larger negative contribution from $\sigma_{\mathcal{AA}}(\omega)$ that competes with the purely positive contribution from $\sigma_{\mathcal{B}}(\omega)$. In panel c, spectral interference is at its strongest and splits the original absorption spectrum from panel a into two distinct regions about $\sim \omega_{eg}^{\rm av}$. Figure~\ref{fig:peak_splitting_interference_GNCT}d shows both the $\sigma_{\mathcal{AA}}(\omega)$ (pink) and $\sigma_{\mathcal{B}}(\omega)$ (purple) contributions that split the total spectrum (shaded in blue) shown in Figure~\ref{fig:peak_splitting_interference_GNCT}c. While $\sigma_{\mathcal{B}}(\omega)$ provides the majority of the spectral intensity to the blue of the splitting, $\sigma_{\mathcal{A}\mathcal{A}}(\omega)$ primarily provides positive intensity to the red of the split and a significant negative contribution centered around $\omega_{eg}^{\rm av}$ that ultimately causes the split. 

This analysis thus provides fundamental insight into two distinct mechanisms of spectral splitting that our GNCT reveals. The first arises from timescale-separated nuclear motions that modulate the transition dipole while the second arises from spectral interference in the strongly correlated limit. 

\subsection{Negative spectral weight in the overcorrelated limit}
\label{section:beyond-physics}

This limit arises when the atomistic simulations predict an $|\mathbf{L}_{\rm MD}(\omega)|$ that presents stronger correlation than could be expected from $J(\omega)$ and $K(\omega)$, leading to a value of $\theta$ that is greater than that obtained from $|\mathbf{L}_{\rm JK} (\omega)|$. Such cases are unphysical and indicate a breakdown of our Gaussian theory where spectral interference exceeds the contribution of purely positive terms in $\sigma_\smlsub{\rm GNCT} (\omega)$, resulting in spectra with negative spectral weight. We provide a metric to track this overestimation, $r_{\theta} = \theta_{\rm MD} / \theta_{\rm JK}$, with a physically reasonable range between 0 and 1.

\begin{figure}[t]
    \centering
\includegraphics[width=1.0\columnwidth]{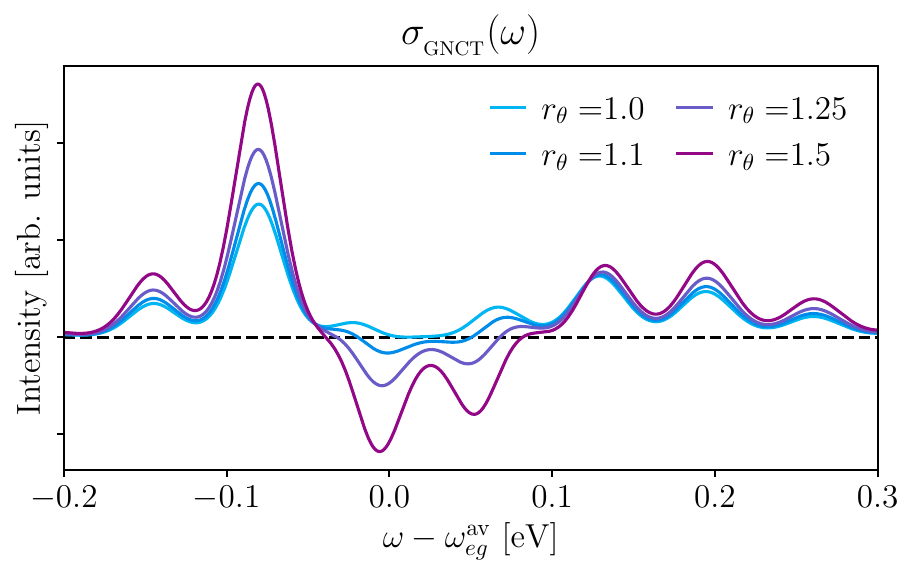}
\caption{GNCT spectra display unphysical (negative) intensity as $r_\theta$ exceeds unity. This negative intensity arises from the overestimated ability (cross correlation) of nuclear motions to simultaneously modulate transition dipoles and energy gaps compared to their ability to modulate these separately.}
\label{fig:unphysical-cross correlation}
\end{figure}

We interrogate the impact of unphysical values of $r_{\theta}$ on spectra in Fig.~\ref{fig:unphysical-cross correlation}. This figure shows spectra predicted by our GNCT with values of $r_{\theta}$ from 1 to 1.5. We begin from the same spectrum in Fig.~\ref{fig:peak_splitting_interference_GNCT}c, for $\theta = 1$ and $\varphi = 1$, then scale the $\sigma_{\mathcal{AA}(\omega)}$ term to increase the $r_{\theta}$ value. Here, as $r_{\theta}$ increases above 1, the GNCT-spectra dip below zero (denoted by the black line), indicating that $r_\theta = 1$ provides an upper bound for physically valid spectra.

\section{Conclusion}

In summary, we introduce a novel Gaussian theory (GNCT) to simulate the optical spectra of chromophores in complex condensed phase environments beyond the Condon approximation. Our theory exactly captures the quantum dynamics of electronic excitations and their nuclear motion-induced fluctuations while offering a simple protocol to parameterize the theory based on atomistic MD simulations and electronic structure calculations that faithfully represent molecular structure, interactions, and motions. Our approach has a negligible computational overhead compared to the state-of-the-art second-order cumulant within the Condon approximation (GCT) and yields excellent agreement with benchmarks for phenolate in vacuum and cyclohexane, and experiments for phenolate in water. By formulating our theory in terms of Gaussian fluctuations about the thermal averages of the energy gap and transition dipole rather than low-order Taylor expansions of the PESs and the transition dipole around reference nuclear configurations (as is the case with microscopically harmonic theories, such as the FCHT method), our theory works for generally anharmonic PESs and everywhere from weak to strong non-Condon fluctuations. Our GNCT yields three distinct spectral densities that offer physical insight into the nuclear motions that determine the optical spectrum by modulating the energy gap, transition dipole, or both. In addition, our GNCT reveals two unique mechanisms for emergent peak splitting from a single electronic transition: (i) separation of timescales in the nuclear modes that modulate the transition dipole moment and (ii) spectral interference arising from correlated fluctuations in the energy gap and transition dipole moment. To aid in spectral analysis, we have introduced the non-Condon and cross-interference factors, which quantify the importance of non-Condon effects and the likelihood of spectral interference associated with spectral peak-splitting. We have also leveraged physically motivated spectral densities to interrogate the range of behaviors that our GNCT yields in the limits of weak versus strong non-Condon and interference effects. Our GNCT thus offers a robust, physically transparent, and efficient tool to simulate optical spectra of complex chromophores in the condensed phase from first principles while fully accounting for the influence of dark transitions and environmental polarization and broadening effects.

\section{Supplementary Material}
See supplementary material for a detailed derivation of the GNCT developed in this work, computational details on FCHT and MD-based GNCT calculations, normal mode assignments of peaks in the phenolate spectral densities, and model spectral density analysis illustrating how spectral interference effects can cause intensity shifts and peak splittings in computed linear spectra.

\begin{acknowledgments}
A.~M.~C.~was supported by an Early Career Award in CPIMS program in the Chemical Sciences, Geosciences, and Biosciences Division of the Office of Basic Energy Sciences of the U.S.~Department of Energy under Award DE-SC0024154. 
T.~J.~Z.~acknowledges support through startup funding provided by Oregon State University. 
\end{acknowledgments}

\section{Data availability}
Raw data for all atomistic simulations involving the phenolate ion in vacuum and in solution, including input files for QM/MM MD simulations, computation of vertical excitation energies using TDDFT, and the calculation of normal modes for FCHT spectra, are available under the following persistent URI: 10.5281/zenodo.8292626. The same URI also contains raw data for all spectra and spectral densities for phenolate presented in the main text and the SI and all input files necessary to reproduce the spectra using the MolSpeckPy package, a spectroscopy python package freely available on GitHub (https://github.com/tjz21/Spectroscopy\_python\_code).

%

\clearpage

\setcounter{section}{0}
\setcounter{equation}{0}
\setcounter{figure}{0}
\setcounter{table}{0}
\setcounter{page}{1}

\renewcommand{\theequation}{S\arabic{equation}}
\renewcommand{\thefigure}{S\arabic{figure}}
\renewcommand{\thepage}{S\arabic{page}}
\renewcommand{\bibnumfmt}[1]{$^{\mathrm{S#1}}$}
\renewcommand{\citenumfont}[1]{S#1}

\title{Supplemental material for ``Breaking through the Condon limit: Theory of optical spectroscopy in the condensed phase from atomistic simulations"}
{\maketitle}

\onecolumngrid

\section{Derivation of the linear optical response functions for the non-Condon Brownian oscillator model}\label{SI-section:non-Conondon-optical-response-derivation-via-path-integrals-GNCT-JCP}

Here, we employ exact path integral techniques to derive analytic expressions for the various components of the total linear optical response function introduced in the main text (Eqs.~\eqref{eq:A-term-from-path-integral-GNCT-JCP} and \eqref{eq:B-n-AA-term-from-path-integral-GNCT-JCP}) for our non-Condon Brownian Oscillator model. Importantly, we provide the final response functions in terms of a family of spectral densities that can be easily constructed from classical MD simulations combined with electronic structure calculations of the vertical energy gaps, $\omega_{eg}^0$, and the transition dipoles. 

For simplicity of notation, we introduce the position matrix element, connecting states $\bra{x}$ and $\ket{y}$, of the imaginary time propagator for mass-weighted harmonic oscillators \cite{FeynmanHibbsPathIntegral-sup, schulman2012techniques-sup} in which we set $\hbar =1$,
\begin{equation}\label{harmonic_path_integral}
\begin{split}
    \bra{x}\erm^{-\beta \opr{H}}\ket{y} &=\sqrt{\frac{\omega}{2\pi\sinh(\beta\omega)}}\exp\lr{\{}{-\frac{\omega}{2\sinh(\beta\omega)}\Big[(x^2+y^2)\cosh(\beta\omega)-2xy\Big]}{\}}\\ &=\sqrt{\frac{\omega}{2\pi\sinh(\beta\omega)}}\exp\lr{\{}{-\frac{\omega}{4}\left[\frac{(x+y)^2}{\coth(\beta\omega/2)}+\frac{(x-y)^2}{\tanh(\beta\omega/2)}\right]}{\}}.
\end{split}
\end{equation}

\subsection{Response functions capturing Condon and non-Condon effects}

We turn our attention to obtaining an analytical expression for the Condon (C), non-Condon (NC), and mixed-C/NC contributions to the total linear optical response function (Eq.~(14) in the main text), 
\begin{equation} \label{eq:non-condon-optical-response-fxn}
\begin{split}
    \chi^{(1)}_\smlsub{\rm GNCT}(t) &=\left|\ev{\tdpo{ge}}\right|^2\ev{\opr{X}(t)}+ \ev{\tdpo{eg}}\mdot \ev{\delta \hat{\boldsymbol{\mu}}_{ge}(t)\opr{X}(t)}\\ &\hspace{.5cm}+ \ev{\tdpo{ge}}\mdot\ev{\opr{X}(t) \delta \tdpo{eg}} + \ev{\delta \tdpo{ge}(t) \opr{X}(t) \delta \hat{ \boldsymbol{\mu}}_{eg}}
    \\&\equiv \chi^{_{\text{C}}}(t) + 
    \chi_{_\text{NC}}^{_{\text{C}}}(t) + \chi^{_\text{NC}}(t),
\end{split}
\end{equation}
with,
\begin{subequations}\label{fundamental_TCF_functions}
\begin{align}
    \chi^{_{\text{C}}}(t) &= \left|\ev{\tdpo{ge}}\right|^2\ev{\opr{X}(t)}\label{C-full-object},\\
    \chi_{_\text{C}}^{_{\text{NC}}}(t)& =\ev{\tdpo{eg}}\mdot\ev{\delta\opr{\boldsymbol{\mu}}_{ge}(t)\opr{X}(t)}+ \ev{\tdpo{ge}}\mdot\ev{\opr{X}(t)\delta\tdpo{eg}},\label{CNC-full-object}\\
    \chi^{_\text{NC}}(t) &=\ev{\delta\tdpo{ge}(t)\opr{X}(t)\delta\opr{\boldsymbol{\mu}}_{eg}}\label{NC-full-object}.
\end{align}
\end{subequations}
Here $\ev{\cdots}\equiv Z^{-1}\mathrm{Tr}_{\rm nuc}[\erm^{-\beta H_g} ...]$ denotes an equilibrium average with respect to the ground state PES, $H_g$, and $Z = \mathrm{Tr}_{\rm nuc}[\erm^{-\beta H_g}]$. In Eq.~\eqref{fundamental_TCF_functions}, we summarize all dynamical objects required by our theoretical treatment of non-Condon effects. We note that the respective correlation functions composing the mixed-C/NC response, $\chi_{_\text{C}}^{_{\text{NC}}}(t)$, are vector-valued quantities, but the overall contribution of the mixed-C/NC term is a scalar. 

Noting that in our non-Condon Brownian Oscillator model, all harmonic oscillators are independent and noninteracting, we can re-express the above response functions into sums and products over the distinct oscillators. For instance, propagators (in real or imaginary time) can be decomposed into a product over particle-specific propagators, 
\begin{equation}
    \erm^{-\beta H_g} = \prod_{k}\erm^{-\beta H_{g,k}},
\end{equation}
meaning that 
\begin{subequations}
\begin{align}
    \opr{X}(t) &= \prod_{k} \opr{X}_k(t),\\
    \frac{\erm^{-\beta H_g}}{Z} &= \prod_{k} \frac{\erm^{-\beta H_{g, k}}}{Z_k}.
\end{align}
\end{subequations}
In turn, the fluctuations of the transition dipoles can also be written as a sum over independent oscillators (Eq.~\eqref{eq:statistical_dipole_GNCT_JCP} in the main text). This allows us to rewrite the response functions in
Eq.~(\ref{fundamental_TCF_functions}) as products of response functions corresponding to individual oscillators,
\begin{subequations}\label{fundamental_TCF_functions-per-oscillator}
\begin{align}
    \ev{\opr{X}(t)} &= \prod_k\ev{\opr{X}_k(t)}_k, \label{C-full-object-decomp}\\
    \ev{\delta\opr{\boldsymbol{\mu}}_{ge}(t)\opr{X}(t)} 
    &= \ev{\opr{X}(t)} \sum_{k} \boldsymbol{\alpha}_k \ev{\opr{q}_k (t)\opr{X}_k(t)}_k \ev{\opr{X}_k(t)}_k^{-1}, \label{CNC-full-object-1-decomp} \\
    \ev{\opr{X}(t)\delta\tdpo{eg}} 
    &= \ev{\opr{X}(t)}\sum_{k} \boldsymbol{\alpha}_k \ev{\opr{X}_k(t)\opr{q}_k (0)}_k  \ev{\opr{X}_k(t)}_k^{-1}, \label{CNC-full-object-2-decomp} \\
    \ev{\delta\tdpo{ge}(t)\opr{X}(t)\delta\opr{\boldsymbol{\mu}}_{eg}} 
    &= \ev{\opr{X}(t)} \Bigg[\sum_{k} \left|\boldsymbol{\alpha_k}\right|^2 \ev{\opr{q}_k(t) \opr{X}_k(t) \opr{q}_k (0)}_k  \ev{\opr{X}_k(t)}_k^{-1} \label{NC-full-object-decomp} \\
    & \hspace{.5cm}+ \sum_{\substack{j,k \\ j \neq k}} \boldsymbol{\alpha}_k \mdot \boldsymbol{\alpha}_j \ev{\opr{q}_k (t)\opr{X}_k(t)}_k \ev{\opr{X}_j(t)\opr{q}_j (0)}_j \ev{\opr{X}_k(t)}_k^{-1}\ev{ \opr{X}_j(t)}^{-1}_{j}\Bigg].  \nonumber
\end{align}
\end{subequations}
Equation~\eqref{fundamental_TCF_functions-per-oscillator} shows that one needs to calculate only one-oscillator dynamical quantities to encode the total non-Condon linear optical response function,

\begin{subequations}\label{single_oscillator_TCFs}
\begin{align}
    \tilde{\chi}^{_\text{C}} (t) &\equiv \ev{\opr{X}_k (t)}_k, \label{C-single-oscillator} \\
    \tilde{\chi}^{_\text{NC}}_{_\text{C}}(t) &\equiv \ev{\opr{q}_k (t) \opr{X}_k (t)}_k, \label{CNC-single-oscillator-1} \\
    \tilde{\chi}'^{_\text{NC}}_{_\text{C}}(t) & \equiv \ev{\opr{X}_k(t) \opr{q}_k(0)}_k, \label{CNC-single-oscillator-2} \\
    \tilde{\chi}^{_\text{NC}}(t) &\equiv \ev{\opr{q}_k(t) \opr{X}_k(t) \opr{q}_k(0)}_k. \label{NC-single-oscillator}
\end{align}
\end{subequations}
We now turn to deriving closed-form expressions for each dynamical object. 

\subsection{Derivation of one-oscillator dynamical objects}
\label{SI-sec:derivation-of-one-oscillator-dynamical-objects}

We set to derive closed expressions for the dynamical objects in Eq.~\eqref{single_oscillator_TCFs}. We begin with Eq.~\eqref{C-single-oscillator}, corresponding to the $k^{th}$ time-ordered exponentiated energy gap,
\begin{equation}\label{C_TCF_solo}
\begin{split}
    \tilde{\chi}^{_\text{C}} (t) &= \ev{\opr{X}_k (t)}_k\\
    &= Z^{-1}_k \text{Tr}_{{\rm nuc}, k} \left[\erm^{-\beta_g \opr{H}_{g,k}} \erm^{-\beta_e \opr{H}_{e,k}} \right],
\end{split}
\end{equation}
where,
\begin{equation}
    Z_k = \frac{1}{2 \sinh(\beta \omega_k / 2)},
\end{equation}
and 
\begin{subequations}
\begin{align}
    \beta_g &= \beta - it \\
    \beta_e &= it.
\end{align}
\end{subequations}
While a Gaussian treatment of this term, which amounts to the spectrum in the absence of non-Condon effects, has already been explained in previous work \cite{mukamel1985fluorescence-sup, Mukamel-book-sup}, we re-introduce its solution here to set the notation that we use in our subsequent derivations. Since we have formally established that one needs only to focus on the $k^{th}$ response function characteristic to each component of the linear optical response function, we drop individual indices in our path integral manipulations. 

To obtain a closed form Eq.~\eqref{C_TCF_solo} using the path integral formalism, we introduce the unitary displacement operator, $\opr{D} = \erm^{-i\opr{p}d}$, which shifts the values of a position operator (of any dimension) by a finite amount. We use this operator to translate the Hamiltonian on the excited state PES so as to center its minimum at the origin,
\begin{equation}
\begin{split}
    \erm^{-\beta_e \opr{H}_{e}} &= \opr{D} \adj{\opr{D}} \erm^{-\beta_e \opr{H}_{e}}\opr{D} \adj{\opr{D}}\\
    &= \erm^{-i \omega_{eg}^{0} t} \opr{D} \erm^{-\beta_e \opr{H}_{g}}\adj{\opr{D}}.
\end{split}
\end{equation}
Noting that $\adj{\opr{D}} \ket{q} = \ket{q - d}$, we expand Eq.~\eqref{C_TCF_solo} in the path integral framework,
\begin{equation}\label{C_term_PI_intro}
\begin{split}
   \tilde{\chi}^{_\text{C}}(t)
   & = Z^{-1} \ing{}{}{r} \ing{}{}{q} \bra{q} \erm^{-\beta_g\opr{H}_g} \ket{r} \bra{r} \erm^{-\beta_e \opr{H}_e} \ket{q} 
   \\ &= \frac{\erm^{-i\omega_{eg}^0t}}{Z} \ing{}{}{r} \ing{}{}{q} \bra{q} \erm^{-\beta_g\opr{H}_g} \ket{r} \bra{r-d} \erm^{-\beta_e \opr{H}_g} \ket{q-d}
   \\ & = \frac{A(\beta_g,\beta_e)}{Z} \erm^{-i\omega_{eg}^0 t - d^2 \vartheta^e} \ing{}{}{r} \ing{}{}{q} 
   \erm^{ -\frac{1}{4} \left[(q+r)^2 \vartheta^+ - 4d(q+r) \vartheta^e  + (q-r)^2 \kappa^+ \right]},
\end{split}
\end{equation}
where we have introduced,
\begin{subequations}
\begin{align}
    A_k(\beta_g,\beta_e)&=\frac{\omega_k}{2\pi}\Big[\sinh(\beta_g\omega_k/2)\sinh(\beta_e\omega_k/2)\Big]^{-\tfrac{1}{2}},\\
    \vartheta_k^e &= \omega_k\tanh(\beta_e\omega_k/2),\\
    \vartheta_k^+ &= \tanh(\beta_g \omega_k/2) + \tanh(\beta_e \omega_k/2) \nonumber \\
    &= \omega_k \frac{\sinh(\beta \omega_k / 2)}{\cosh(\beta_g \omega_k /2) \cosh(\beta_e \omega_k /2)},\\
    \kappa^+_k &= \coth(\beta_g \omega_k /2) + \coth(\beta_e \omega_k /2) \nonumber \\ &= \omega_k\frac{\sinh(\beta\omega_k/2)}{\sinh(\beta_g\omega_k/2)\sinh(\beta_e\omega_k/2)}.
\end{align}
\end{subequations}
Performing the integral in Eq.~\eqref{C_term_PI_intro}, we find
\begin{equation}\label{C_TCF_solo_soln}
    \tilde{\chi}^{_\text{C}}(t) = \erm^{-i \omega_{eg}^{\textrm{av}, k} t - g_{2,k}(t)},
\end{equation}
where we have expanded the adiabatic energy gap in terms of,
\begin{equation}
    \omega_{eg}^{0,k} = \omega_{eg}^{\textrm{av},k} + \frac{1}{2}\omega_k^2 d_k^2.
\end{equation}
The function $g_{2, k}(t)$ is the second-order cumulant with respect to the fluctuations in the energy gap between the ground and excited state PESs,
\begin{equation}
    \begin{split}
        g_{2, k}(t) &= \frac{1}{2}\omega_kd_k^2\Omega(\beta,\omega_k,t),
    \end{split}
\end{equation}
with
\begin{equation}\label{SIeq:energy-gap-response-integrand}
    \Omega(\beta, \omega_k, t) \equiv \coth(\beta \omega_k/2)\big[1-\cos(\omega_k t) \big]+i\big[\sin(\omega_kt)-\omega_j t\big].
\end{equation}
With the notation presented here, we are now in a position to derive the three remaining one-oscillator dynamical objects encoding the non-Condon elements of our theory. 

We now derive closed expressions for $\tilde{\chi}^{_\text{NC}}_{_\text{C}}(t)$ and $\tilde{\chi}'^{_\text{NC}}_{_\text{C}}(t)$ [see Eq.~\eqref{single_oscillator_TCFs}], which we show to be equivalent, allowing one to express $\chi^{_\text{NC}}_{_\text{C}} (t)$ as a single term rather than the sum of two. Starting with $\tilde{\chi}^{_\text{NC}}_{_\text{C}}(t)$, which serves as the first non-Condon contribution to the linear optical response function:
\begin{equation}\label{C/NC_1_TCF_PI}
\begin{split}
   \tilde{\chi}^{_\text{NC}}_{_\text{C}}(t) &= \ev{\opr{q} (t) \opr{X} (t)} 
    \\&=Z^{-1} \ing{}{}{r} \ing{}{}{q} \bra{q} \erm^{-\beta_g \opr{H}_g} \opr{q} \ket{r} \bra{r} \erm^{-\beta_e \opr{H}_e} \ket{q} 
   \\ &=\frac{\erm^{-i \omega_{eg}^0 t}}{Z} \ing{}{}{r} \ing{}{}{q} r \bra{q} \erm^{-\beta_g \opr{H}_g} \ket{r} \bra{r - d} \erm^{-\beta_e \opr{H}_g} \ket{q - d}
   \\ & = \frac{A(\beta_g,\beta_e)}{Z} \erm^{-i\omega_{eg}^0 t - d^2 \vartheta^e} \ing{}{}{r} \ing{}{}{q} r
   \erm^{ -\frac{1}{4} \left[(q+r)^2 \vartheta^+ - 4d(q+r) \vartheta^e  + (q-r)^2 \kappa^+ \right]}
    \\& = \frac{1}{2} \erm^{- i \omega_{eg}^{\textrm{av}, k} t - g_{2, k} (t)} \Theta (\beta, \omega_k, t),
\end{split}
\end{equation}
where
\begin{equation}\label{SIeq:cross-response-integrand}
\begin{split}
    \Theta (\beta, \omega_k, t) &\equiv 1  - \cos(\omega_k t) + i \coth(\beta \omega_k / 2) \sin(\omega_k t) \\
    &= \frac{i}{\omega_k} \pdiv{}{t} \Omega(\beta, \omega_k, t).
\end{split}
\end{equation}

One can similarly show that $\tilde{\chi}'^{_\text{NC}}_{_\text{C}}(t)$ takes the same form,
\begin{equation}\label{C/NC_2_PI}
\begin{split}
    \tilde{\chi}'^{_\text{NC}}_{_\text{C}}(t) & = \frac{A(\beta_g,\beta_e)}{Z} \erm^{-i\omega_{eg}^0 t - d^2 \vartheta^e} \ing{}{}{r} \ing{}{}{q} q
   \erm^{ -\frac{1}{4} \left[(q+r)^2 \vartheta^+ - 4d(q+r) \vartheta^e  + (q-r)^2 \kappa^+ \right]}\\
   &= \frac{1}{2} \erm^{- i \omega_{eg}^{\textrm{av}, k} t - g_{2, k} (t)} \Theta (\beta, \omega_k, t).
\end{split}
\end{equation}
Hence, $\tilde{\chi}^{_\text{NC}}_{_\text{C}}(t) = \tilde{\chi}'^{_\text{NC}}_{_\text{C}}(t)$.

We now turn to the last one-oscillator dynamical object, which constitutes the second non-Condon contribution to the linear optical response function:
\begin{equation}\label{SIeq:NC_TCF_PI}
\begin{split}
    \tilde{\chi}^{_\text{NC}}(t) &= Z^{-1} \ing{}{}{r} \ing{}{}{q} \bra{q} \erm^{-\beta_g \opr{H}_g} \opr{q} \ket{r} \bra{r} \erm^{-\beta_e \opr{H}_e} \opr{q} \ket{q}
    \\ &=\frac{\erm^{-i \omega_{eg}^0 t}}{Z} \ing{}{}{r} \ing{}{}{q} r q \bra{q} \erm^{-\beta_g \opr{H}_g} \ket{r} \bra{r - d} \erm^{-\beta_e \opr{H}_g} \ket{q - d}
   \\ & = \frac{A(\beta_g,\beta_e)}{Z} \erm^{-i\omega_{eg}^0 t - d^2 \vartheta^e} \ing{}{}{r} \ing{}{}{q} r q
   \erm^{ -\frac{1}{4} \left[(q+r)^2 \vartheta^+ - 4d(q+r) \vartheta^e  + (q-r)^2 \kappa^+ \right]}
   \\ &= \erm^{- i \omega_{eg}^{\textrm{av}, k} t - g_{2, k} (t)} \left[\frac{1}{2 \omega_k} \Phi(\beta, \omega_k, t) + \frac{1}{4}d^2_k\Theta^2(\beta, \omega_k, t) \right],
\end{split}
\end{equation}
where,
\begin{equation}\label{SIeq:dipole-response-integrand}
\begin{split}
    \Phi(\beta, \omega_k, t) &\equiv \cos(\omega_k t)\coth(\beta \omega_k / 2) - i\sin(\omega_k t) 
    \\ &= \frac{1}{\omega_k^2} \pdiv{^2}{t^2} \Omega(\beta, \omega_k, t).
\end{split}
\end{equation}
We can now construct $\chi^{_\text{C}}(t)$, $\chi^{_\text{NC}}_{_\text{C}}(t)$, and $\chi^{_\text{NC}}(t)$ from the corresponding one-oscillator objects. 

\subsubsection{The pure Condon response function}

Returning to the multi-oscillator case, we substitute  Eq.~\eqref{C_TCF_solo_soln} into Eqs.~\eqref{C-full-object-decomp} and \eqref{C-full-object}, to recover the second-order cumulant form of the optical response of the BOM,
\begin{equation}\label{SIeq:C_optical_response_final_expression}
    \chi^{_\text{C}}(t) = \left| \ev{\tdpo{ge}} \right|^2 \erm^{-i \omega_{eg}^\textrm{av}t-g_2(t)},
\end{equation}
where
\begin{subequations}
\begin{align}
    g_2(t) &= \frac{1}{2} \sum_{j} \omega_kd_k^2\Omega(\beta,\omega_k,t) ,\\
    \omega_{eg}^\textrm{av} &= \sum_{k}\omega_{eg}^{\textrm{av},k}.
\end{align}
\end{subequations}
Here the full expression for the second-order cumulant, $g_2(t)$, and average frequency of the adiabatic energy gap, $\omega_{eg}^\textrm{av}$, include contributions from all oscillators in the system.

\subsubsection{The mixed Condon/non-Condon (C/NC) response function}

By substituting Eqs.~\eqref{C/NC_1_TCF_PI} and \eqref{C/NC_2_PI} into Eqs.~\eqref{CNC-full-object-1-decomp} and \eqref{CNC-full-object-2-decomp} the mixed-C/NC response in Eq.~\eqref{CNC-full-object} is,
\begin{equation}\label{SIeq:CNC_optical_response_final_expression}
\begin{split}
    \chi_{_\text{NC}}^{_\text{C}} (t) &= -\frac{1}{2} \erm^{-i \omega_{eg}^\textrm{av}t-g_2(t)}  \left( \ev{\tdpo{ge}} + \ev{\tdpo{eg}} \right) \cdot \sum_k \boldsymbol{\alpha}_k d_k \Theta(\beta, \omega_k, t)
    \\ &= - \erm^{-i \omega_{eg}^\textrm{av}t-g_2(t)}  \text{Re} \left\{ \ev{\tdpo{ge}} \right\} \cdot \sum_k \boldsymbol{\alpha}_k d_k \Theta(\beta, \omega_k, t),
\end{split}
\end{equation}
where we have used the fact that $\ev{\tdpo{eg}} = \ev{\tdpo{ge}}^*$.  

\subsubsection{The pure non-Condon response function}

We obtain the pure-NC contribution to the optical response by substituting Eq.~\eqref{SIeq:NC_TCF_PI} into Eqs.~\eqref{NC-full-object-decomp} and \eqref{NC-full-object},
\begin{equation}\label{SIeq:NC_optical_response_final_expression}
\begin{split}
    \chi^{_\text{NC}}(t) &= \erm^{-i \omega_{eg}^\textrm{av}t-g_2(t)}\Bigg[\frac{1}{2}\sum_k \frac{\left| \boldsymbol{\alpha}_k \right|^2}{\omega_k}\Phi(\beta, \omega_k, t) + \frac{1}{4}\sum_j \left|\boldsymbol{\alpha}_j \right|^2 d_j^2 \Theta^2(\beta, \omega_j, t) \\&\hspace{2.5cm}+ \frac{1}{4}\sum_{\substack{j,k \\ j\neq k}} \boldsymbol{\alpha}_j \cdot \boldsymbol{\alpha}_k d_j d_k \Theta(\beta, \omega_j, t) \Theta(\beta, \omega_k, t) \Bigg]\\
    &= \erm^{-i \omega_{eg}^\textrm{av}t-g_2(t)}\Bigg[\frac{1}{2}\sum_k \frac{\left| \boldsymbol{\alpha}_k \right|^2}{\omega_k}\Phi(\beta, \omega_k, t) \\
    &\hspace{2.5cm} + \frac{1}{4}\sum_{j,k} \boldsymbol{\alpha}_j \cdot \boldsymbol{\alpha}_k d_j d_k \Theta(\beta, \omega_j, t) \Theta(\beta, \omega_k, t) \Bigg].
\end{split}
\end{equation}

\subsubsection{Bringing it all together: The full non-Condon linear optical response function}

We have thus obtained all necessary expressions to define the non-Condon linear optical response function. Taking the sum of all individual contributions from Eqs.~\eqref{SIeq:C_optical_response_final_expression},~\eqref{SIeq:CNC_optical_response_final_expression} and~\eqref{SIeq:NC_optical_response_final_expression}, we obtain,
\begin{equation} \label{SIeq:non-Condon-response-discrete-limit}
\begin{split}
    \chi_\smlsub{\rm GNCT}^{(1)}(t) &= \erm^{-i\omega_{eg}^\textrm{av}t-g_2(t)}\Big[\left|\ev{\tdpo{eg}}\right|^2 + 2\text{Re}\lr{\{}{\ev{\tdpo{eg}}}{\}}\cdot\boldsymbol{\mathcal{A}}(t)+\boldsymbol{\mathcal{A}}(t)\cdot\boldsymbol{\mathcal{A}}(t)+\mathcal{B}(t)\Big].
\end{split}
\end{equation}
Here, the vector- and scalar-valued functions composing the non-Condon linear optical response function are,
\begin{subequations}\label{SIeq:A_and_B_fxns_dicrete}
\begin{align}
    \boldsymbol{\mathcal{A}}(t) &= \frac{1}{2}\sum_k \boldsymbol{\alpha}_k d_k \Theta(\beta, \omega_k, t), \\
    \mathcal{B}(t) &= \frac{1}{2}\sum_k \frac{\left| \boldsymbol{\alpha}_k \right|^2}{\omega_k}\Phi(\beta, \omega_k, t).
\end{align}
\end{subequations}
Equation~\eqref{SIeq:non-Condon-response-discrete-limit} represents the \textit{exact} non-Condon linear optical response function for our minimal model of a chromophore whose energy gap and transition dipole fluctuations display Gaussian statistics, which can be expected to hold for many systems in the condensed phase. 

\section{The Non-Condon spectrum arises from the convolution of $\boldsymbol{\mathcal{A}}(\omega)$ and $\mathcal{B}(\omega)$ with the Condon lineshape}
\label{SI-section:spectrum-convolution-GNCT-JCP}

In Sec.~\ref{subsection:GNCT-spectrum-from-GCT-spectrum-KL-convolution-GNCT-JCP} of the main text, we show that the non-Condon effects that control the GCT lineshape arise from the convolution of the spectral densities $K(\omega)$ and $\mathbf{L}(\omega)$ with $\sigma_\smlsub{\rm GCT}(\omega)$. Here, we derive the three expressions that contribute non-Condon spectral features in our GNCT. This rewriting of the expressions is important as it lays the foundation for our analysis in Sec.~S\ref{SI-sec:spectral-interference-GNCT-JCP} of spectral interference in terms of the coefficients $\{ \boldsymbol{\alpha}_j d_j \}$. This analysis identifies the specific spectral features to which $\mathbf{\sigma_{\mathcal{A}}} (\omega)$ and $\sigma_{\mathcal{AA}}(\omega)$ can give rise.

To obtain closed-form expressions of $\boldsymbol{\sigma_{\mathcal{A}}}(\omega)$, $\sigma_{\mathcal{AA}}(\omega)$ and $\sigma_{\mathcal{B}}(\omega)$ as convolutions of $K(\omega)$ and $\mathbf{L}(\omega)$ with $\sigma_\smlsub{\rm GCT}(\omega)$, we invoke the convolution theorem to write,  
\begin{subequations}
\begin{align}
    \boldsymbol{\sigma_{\mathcal{A}}}(\omega) &= \int_{-\infty}^{\infty} \text{d}t \, \erm^{i \omega t} \boldsymbol{\mathcal{A}} (t) \, \erm^{-i \omega_{eg}^{\rm av} t - g_2(t)} \nonumber \\&= \frac{1}{2 \pi } \int_{-\infty}^{\infty} \text{d} \Omega \, \boldsymbol{\mathcal{A}}(\Omega) \, \sigma_\smlsub{\rm GCT}(\omega - \Omega),\label{SIeq:non-Condon-A-spectrum-initial-convolution-GNCT-JCP} \\
    \sigma_{\mathcal{AA}} (\omega) &= \int_{-\infty}^{\infty} \text{d}t \, \erm^{i \omega t} \mathcal{A}^2 (t) \, \erm^{-i \omega_{eg}^{\rm av} t - g_2(t)} \nonumber \\&= \frac{1}{4 \pi^2 } \int_{-\infty}^{\infty} \text{d} \Omega \int_{-\infty}^{\infty} \text{d} \Omega' \, \mathcal{A}(\Omega) \mathcal{A}(\Omega-\Omega') \, \sigma_\smlsub{\rm GCT}(\omega - \Omega - \Omega'),\label{SIeq:non-Condon-AA-spectrum-initial-convolution-GNCT-JCP} \\
    \sigma_{\mathcal{B}} (\omega) &= \int_{-\infty}^{\infty} \text{d}t \, \erm^{i \omega t} \mathcal{B} (t) \, \erm^{-i \omega_{eg}^{\rm av} t - g_2(t)} \nonumber \\&= \frac{1}{2 \pi } \int_{-\infty}^{\infty} \text{d} \Omega \, \mathcal{B}(\Omega) \, \sigma_\smlsub{\rm GCT}(\omega - \Omega),\label{SIeq:non-Condon-B-spectrum-initial-convolution-GNCT-JCP} 
\end{align}
\end{subequations}
We now obtain explicit expressions for $\boldsymbol{\mathcal{A}}(\omega)$ and $\mathcal{B}(\omega)$ in terms of the two spectral densities introduced in Sec.~\ref{sec:non-Condon-coefficients-from-simulation}. Beginning with Eq.~\eqref{SIeq:non-Condon-A-spectrum-initial-convolution-GNCT-JCP}, by evaluating the Fourier transform of $\boldsymbol{\mathcal{A}}(t)$, we find, 
\begin{equation}\label{SIeq:non-Condon-A-term-Fourier-transform-GNCT-JCP}
\begin{split}
    \boldsymbol{\mathcal{A}} (\omega) &= \frac{1}{\pi} \int_0^{\infty} \text{d}\Omega \, \frac{\mathbf{L}(\Omega)}{\Omega} \int_{-\infty}^{\infty} \text{d}t \, \erm^{i \omega t} \Big[ 1 - \cos(\Omega t) + i \sin(\Omega t) \coth(\beta \Omega / 2)  \Big] \\&= \int_0^{\infty} \text{d}\Omega \, \frac{\mathbf{L}(\Omega)}{\Omega} \Big[ \delta(\omega) + \delta(\omega + \Omega)[ \coth(\beta \Omega / 2) - 1 ] - \delta(\omega-\Omega) [ \coth(\beta \Omega / 2) + 1 ]  \Big] \\& = \int_0^{\infty} \text{d}\Omega \, \frac{\mathbf{L}(\Omega)}{\Omega} \left[ \delta(\omega) - \frac{2}{\Omega} \frac{\mathbf{L}(\Omega)}{1 - \erm^{-\beta \Omega}} \delta(\omega-\Omega) \right]
    \\& \equiv 2 \pi \left[\boldsymbol{\lambda}_{\mathbf{L}} \delta(\omega) - \frac{\boldsymbol{\rho}_{\mathbf{L}}(\omega)}{\omega} \right].
\end{split}
\end{equation}
Here, $\boldsymbol{\lambda}_{\mathbf{L}} = \frac{1}{\pi} \int_{0}^{\infty} \text{d} \omega \, \frac{\mathbf{L}(\omega)}{\omega}$, $\boldsymbol{\rho}_{\mathbf{L}}(\omega) = \frac{1}{\pi} \frac{\mathbf{L}(\omega)}{1 - \erm^{- \beta \omega}}$, and in going from lines two to three, we omitted terms that are zero based on the limit of integration over $\Omega$. The resulting expression for $\boldsymbol{\mathcal{A}}(\omega)$ is composed of the difference between, $\boldsymbol{\lambda}_{\mathbf{L}} \delta(\omega)$, a temperature-independent component fixed at the origin and, $\boldsymbol{\rho}_{\mathbf{L}}(\omega)$, a temperature-dependent component defined over all positive frequencies. 

Next, we follow an analogous derivation and find $\mathcal{B}(\omega)$, 
\begin{equation}\label{SIeq:non-Condon-B-term-Fourier-transform-GNCT-JCP}
\begin{split}
    \mathcal{B}(\omega) &= \frac{1}{\pi} \int_{0}^{\infty} \text{d}\Omega \, K (\Omega) \int_{-\infty}^{\infty} \text{d}t \, \erm^{i \omega t} \Big[\cos(\Omega t) \coth(\beta \Omega / 2) - i \sin(\Omega t) \Big] \\&= \int_{0}^{\infty} \text{d}\Omega \, K (\Omega) \delta(\omega + \Omega)[ \coth(\beta \Omega / 2) - 1 ] + \delta(\omega-\Omega) [ \coth(\beta \Omega / 2) + 1 ]  \Big] \\&= 2 \frac{K(\omega)}{1 - \erm^{-\beta \omega}} \\&\equiv 2 \pi \rho_{K}(\omega).
\end{split}
\end{equation}
Here, $\rho_{K}(\omega) = \frac{1}{\pi} \frac{ K(\omega) }{ 1 - \erm^{- \beta \omega } }$. Not surprisingly, the expression for $\mathcal{B}(\omega)$ follows closely to that found for $\boldsymbol{\mathcal{A}}(\omega)$. Lastly, we substitute these results into Eqs.~\eqref{SIeq:non-Condon-A-spectrum-initial-convolution-GNCT-JCP}-\eqref{SIeq:non-Condon-AA-spectrum-initial-convolution-GNCT-JCP} to obtain forms for the non-Condon contributions to the absorption spectrum in terms of convolutions with the Condon (GCT) lineshape, $\sigma_\smlsub{\rm GCT} (\omega - \Omega)$,
\begin{subequations}
\begin{align}
    \boldsymbol{\sigma_{\mathcal{A}}}(\omega) &= \boldsymbol{\lambda}_{\mathbf{L}} \sigma_\smlsub{\rm GCT}(\omega) - \int_0^{\infty} \text{d} \Omega \, \frac{\boldsymbol{\rho}_{\mathbf{L}}(\Omega)}{\Omega} \sigma_\smlsub{\rm GCT} (\omega - \Omega)\label{SIeq:non-Condon-A-spectrum-final-convolution-GNCT-JCP},\\ 
    \sigma_{\mathcal{AA}}(\omega) &= 
    |\boldsymbol{\lambda}_{\mathbf{L}}|^2 \sigma_\smlsub{\rm GCT} (\omega) - 2 \boldsymbol{\lambda}_{\mathbf{L}} \int_0^{\infty} \text{d} \Omega \frac{\boldsymbol{\rho}_{\mathbf{L}}(\Omega)}{\Omega} \sigma_\smlsub{\rm GCT} (\omega - \Omega)\nonumber
    \\&+ \int_0^{\infty} \text{d} \Omega \int_0^{\infty} \text{d} \Omega' \, \frac{\boldsymbol{\rho}_{\mathbf{L}}(\Omega)}{\Omega} \frac{\boldsymbol{\rho}_{\mathbf{L}}(\Omega-\Omega')}{\Omega-\Omega'} \sigma_\smlsub{\rm GCT} (\omega - \Omega-\Omega')\label{SIeq:non-Condon-AA-spectrum-final-convolution-GNCT-JCP},\\
    \sigma_{\mathcal{B}}(\omega) &=  \int_0^{\infty} \text{d} \Omega \, \rho_{K}(\Omega) \sigma_\smlsub{\rm GCT} (\omega - \Omega)\label{SIeq:non-Condon-B-spectrum-final-convolution-GNCT-JCP}.
\end{align}
\end{subequations}
Using these expressions, we find the non-Condon absorption spectrum to be, 
\begin{equation}
    \sigma_\smlsub{\rm GNCT} (\omega) = \left[ |\langle \hat{\boldsymbol{\mu}}^s_{ge} \rangle|^2 \delta(\omega) - 2 \langle \hat{\boldsymbol{\mu}}^s_{ge} \rangle \cdot \frac{\boldsymbol{\rho}_{\mathbf{L}} (\omega)}{\omega} + \frac{\boldsymbol{\rho}_{\mathbf{L}} (\omega)}{\omega} \ast \frac{\boldsymbol{\rho}_{\mathbf{L}} (\omega)}{\omega} + \rho_K(\omega) \right] \ast \sigma_\smlsub{\rm GCT} (\omega).
\end{equation}
Here, $\ast$ denotes the convolution operation, and $\langle \hat{\boldsymbol{\mu}}^s_{ge} \rangle = \langle \hat{\boldsymbol{\mu}}_{ge} \rangle + \boldsymbol{\lambda}_{\mathbf{L}}$ is the renormalized equilibrium-average transition dipole. Since $\boldsymbol{\lambda}_{\mathbf{L}}$ can enhance or reduce the elements of $\langle \hat{\boldsymbol{\mu}}_{ge} \rangle$, the value of $|\langle \hat{\boldsymbol{\mu}}^s_{ge} \rangle|^2$ can be greater or less than $|\langle \hat{\boldsymbol{\mu}}_{ge} \rangle|^2$, and therefore $\boldsymbol{\lambda}_{\mathbf{L}}$  has the ability to either amplify or attenuate the Condon contribution to the non-Condon lineshape.

\section{Bridging our formal non-Condon Gaussian response functions to molecular simulation}\label{SI-section:spectral-densities-from-molecular-simulation-GNCT-JCP}

The expression for the linear optical response function in Eq.~\eqref{SIeq:non-Condon-response-discrete-limit} exactly treats the quantum dynamics of our Gaussian non-Condon model of a chromophore subject to pure dephasing and offers a formal connection between the energy gap displacements, $d_j$, the transition dipole fluctuations, $\boldsymbol{\alpha}_j$, and the spectrum. However, while the coefficients $\boldsymbol{\alpha}_j$ and $d_j$ parameterize the coupling strength between the nuclear degrees of freedom and the electronic transition, Eq.~\eqref{SIeq:non-Condon-response-discrete-limit} does not provide a protocol for obtaining these quantities from molecular simulations. Here, we show how to obtain $\boldsymbol{\alpha}_j$ and $d_j$ by connecting them to spectral densities, i.e., Fourier transforms of quantum auto- and cross correlation functions of the energy gap and transition dipole fluctuations, judiciously chosen to recover these coefficients. In particular, we introduce two new spectral densities, one of the fluctuations of the transition dipole and a second that quantifies the correlation between the fluctuations of the transition dipole and the energy gap. Our expressions can be used with exact or approximate quantum dynamics schemes (e.g., quantum-classical path integrals,\cite{makrictpi-sup} open-chain path integrals,\cite{tuckerman2018open-sup} centroid path integral,\cite{cao1994formulation-sup, jang1999cmd-sup} and ring-polymer path integral\cite{mano2004quantum-sup, markland2018nuclear-sup} methods). Because of the broad applicability of the MD toolbox for atomistic simulation, here we provide a straightforward means to employ \textit{classical} MD simulations to capture Gaussian non-Condon spectra for complex chemical systems. To achieve this, we approximate the quantum Kubo-transformed correlation functions with their classical analogs and employ the Fourier space detailed balance conditions \cite{bader1994quantum-sup, Egorov1999-sup, Kim2002b-sup, mano2004quantum-sup, ramirez2004quantum-sup} to provide access to the target spectral densities encoding $\boldsymbol{\alpha}_j$ and $d_j$. Thus, the protocol here offers a direct means to employ MD simulations and electronic structure calculations to construct the optical spectra of chromophores in the condensed phase with non-Condon effects.
We begin by introducing the quantum auto and cross correlation functions of the energy gap and transition dipole fluctuations, 
\begin{equation}\label{SIeq:quantum-dipole-auto-tcf-GNCT-JCP}
\begin{split}
    C_{\delta\mu \delta\mu}(t) &= \ev{\delta \tdpo{}(t) \mdot \delta\tdpo{}(0)} \\
    &= \frac{1}{2}\sum_k \frac{\left|\boldsymbol{\alpha}_k \right|^2}{\omega_k}\big[\coth(\beta \omega_k / 2) \cos(\omega_k t) - i \sin(\omega_k t)\big]\\
    & \equiv C'_{\delta\mu \delta\mu}(t) + i C''_{\delta\mu \delta\mu}(t)
\end{split}
\end{equation}
and
\begin{equation}\label{SIeq:quantum-egap-dipole-cross-tcf-GNCT-JCP}
\begin{split}
    \mathbf{C}_{\delta U \delta\mu}(t)&=\ev{\delta \opr{U}(t)\delta\tdpo{}(0)} \\
    &= \frac{1}{2} \sum_k \boldsymbol{\alpha}_k \omega_k d_k \big[\coth(\beta \omega_k / 2) \cos(\omega_k t) - i \sin(\omega_k t)\big]\\
    & \equiv \mathbf{C}'_{\delta U \delta\mu}(t) + i \mathbf{C}''_{\delta U \delta\mu}(t).
\end{split}
\end{equation}
Here,
\begin{equation}
    \delta \opr{U}(t) = \sum_k \omega^2_kd_k\opr{q}_k(t)
\end{equation}
is the time-dependent fluctuation of the energy gap between the ground and excited state PESs, where the Fourier transform of $C^{''}_{\delta U \delta U} (t)$ provides the well-known energy gap spectral density \cite{Mukamel-book-sup, Zuehlsdorff2019-sup},
\begin{equation}
    J(\omega) = \frac{\pi}{2} \sum_k \omega_k^3 d^2_k \delta(\omega-\omega_k).
\end{equation}

Motivated by the detailed balance condition, which allows one to construct the full frequency representation of a quantum mechanical equilibrium time correlation function from the frequency representation of its imaginary component, we write,
\begin{equation}\label{SIeq:K-auxilliary-spec-dens}
\begin{split}
    C''_{\delta\mu \delta\mu}(\omega) &= i \ing{-\infty}{\infty}{t} \erm^{i \omega t} C''_{\delta \mu}(t)
    \\&= \frac{\pi}{2}\sum_j\frac{|\boldsymbol{\alpha}_j|^2}{\omega_j}\Big[ \delta(\omega-\omega_j)-\delta(\omega+\omega_j) \Big]
    \\&\equiv K(\omega)-K(-\omega),
\end{split}
\end{equation}
and 
\begin{equation}\label{SIeq:L-auxilliary-spec-dens}
\begin{split}
    \mathbf{C}''_{ \delta U\delta\mu}(\omega) &= i \ing{-\infty}{\infty}{t} \erm^{i \omega t} \mathbf{C}''_{\delta U \delta \mu} (t)
    \\&= \frac{\pi}{2}\sum_j\boldsymbol{\alpha}_j\omega_jd_j\Big[\delta(\omega-\omega_j)-\delta(\omega+\omega_j)\Big]
    \\&\equiv \mathbf{L}(\omega)-\mathbf{L}(-\omega).
\end{split}
\end{equation}
Here, $C^{''}_{\delta A \delta B} (t) \equiv {\rm Im} \langle \delta \hat{A}(t) \hat{B}(0) \rangle$ is the imaginary part of the quantum time correlation function of interest. The auxiliary spectral densities that we introduced in Eqs.~\eqref{SIeq:K-auxilliary-spec-dens} and \eqref{SIeq:L-auxilliary-spec-dens}, $K(\omega)$ and $\mathbf{L}(\omega)$, provide the key to uniting our quantum dynamical theory with numerical data obtainable from classical molecular dynamics simulations. 

To establish the specific connection to classical dynamics, we employ the relationship between the frequency representation of the Kubo-transformed quantum equilibrium time correlation function and its classical analog. Using the Kubo formalism allows us to write Eqs.~\eqref{SIeq:quantum-dipole-auto-tcf-GNCT-JCP} and \eqref{SIeq:quantum-egap-dipole-cross-tcf-GNCT-JCP} as,
\begin{subequations}\label{SIeq:Kubo-aux-spec-dens}
\begin{align}
    K(\omega) &\approx \theta(\omega) \frac{\beta \omega}{2} \ing{-\infty}{\infty}{t} \erm^{i \omega t} C^\textrm{cl}_{ \delta\mu \delta\mu }(t), \\
    \mathbf{L}(\omega) &\approx  \theta(\omega) \frac{\beta \omega}{2} \ing{-\infty}{\infty}{t} \erm^{i \omega t} \mathbf{C}^\textrm{cl}_{\delta U \delta\mu}(t).
\end{align}
\end{subequations}
Here, $\theta(\omega)$ is the Heaviside step function, $C^\textrm{cl}_{\delta\mu \delta\mu}(t)$ and $\mathbf{C}_{\delta U \delta\mu}^\textrm{cl}(t)$ are the purely real and symmetric classical auto- and cross- equilibrium time correlation functions of the fluctuations in the transition dipole and PES energy gap. The quantum harmonic correction factor, $\tfrac{\beta \omega}{2}$, arises from approximating the Kubo-transformed correlation function with the classical one and using the Fourier space connection between the Kubo-transformed correlation function and the full quantum one \cite{bader1994quantum-sup, Egorov1999-sup, Kim2002b-sup, mano2004quantum-sup, ramirez2004quantum-sup}. 

We now reconstruct the vector- and scalar-valued functions (Eq.~\eqref{SIeq:A_and_B_fxns_dicrete}) using the spectral densities that define the strength of coupling between the solvent modes and the fluctuations in the energy gap and the transition dipole of the chromophore in the condensed phase,
\begin{subequations}
\begin{align}
    g_2(t) &= \frac{1}{\pi} \ing{0}{\infty}{\omega} \frac{J(\omega)}{\omega^2} \Omega(\beta, \omega, t), \label{SIeq:g2-continuous} 
    \\ \boldsymbol{\mathcal{A}}(t)&=\frac{1}{\pi}\ing{0}{\infty}{\omega}\frac{\mathbf{L}(\omega)}{\omega}\Theta(\beta,\omega,t), \label{SIeq:A-continuous} \\
    \mathcal{B}(t) &= \frac{1}{\pi}\ing{0}{\infty}{\omega}K(\omega)\Phi(\beta,\omega,t) \label{SIeq:B-continuous}.
\end{align}
\end{subequations}
These expressions for $g_2(t)$, $\boldsymbol{\mathcal{A}}(t)$, and $\mathcal{B}(t)$ fully connect our Gaussian theory of a non-Condon BOM to molecular dynamics simulations.

\section{Computational methods}\label{SI-section:computational-details-GNCT-JCP}

\subsection{FCHT Calculations}

For phenolate in vacuum, we use the standard Franck-Condon Herzberg-Teller (FCHT) approach based on a harmonic approximation to the ground and excited state PESs. To construct the model PES, we perform ground and exited state geometry optimizations using the TeraChem software package\cite{Ufimtsev2009-sup,Isborn2011-sup} with the 6-31+G*\cite{Dunning1990-sup} basis set and CAM-B3LYP\cite{Yanai2004-sup} DFT functional. We then compute vibrational normal modes and frequencies around the respective minima using a finite difference scheme and additionally construct the first derivative of the transition dipole moment with respect to the normal mode coordinates in the excited state. We then use the normal modes and the coordinates of the two minima to compute the displacement vector between the ground and excited state minima, and the Duschinsky rotation relating ground and excited state normal modes. This information fully specifies a model Hamiltonian for the chromophore, where the ground and excited state PESs are approximated to be harmonic around their minima and the transition dipole moment is expressed through a Taylor expansion to first order around the excited state minimum.  

The FCHT spectrum shown in the main text is generated using the MolSpeckPy code that is freely available on GitHub.\cite{molspecpy-sup} The spectrum is computed by evaluating the exact finite temperature response function for the harmonic Hamiltonian directly in the time domain, avoiding computationally costly sum-over-states expressions in the frequency domain\cite{Baiardi2013-sup, deSouza2018-sup}. 

To directly compare spectral lineshapes computed in the FC scheme in vacuum to GCT and GNCT lineshapes in vacuum and cyclohexane (Figs.~1 and 3 in the main text), we artificially broadened all FC-based vacuum spectra presented in this work by including a bath of low-frequency BOM modes\cite{Zuehlsdorff2019-sup} characterized by a Debye spectral density of the form
\begin{equation}\label{eq:debye-spectral-density-for-solvent}
    {J}_{\textrm{env}}(\omega) = 2\lambda_{\textrm{env}} \omega_c \frac{ \omega }{\omega_c^2 + \omega^2}.
\end{equation}
A cutoff frequency of $\omega_c=22$~cm$^{-1}$ and an effective environmental reorganization energy of $\lambda_\textrm{env}=0.00015$~Ha are used throughout. 

\subsection{MD Trajectories}

We constructed all correlation functions from molecular dynamics trajectories of the phenolate both in vacuum and in explicit solvent environments. We generated \emph{ab initio} molecular dynamics (AIMD) trajectories using TeraChem with the same functional and basis set as used in the geometry optimizations. The trajectories consisted of 22~ps of NVT dynamics at 300~K with a time step of 0.5~fs; we held the temperature constant using a Langevin thermostat with a collision frequency of 1~ps$^{-1}$. We discarded the first 2~ps of the trajectory (i.e., we did not use these in constructing correlation functions) to allow the system to equilibrate.

For the trajectories of phenolate in cyclohexane and water, we used the mixed quantum mechanical/molecular mechanical (QM/MM)\cite{Warshel1972-sup} interface between Amber20\cite{amber-sup} and TeraChem\cite{Isborn2012-sup} to model a 30~\AA\, sphere of the solvent molecules in open boundary conditions. Water molecules used the TIP3P\cite{TIP3P-sup} force field (FF), while cyclohexane and phenolate used force fields generated using AmberTools. We first equilibrated the systems using only (FF) MD for 50~ps, and subsequently treated phenolate quantum mechanically while describing the solvent environment through the classical FF, using QM/MM MD, for an additional 22~ps. 

\subsection{Constructing Classical Correlation Functions from MD Data}

We sampled the latter 20~ps of each AIMD or QM/MM MD trajectory every 2~fs for a total of 10,000 snapshots, with the first 2~ps of trajectory being discarded to allow for equilibration of the QM system. The first excited state was calculated at each snapshot using time-dependent density-functional theory (TDDFT) as implemented in the TeraChem package\cite{Isborn2012-sup}, using the same functional and basis set as in the geometry optimizations and frequency calculations used to generate the FCHT spectrum. For the trajectories with explicit solvent, in the case of cyclohexane, only phenolate was treated at the QM level and the effect of the surrounding solvent molecules was included in the calculation through classical point charges. For phenolate in water, we performed two calculations: 1) a calculation where we accounted for the solvent environment purely at the MM-level through classical point charges; and 2) a calculation where we treated all solvent molecules with a center of mass within 6~\AA\, of any atom of phenolate fully at the QM level when computing excitation energies, to better describe environmental polarization effects\cite{Zuehlsdorff2018b-sup}. We performed all excited state calculations using full TDDFT, i.e., without resorting to the Tamm-Dancoff approximation\cite{Hirata1999-sup}.

We then used the excitation energies of the S$_1$ state to calculate the classical energy autocorrelation function described in the main text. As a vector quantity calculated relative to the nuclear coordinates of the QM system and not a physical observable (only the square of the transition dipole, proportional to the oscillator strength of the transition, is a physical observable), the transition dipole moments as calculated by TeraChem required further processing before they could yield well-behaved correlation functions.

Although the computed value of the transition dipole moment is sensitive to both the rotational and vibrational nuclear degrees of freedom, rotational contributions to the correlation functions are unnecessary in capturing vibronic spectroscopic effects. We thus removed rotational effects by first transforming each snapshot along the MD trajectory to the Eckart frame, or, equivalently, by rotating the transition dipole moments to the Eckart frame after they have been calculated. Transformation to the Eckart frame is equivalent to minimizing the mass-weighted square deviation of the nuclear coordinates of the chromophore in one snapshot with respect to its coordinates in some reference geometry, which we take to be the first snapshot of the trajectory. Krasnoshcheckov and coworkers\cite{Krasnoshchekov2014-sup} additionally showed, after translating both geometries to their center of mass, that by expressing the desired rotation as a unit quaternion, the required rotation can be trivially computed as an eigenvalue problem. Taking the rotation of a vector $\bf{v}$ around an axis and by an angle determined by quaternion $\bf{q}$ to be $[0,{\bf{v}}'] = {\bf{q}} [0,{\bf{v}}] {\bf{q}^{-1}}$, the quaternion needed to rotate the geometry of an MD snapshot to the Eckart frame defined by some reference geometry is the eigenvector with the smallest eigenvalue of the symmetric matrix $\bf{C}$
\begin{equation}
    {\bf{C}} = \sum_i m_i \begin{pmatrix}
        C_{i,11}& C_{i,12}& C_{i,13}& C_{i,14} \\
                & C_{i,22}& C_{i,23}& C_{i,24} \\
                &         & C_{i,33}& C_{i,34} \\
                &         &         & C_{i,44} \\
    \end{pmatrix},
\end{equation}
with elements
\begin{subequations}
\begin{align*}
    C_{i,11} &= (x_{i,\mathrm{ref}} - x_{i,\mathrm{MD}})^{2} + (y_{i,\mathrm{ref}} - y_{i,\mathrm{MD}})^{2} + (z_{i,\mathrm{ref}} - z_{i,\mathrm{MD}})^{2};\\
    C_{i,22} &= (x_{i,\mathrm{ref}} - x_{i,\mathrm{MD}})^{2} + (y_{i,\mathrm{ref}} + y_{i,\mathrm{MD}})^{2} + (z_{i,\mathrm{ref}} + z_{i,\mathrm{MD}})^{2};\\
    C_{i,33} &= (x_{i,\mathrm{ref}} + x_{i,\mathrm{MD}})^{2} + (y_{i,\mathrm{ref}} - y_{i,\mathrm{MD}})^{2} + (z_{i,\mathrm{ref}} + z_{i,\mathrm{MD}})^{2};\\
    C_{i,44} &= (x_{i,\mathrm{ref}} + x_{i,\mathrm{MD}})^{2} + (y_{i,\mathrm{ref}} + y_{i,\mathrm{MD}})^{2} + (z_{i,\mathrm{ref}} - z_{i,\mathrm{MD}})^{2};\\
    C_{i,12} &= (y_{i,\mathrm{ref}} - y_{i,\mathrm{MD}})(z_{i,\mathrm{ref}} + z_{i,\mathrm{MD}}) - (y_{i,\mathrm{ref}} + y_{i,\mathrm{MD}})(z_{i,\mathrm{ref}} - z_{i,\mathrm{MD}});\\
    C_{i,13} &= (z_{i,\mathrm{ref}} - z_{i,\mathrm{MD}})(x_{i,\mathrm{ref}} + x_{i,\mathrm{MD}}) - (z_{i,\mathrm{ref}} + z_{i,\mathrm{MD}})(x_{i,\mathrm{ref}} - x_{i,\mathrm{MD}});\\
    C_{i,14} &= (x_{i,\mathrm{ref}} - x_{i,\mathrm{MD}})(y_{i,\mathrm{ref}} + y_{i,\mathrm{MD}}) - (x_{i,\mathrm{ref}} + x_{i,\mathrm{MD}})(y_{i,\mathrm{ref}} - y_{i,\mathrm{MD}});\\
    C_{i,23} &= (x_{i,\mathrm{ref}} - x_{i,\mathrm{MD}})(y_{i,\mathrm{ref}} - y_{i,\mathrm{MD}}) - (x_{i,\mathrm{ref}} + x_{i,\mathrm{MD}})(y_{i,\mathrm{ref}} + y_{i,\mathrm{MD}});\\
    C_{i,24} &= (z_{i,\mathrm{ref}} - z_{i,\mathrm{MD}})(x_{i,\mathrm{ref}} - x_{i,\mathrm{MD}}) - (z_{i,\mathrm{ref}} + z_{i,\mathrm{MD}})(x_{i,\mathrm{ref}} + x_{i,\mathrm{MD}});\\
    C_{i,34} &= (y_{i,\mathrm{ref}} - y_{i,\mathrm{MD}})(z_{i,\mathrm{ref}} - z_{i,\mathrm{MD}}) - (y_{i,\mathrm{ref}} + y_{i,\mathrm{MD}})(z_{i,\mathrm{ref}} + z_{i,\mathrm{MD}}).
\end{align*}
\end{subequations}
Here the subscript $i$ indicates the indexing of atoms in the chromophore, while the subscripts $\mathrm{ref}$ and $\mathrm{MD}$ indicate the reference and MD snapshot geometries respectively; $m$ is the mass; and $x$, $y$, and $z$ are the spatial coordinates of the specified atom. For all solvated systems, only the chromophore atoms are used when computing the transformation to the Eckart frame. 

\begin{figure*}[b]
\includegraphics[width=.75\columnwidth]{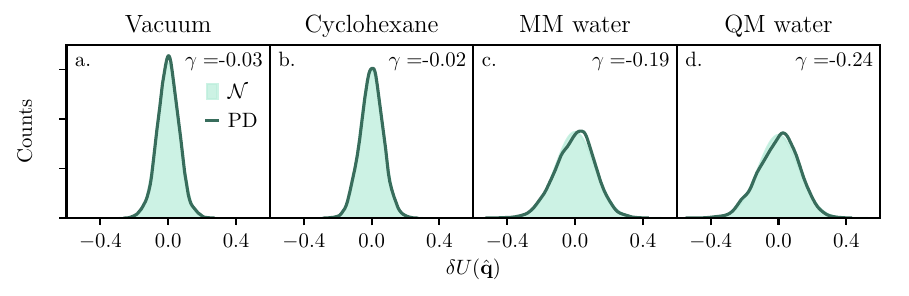}
\caption{Probability distributions (PDs) showing the energy gap fluctuations (solid lines) for phenolate in vacuum, cyclohexane, MM-water, and QM-water. Here, the PDs are superimposed over shaded regions representing normal Gaussian distributions ($\mathcal{N}$) generated using the standard deviation from the corresponding MD trajectory data. We employ the skewness values ($\gamma$) of the transition dipole fluctuations from each trajectory as a measure of how well the MD-based energy gap fluctuations follow Gaussian statistics.}
\label{SI-fig:egap-statistics}
\end{figure*}

Having transformed each transition dipole moment to a consistent frame of reference, there remains the issue that the sign of this vector quantity is arbitrary, as only its square is a physical observable. We are therefore needed to choose the sign for each transition dipole such that the chromophore's transition dipole moment fluctuates smoothly in time across the trajectory, without discontinuities. To do this, we compare the value calculated for the transition dipole moment at each snapshot with the value it is expected to have based on previous snapshots based on some form of extrapolation (see below). We then choose the sign which minimizes the residue between the two. 

There are various ways to approximately extrapolate the next value of the transition dipole moment. For example, one can assume that the transition dipole in one snapshot should have a similar value to that in the previous one. This is often sufficient. In this work, we see reduced noise in correlation functions when using a second-order extrapolation (where each extrapolation is based on a quadratic fit to the three preceding data points), as it better accounts for rapid changes in the value of the transition dipole moment.

\section{Phenolate trajectory statistics: Energy gap and transition dipole distributions in vacuum, cyclohexane and water}\label{SI-section:statistics-of-MD-trajectories-GNCT-JCP-GNCT-JCP}

\begin{figure*}[b]
\includegraphics[width=.75\columnwidth]{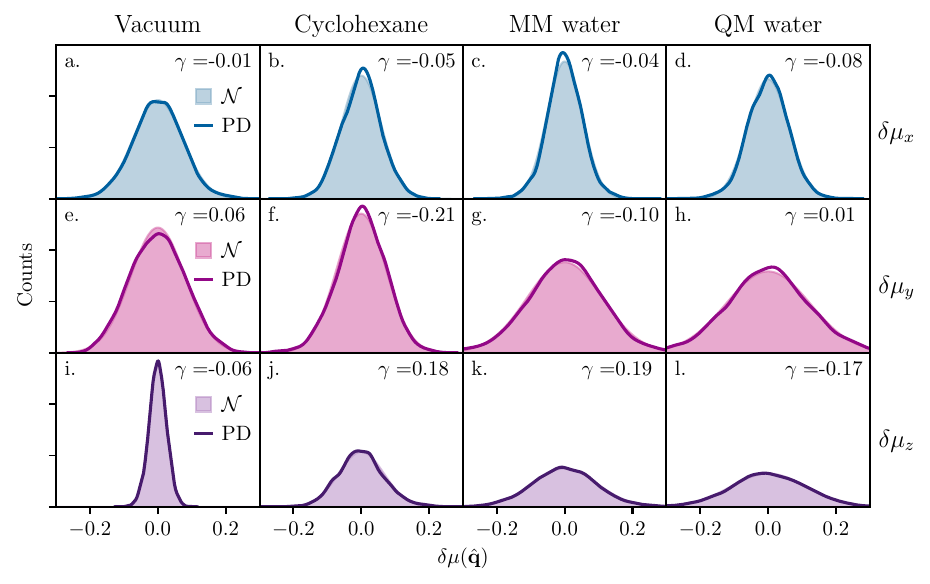}
\caption{Probability distributions (PDs) showing the transition dipole fluctuations in $x$ (blue lines), $y$ (fuchsia lines), and $z$ (purple lines) directions for phenolate in vacuum, cyclohexane, MM-water, and QM-water. Here, the PDs shown as solid lines are superimposed over shaded regions representing the normal Gaussian distributions ($\mathcal{N}$) generated using the standard deviation from the corresponding MD trajectory data. We employ the skewness values ($\gamma$) of the transition dipole fluctuations from each trajectory as a measure of how well the MD-based transition dipole fluctuations follow Gaussian statistics.}
\label{SI-fig:transition-dipole-statistics}
\end{figure*}

In this manuscript, we have treated statistical non-Condon effects by recognizing that the energy gap and transition dipole of a chromophore fluctuate due to \textit{both} external nuclear motions (i.e., the solvent) and the intramolecular motions of the chromophore. At the heart of our GNCT is the invocation of the central limit theorem, allowing us to posit that as the number of nuclear motions inducing fluctuations of the energy gap and transition dipole approach the macroscopic limit it is likely that these fluctuations display Gaussian statistics. Figures~\ref{SI-fig:egap-statistics} and \ref{SI-fig:transition-dipole-statistics} test this assumption by comparing ideal normal probability distributions ($\mathcal{N}$) generated using the MD-derived standard deviations of the energy gap and transition dipole fluctuations for phenolate in vacuum, cyclohexane, classical (MM) water, and quantum mechanical (QM) water to the \textit{direct} construction of their probability distributions. 

Beginning with Fig.~\ref{SI-fig:egap-statistics}, all PDs agree well with the normal distributions of the energy gap fluctuations for phenolate in vacuum (a), cyclohexane (b), MM water (c) and QM water (d). Here, the corresponding skewness values, $\gamma$, calculated for the MD-based probability distributions lie under 0.3 (with $\gamma=0$ corresponding to \textit{exactly} Gaussian fluctuations), indicating that Gaussian statistics are sufficient to describe the energy gap fluctuations of phenolate across these environments. Interestingly, the choice of solvent and the level of theory employed to model the solvent significantly impact the value of $\gamma$. Panels a to b, corresponding to phenolate simulated in vacuum versus cyclohexane, have similarly small values for $\gamma$, $-0.03$ and $-0.02$. Further, when going from phenolate in MM-water to phenolate in QM-water, one observes increasingly non-Gaussian distributions, with $\gamma = -0.19$ and $-0.24$, respectively. The reason for these skewness values is simple: the ability of the hydroxyl group on phenolate to hydrogen bond to neighboring water molecules likely introduces very strong interactions in which specific nuclear motions couple strongly to the fluctuations of the energy gap (See Fig.~\ref{SIfig:phenol_water_sds}). 

Figure~\ref{SI-fig:transition-dipole-statistics} considers the statistics for the components of the transition dipole fluctuations $\delta \mu_x(\hat{\mathbf{q}})$ (a-d), $\delta \mu_y(\hat{\mathbf{q}})$ (e-h) and $\delta \mu_z(\hat{\mathbf{q}})$ (i-l) calculated for phenolate in vacuum, cyclohexane, MM-water, and QM-water. Since we write $\delta \hat{\boldsymbol{\mu}} (\hat{\mathbf{q}})$ as a sum over Gaussian variables that are scaled in three dimensions, we calculate the distributions in Fig.~\ref{SI-fig:transition-dipole-statistics} for the $x$, $y$ and $z$ transition dipole fluctuations in each trajectory separately. What is remarkable about these results is that the skewness factors always remain bound at low values, $|\gamma|<0.3$, indicating that Gaussian statistics are sufficient to describe these fluctuations, even when $\delta \mu_y(\hat{\mathbf{q}})$ and $\delta \mu_z(\hat{\mathbf{q}})$ display skewness values that are somewhat larger, approaching the bound of $|\gamma|=0.3$. Hence, our Gaussian theory for non-Condon effects in the optical spectra of phenolate in vacuum, cyclohexane, and water holds. 

\section{Normal mode analysis of computed spectral densities}\label{SI-sec:spectral-density-normal-mode-analysis-GNCT-JCP}

\subsection{Phenolate in Vacuum}

\begin{figure}[h]
    \centering
    \includegraphics[width=1\textwidth]{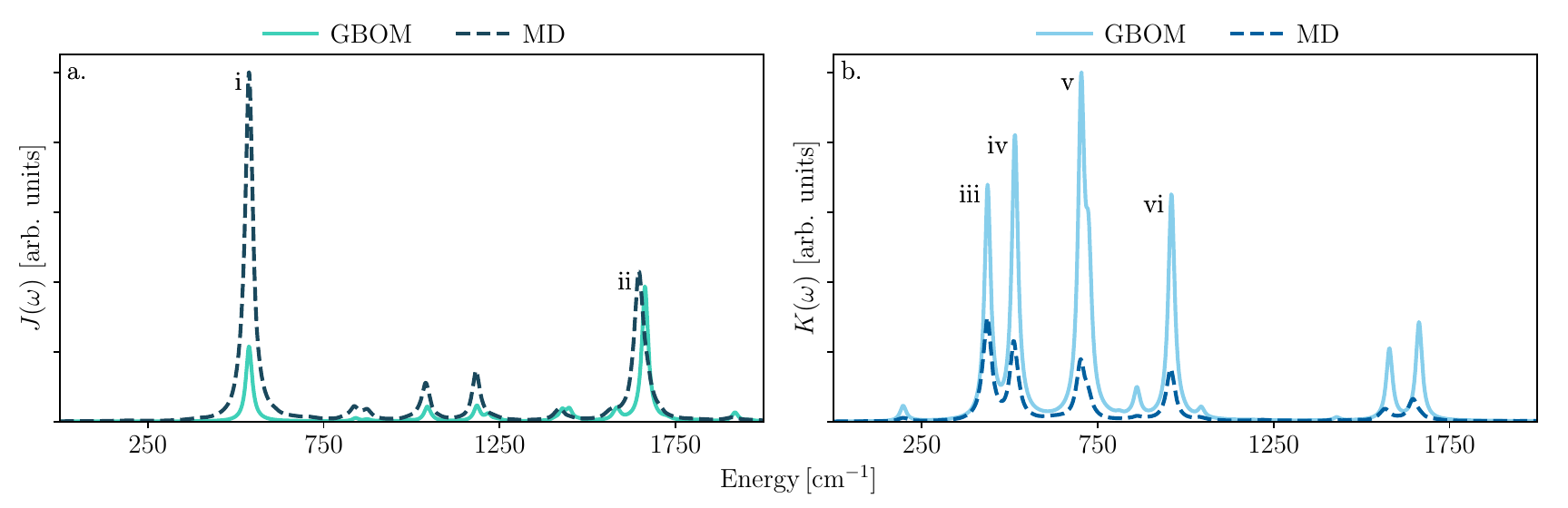}
    \caption{a) $J(\omega)$ and b) $K(\omega)$ of the phenolate ion, as computed from a GBOM parameterized from normal mode calculations as well as an MD calculation in vacuum. The most prominent features are assigned to ground-state normal modes.}
    \label{SIfig:phenol_vac_sds}
\end{figure}

We find that peak positions in the spectral densities of phenolate in vacuum, shown in Fig.~\ref{SIfig:phenol_vac_sds}, closely align with the normal modes calculated analytically for the ground state optimized geometry in Gaussian\cite{gdv-sup} at the 6-31+G*\cite{Dunning1990-sup}/CAM-B3LYP\cite{Yanai2004-sup} level of theory. The frequencies which couple most strongly to the energy gap and transition dipole moments have their corresponding normal modes shown in  Fig.~\ref{SIfig:phenol_modes} i-vi. Modes i and ii are symmetric in-plane ring deformations and C-C-H bends (A1), while modes iii-vi are symmetric out-of-plane bending modes. Here, mode iii maintains the C$_2$ symmetry axis around the C=O bond and oscillates between two possible boat conformations (A2), while modes iv, v and vi maintain the out-of-plane $\sigma_v$ symmetry (B1). Mode iv also creates a boat conformation in the carbon ring, but modes v and vi create chair conformations. As out-of-plane modes, both the A2 and B1 modes break the aromaticity of the $\pi$-system, allowing for larger changes in electron density in the excited state and thus a large change in transition dipole moment.

One can directly compute spectral densities $J(\omega)$ and $K(\omega)$ from the exact analytical correlation functions of the Generalized Brownian Oscillator Model (GBOM) Hamiltonian\cite{Zuehlsdorff2019}, a harmonic Hamiltonian that includes Duschinsky mode mixing and frequency changes upon excitation. The GBOM is the underlying Hamiltonian of the FCHT method. This allows us to probe the effect of the harmonic approximation of the PES inherent in the FCHT approach for a small, rigid molecule. Figure~\ref{SIfig:phenol_vac_sds} shows these spectral densities. While peak positions between the GBOM spectral densities and those constructed directly from MD are in good agreement, suggesting that anharmonic corrections are small, high-frequency peaks such as the mode corresponding to the symmetric in-plane ring deformation (ii) shift to lower frequency in the MD-derived spectral density. This confirms that the explicit sampling of the PES accounts for anharmonic effects, even in a molecule as small and rigid as phenolate. Additionally, peak intensities in the GBOM-derived spectral densities are markedly different from the MD-sampled counterparts, with $J(\omega)$ exhibiting higher intensity in the MD sampling, especially for the in-plane deformation mode i, whereas $K(\omega)$ exhibits lower intensity in the MD-sampling across all modes. This finding shows that even for a small, rigid molecule, the GNCT accounts for non-Condon effects in a substantially different way than FCHT, and that the explicit sampling of the PES captures effects that the Taylor expansion of the transition dipole around its minimum misses.    
\begin{figure}[t]
    \centering
    \includegraphics[width=1\textwidth]{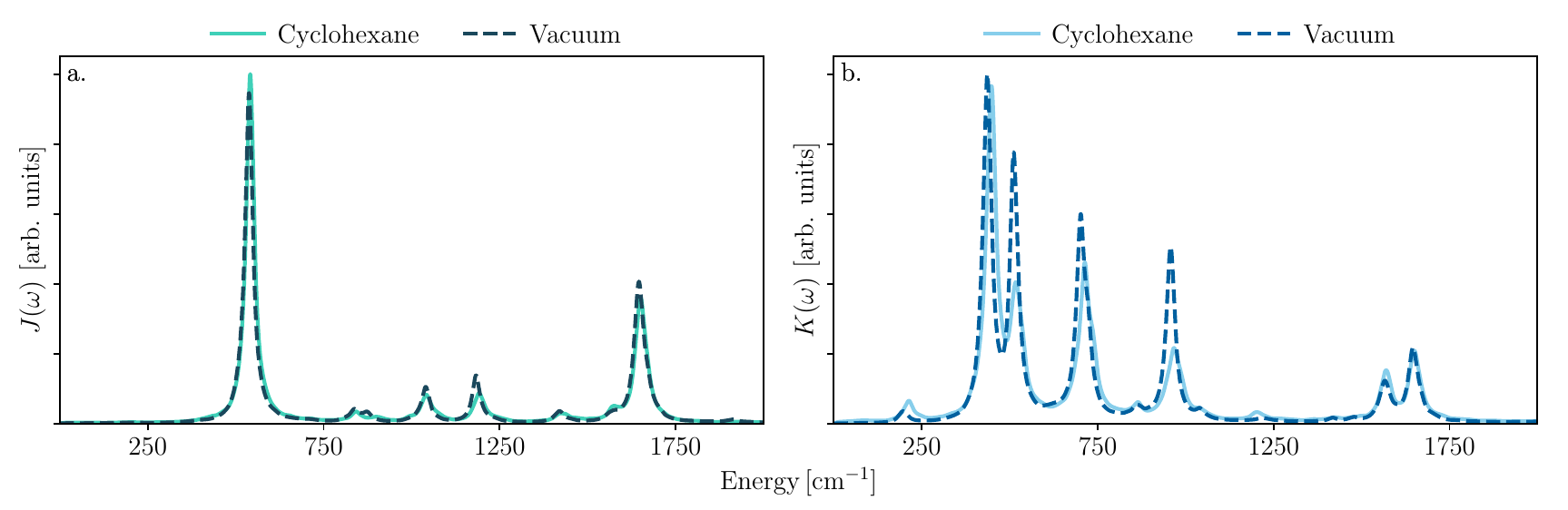}
    \caption{MD-based dpectral densities of the energy gap $J(\omega)$ (a) and transition dipole $K(\omega)$ (b) fluctuations for the phenolate ion in vacuum (dashed) and in cyclohexane (solid).}
    \label{SIfig:phenol_cyclohex_sds}
\end{figure}

\subsection{Phenolate in Cyclohexane}

Figure~\ref{SIfig:phenol_cyclohex_sds} shows the dipole and energy gap spectral densities computed for phenolate in cyclohexane compared to those computed for the isolated molecule in vacuum. The spectral densities are in good agreement, and only minor differences in peak intensities and positions are evident. Normal mode assignments are therefore identical in cyclohexane and in vacuum (see SI Fig.~\ref{SIfig:phenol_vac_sds}). This result confirms that the solute-solvent interactions are very weak, as can be seen in the close agreement of the two lineshapes in the GNCT approach shown in the main text. We further note that the spectral densities shown in Fig.~\ref{SIfig:phenol_cyclohex_sds} are in much closer agreement than the spectral densities for the vacuum MD simulation than the harmonic GBOM Hamiltonian shown in Fig.~\ref{SIfig:phenol_vac_sds}, confirming that solvent effects induced by cyclohexane are significantly weaker than anharmonic effects of the full chromophore PES sampled in our GNCT through MD. 
\begin{figure}[h]
    \centering
    \includegraphics[width=1\textwidth]{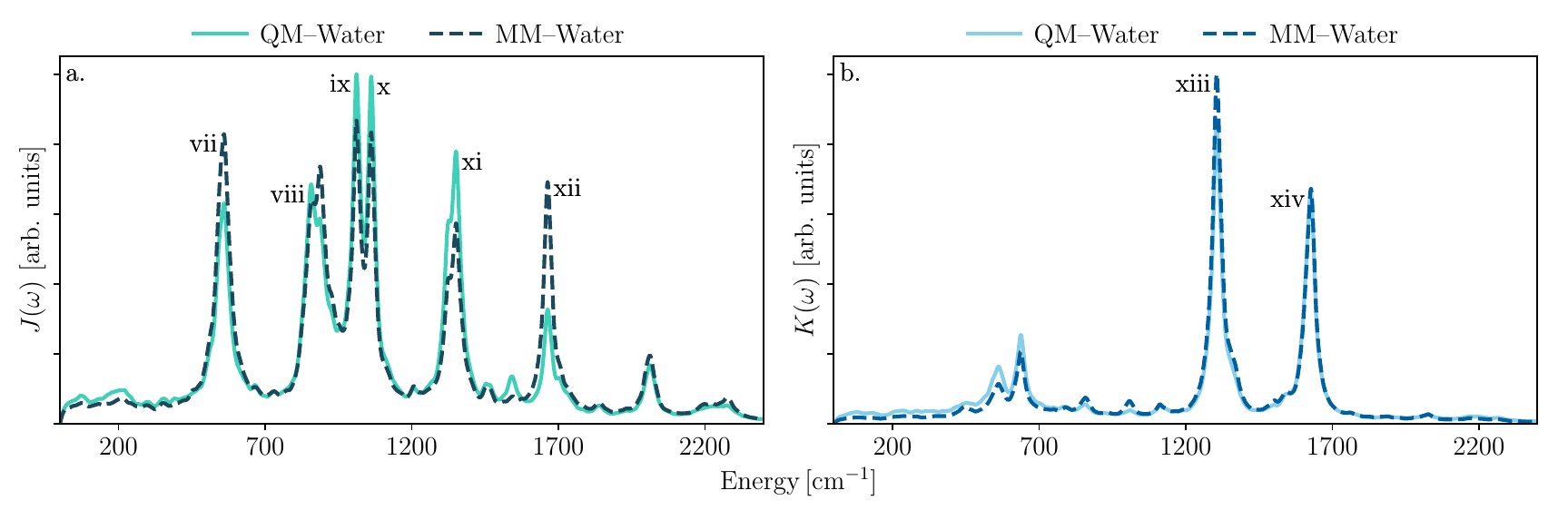}
    \caption{a) $J(\omega)$ and b) $K(\omega)$ of the phenolate ion as computed from MD simulations in pure MM point charge and 6~\AA~QM water solvent representations. The most prominent features are assigned to ground-state normal modes.}
    \label{SIfig:phenol_water_sds}
\end{figure}

\subsection{Phenolate in Water}

\begin{figure}[b]
    \centering
    \includegraphics[width=0.9\textwidth]{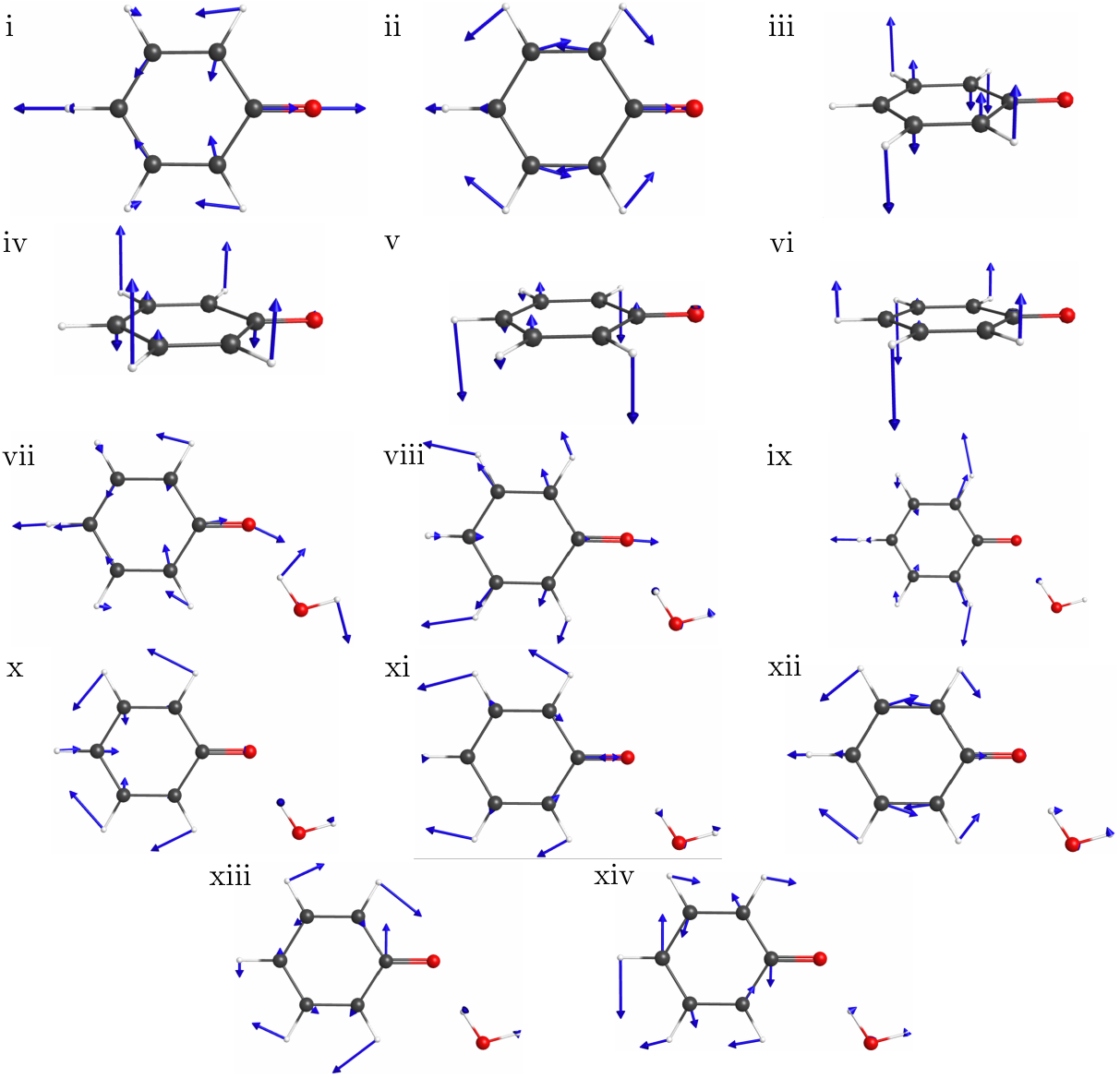}
    \caption{Important normal modes calculated for phenolate in vacuum for modes i-vi, and phenolate in PCM with an H-bonded water for modes vii-xiv. Each labeled mode corresponds to one of the peaks labeled in Figure~\ref{SIfig:phenol_vac_sds} and Figure~\ref{SIfig:phenol_water_sds}. Normal mode visualizations were generated using WebMO\cite{WebMO2022-sup}.
    }
    \label{SIfig:phenol_modes}
\end{figure}

Figure~\ref{SIfig:phenol_water_sds} assigns the peaks in the spectral densities of phenolate in both MM and QM solvent environments to the normal modes of the optimized ground state of phenolate at the 6-31+G*/CAM-B3LYP level of theory with a single water molecule hydrogen-bonded to the oxygen. We calculated these normal modes using the Polarizable Continuum Model (PCM)\cite{Cammi2005-sup, Mennucci2012-sup} with the dielectric constant of water in Gaussian. The frequencies most strongly coupled to the energy gap and transition dipole moment fluctuations have their corresponding normal modes shown in  Figure~\ref{SIfig:phenol_modes} modes vii-xiv. Like in vacuum, the strongest coupling to the S$_0\rightarrow$S$_1$ energy gap is from the symmetric in-plane ring deformations and C-C-H bending modes (A1 when considering only phenolate), but more modes have a strong coupling (modes vii-xii) compared to phenolate in vacuum. Each of modes vii-xii has an analogous mode and peak in Figure~\ref{SIfig:phenol_vac_sds}a, with mode vii corresponding to mode i and mode xii to mode ii, but modes analogous to modes viii-xi are very weakly coupled compared with i and ii. This can be understood as a result of the strong interaction of the phenolate with H-bonded water in mode vii, as well as some interaction in mode xii, which breaks the A1 symmetry; this allows modes viii-xi to become relatively more important. This effect becomes stronger when the surrounding solvent environment is treated quantum mechanically compared with force field treatment and the intensity of peaks vii and xii decreases further with the QM solvent model due to stronger interaction.

In an explicit solvent environment, the transition dipole moments are still weakly coupled to modes analogous to modes iii-vi, but these are no longer as important as modes xiii and xiv. Modes xiii and xiv are asymmetric in-plane ring deformations and C-C-H bending modes (B2 when considering only phenolate), but especially interesting are the strong motions of the carbon bound to the oxygen and the carbon and hydrogen $para$ to it. Modes xiii and xiv are the only important modes in which these atoms move significantly from the molecular axis, causing symmetry-breaking C-C=O and C-C-H bending modes that create a dipole moment that is unfavorable in vacuum, but becomes stabilized by the screening effect of the solvent. The modes analogous to modes xiii and xiv are present in vacuum as well, but xiii does not couple at all to the dipole moment and xiv couples more weakly. The second-highest-energy peak in Figure~\ref{SIfig:phenol_vac_sds}b corresponds to mode xiv.

\section{Spectral interference effects: How cross correlated fluctuations shift intensity and split peaks}

In Sec.~\ref{section:phenolate-cross correlation-effects-GNCT-JCP} of the main text, we introduce the correlated-interference factor, $\theta$, that quantifies the extent to which spectral interference controls lineshapes. In what follows, we show how the Cauchy-Schwartz inequality allows us to derive $\theta$,  relate $J(\omega)$, $K(\omega)$ and $|\mathbf{L(\omega)}|^2$, and further, how $\theta$ measures the ability of $\boldsymbol{\sigma_{\mathcal{A}}}(\omega)$ and $\sigma_{\mathcal{AA}}(\omega)$ contribute regions of positive and negative intensity as they are parameterized by $\mathbf{L}(\omega)$. We then establish a \textit{rigorous bound} on the magnitude of $|\mathbf{L}^2(\omega)|$ in which we provide the means to calculate the physically valid margin of spectral interference that is possible for a given chemical system. In Sec.~\ref{SI-sec:model-spectral-density-build-GNCT-JCP}, we return to the physically-motivated spectral densities employed in Sec.~\ref{sec:parameter-regimes-at-extrema} of the main text and use these spectral densities in Sec.~\ref{SI-section:symmetries-of-spectral-spectral-interference-GNCT-JCP} to illustrate how spectral interference begins to center at $\omega_{eg}^{\rm av}$ as $\theta \rightarrow 1$. Lastly, in Sec.~\ref{SI-section:symmetries-of-spectral-spectral-interference-GNCT-JCP}, we provide a general description of spectral interference that elucidates the shapes of $\boldsymbol{\sigma_{\mathcal{A}}}(\omega)$ and $\sigma_{\mathcal{AA}}(\omega)$.

\subsection{Derivation of the correlated-interference factor}\label{SI-section:correlated-interference-factor-derivation}

Since $\mathbf{L}(\omega)$ quantifies the strength of cross correlation between fluctuations of the energy gap and transition dipole, one can hypothesize that its structure should contain peaks at frequencies where the peaks of $J(\omega)$ and $K(\omega)$ overlap. The extent to which this overlap manifests in $\mathbf{L}(\omega)$ should therefore offer a means to quantify the extent of correlation that can in principle emerge in a spectrum. We suggest a form for the correlated interference factor that ranges between 0, indicating no correlation, and 1, indicating maximal correlation. To achieve this, we begin by invoking the Cauchy-Schwartz inequality to establish that 
\begin{equation}\label{SIeq:CS-inequality-for-J-and-K-GNCT-JCP}
    \int \text{d}\omega \, J(\omega) K(\omega) = \frac{\pi^2}{4} \sum_k |\boldsymbol{\alpha}_k|^2 \omega_k^2 d^2_k \leq \lVert J(\omega) \rVert \lVert K(\omega) \rVert.
\end{equation}
One can also obtain the sum of coefficients on the left side of Eq.~\eqref{SIeq:CS-inequality-for-J-and-K-GNCT-JCP} by integrating $|\mathbf{L}(\omega)|^2$, demonstrating that 
\begin{equation}\label{SIeq:L-power-spectrum-discrete-GNCT-JCP}
    \lVert \mathbf{L}(\omega) \rVert^2   
    = \frac{\pi^2}{4} \sum_k |\boldsymbol{\alpha}_k|^2 \omega_k^2 d_k^2 \leq \lVert J(\omega) \rVert \lVert K(\omega) \rVert,
\end{equation}
Positing the following form for the correlated-interference factor,
\begin{equation}\label{SIeq:CS-inequality-L-to-J-and-K-GNCT-JCP}
    \theta = \frac{\lVert \mathbf{L} (\omega) \rVert^2}{\lVert J (\omega) \rVert \lVert K(\omega) \rVert},
\end{equation}
it becomes clear that Eq.~\eqref{SIeq:CS-inequality-L-to-J-and-K-GNCT-JCP} suggests the following range for the interference factor:
\begin{equation}
    0 \leq \theta \leq 1.
\end{equation}
Indeed, since each component of this ratio is non-negative, $\theta$ can take the \textit{minimum value of 0}, corresponding to the case where the fluctuations of the energy gap and transition dipole are completely uncorrelated. The Cauchy-Schwartz inequality indicates that $\theta$ can take \textit{a maximum value of 1}, indicating perfect correlation between $J(\omega)$ and $K(\omega)$. Yet, as we stated in Sec.~\ref{section:phenolate-cross correlation-effects-GNCT-JCP} of the main manuscript, spectral densities based on MD sampling of the ground state PES for the transition dipole and energy gap fluctuations can lead to $|\mathbf{L}(\omega)|^2 \neq J(\omega) K(\omega)$ and therefore a possible interference factor $\theta > 1$. We argued in Sec.~\ref{section:beyond-physics} that \textit{$\theta > 1$ is unphysical and suggests the breakdown of the Gaussian approximation to be unphysical}. We prove this below.

We prove that $\theta > 1$ is unphysical by demonstrating that $J(\omega)K(\omega) = |\mathbf{L}(\omega)|^2$ in the limit of a perfect Gaussian prescription for the fluctuations with perfect correlation. We begin by recalling that the spectral densities $J(\omega)$, $K(\omega)$ and $\mathbf{L}(\omega)$ arise from the Fourier transform of quantum equilibrium time correlation functions (see Eqs.~\eqref{eq:energy-spectral-density-in-terms-of-tcf}, \eqref{eq:discrete_K} and \eqref{eq:discrete_L}) of the main text and Eqs.~\eqref{SIeq:quantum-dipole-auto-tcf-GNCT-JCP} and Eqs.~\eqref{SIeq:quantum-egap-dipole-cross-tcf-GNCT-JCP}). Employing the convolution theorem, we recast the product $J(\omega) K(\omega)$ as a convolution of their corresponding quantum time correlation functions and set this quantity equal to the Fourier transform of an undetermined time-dependent function, $\mathcal{C}(t)$, 
\begin{equation}\label{SIeq:convolution-JK-spectral-densities-CI-factor-GNCT-JCP}
\begin{split}
    J(\omega) K(\omega) &= - \Theta(\omega) \int_{-\infty}^{\infty} \text{d} t \, \erm^{i \omega t} \int_{-\infty}^{\infty} \text{d} \tau \, C^{''}_{\delta U \delta U} (\tau) C^{''}_{\delta \mu \delta \mu} (t - \tau) \\ &\equiv \int_{-\infty}^{\infty} \text{d} t \, \erm^{i \omega t} \mathcal{C}(t).
\end{split}
\end{equation}
We then find an explicit form for $\mathcal{C}(t)$ in terms of $\{\omega_j, d_j, \boldsymbol{\alpha}_j \}$,
\begin{equation}\label{SIeq:C-JK-convolution}
\begin{split}
    \mathcal{C}(t) &= -\int_{-\infty}^{\infty} \text{d} \tau \, C^{''}_{\delta U \delta U} (\tau) C^{''}_{\delta \mu \delta \mu} (t - \tau) \\ &= -\frac{1}{4}\sum_{j,k} \frac{|\boldsymbol{\alpha}_j|^2}{\omega_j} \omega^3_k d^2_k \int_{-\infty}^{\infty} \text{d} \tau \, \sin(\omega_j \tau) \sin(\omega_k [t-\tau]) \\&= -\frac{1}{4}\sum_{j,k} \frac{|\boldsymbol{\alpha}_j|^2}{\omega_j} \omega^3_k d^2_k \int_{-\infty}^{\infty} \text{d} \tau \, \sin(\omega_j \tau) \Big[ \sin(\omega_k t) \cos(\omega_k \tau) - \cos(\omega_k t) \sin(\omega_k \tau) \Big] \\&= \frac{1}{4}\sum_{j,k} \frac{|\boldsymbol{\alpha}_j|^2}{\omega_j} \omega^3_k d^2_k \cos(\omega_k t) \int_{-\infty}^{\infty} \text{d} \tau \, \sin(\omega_j \tau) \sin(\omega_k \tau) \\&= \frac{\pi}{2}\sum_{j,k} \frac{|\boldsymbol{\alpha}_j|^2}{\omega_j} \omega^3_k d^2_k \cos(\omega_k t) \Big[ \delta(\omega_j - \omega_k) - \delta(\omega_j + \omega_k) \Big] \\ &= \frac{\pi}{4}\sum_{j,k} |\boldsymbol{\alpha}_j|^2 \omega^2_j d^2_j \cos(\omega_j t). 
\end{split}
\end{equation}
We then substitute this result in Eq.~\eqref{SIeq:C-JK-convolution} back into Eq.~\eqref{SIeq:convolution-JK-spectral-densities-CI-factor-GNCT-JCP} to find that, 
\begin{equation}\label{SIeq:time-convolution-autos-tcfs-egap-dipole-CI-factor-GNCT-JCP}
\begin{split}
    J(\omega) K(\omega) &= \Theta(\omega) \mathcal{C} (\omega)
    \\&= \frac{\pi}{4}\sum_{j,k} |\boldsymbol{\alpha}_j|^2 \omega^2_j d^2_j \Theta(\omega) \int_{-\infty}^{\infty} \text{d} t \, \erm^{i \omega t} \cos(\omega_j t) \\ &= \frac{\pi^2}{4}\sum_{j,k} |\boldsymbol{\alpha}_j|^2 \omega^2_j d^2_j \delta(\omega - \omega_j).
\end{split}
\end{equation}

We now perform a similar manipulation for the $\mathbf{L}(\omega)$ power spectrum, which we set equal to the Fourier transform of a second undetermined time-dependent function, $\tilde{\mathcal{C}}(t)$
\begin{equation}\label{SIeq:convolution-L-squared-spectral-density-CI-factor-GNCT-JCP}
\begin{split}
    |\mathbf{L}(\omega)|^2 &= - \Theta(\omega) \int_{-\infty}^{\infty} \text{d} t \, \erm^{i \omega t} \int_{-\infty}^{\infty} \text{d} \tau \, \mathbf{C}^{''}_{\delta U \delta \mu} (\tau) \cdot \mathbf{C}^{''}_{\delta U \delta \mu} (t - \tau) \\ &\equiv \int_{-\infty}^{\infty} \text{d} t \, \erm^{i \omega t} \tilde{ \mathcal{C}}(t),
\end{split}
\end{equation}
and employ a similar strategy to find an explicit form for $\tilde{\mathcal{C}}(t)$ in terms of $\{\omega_j, d_j, \boldsymbol{\alpha}_j \}$,
\begin{equation}\label{SIeq:time-convolution-cross-tcfs-egap-dipole-CI-factor-GNCT-JCP}
\begin{split}
    \tilde{ \mathcal{C}}(t) &= \frac{1}{4} \sum_{j,k} \boldsymbol{\alpha}_j \mdot \boldsymbol{\alpha}_k d_j d_k \omega_j \omega_k \int_{-\infty}^{\infty} \text{d} \tau \, \sin(\omega_j \tau) \sin(\omega_k[t-\tau]) \\&= \frac{\pi}{4} \sum_{j} |\boldsymbol{\alpha}_j|^2 \omega_j^2 d^2_j \cos(\omega_j t).
\end{split}
\end{equation}
We then evaluate the Fourier transform of $\tilde{ \mathcal{C}}(t)$ to conclude that,
\begin{equation}\label{SIeq:equivalence-between-L-squared-and-JK-JCP-GNCT}
    \Theta(\omega) \mathcal{C} (\omega) = \Theta(\omega) \tilde{\mathcal{C}} (\omega).
\end{equation}
Hence, for a system displaying statistical non-Condon effects where nuclear motions induce \textit{Gaussian fluctuations} of the energy gap and transition dipole, the spectral densities that encode the dynamics of \textit{perfectly correlated fluctuations} satisfy the relationship,
\begin{equation}
    |\mathbf{L}(\omega)|^2 = J(\omega) K(\omega),
\end{equation}
implying the previously proposed range of \textit{physically allowed} valued for $\theta$: 
\begin{equation}
    0 \leq \theta \leq 1. 
\end{equation}
Violation of this bound indicates the breakdown of the Gaussian assumption. This result further validates our use of $L_{\rm JK} (\omega)$ to model correlated and anti-correlated peaks in Secs.~\ref{section:tight-tdp} and \ref{section:broad-tdp} in the main text by placing a \textit{rigorous upper bound} on the intensity of peaks in $\mathbf{L}(\omega)$ determined by the overlap between $J(\omega)$ and $K(\omega)$.

Since $\theta$ from an MD-based $\mathbf{L}_{\rm MD}(\omega)$ need not follow $|\mathbf{L}(\omega)|^2 = J(\omega) K(\omega)$, we introduce the \textit{cross correlation ratio} to track amount of interference reduction when $|\mathbf{L}(\omega)|^2 \neq J(\omega) K(\omega)$, 
\begin{equation}\label{SIeq:cross correlation-ratio-CI-factor-GNCT-JCP}
    r_\theta = \frac{\theta_{\rm MD}}{\theta_{\rm JK}}.
\end{equation}
Here $\theta_{\rm MD}$ and $\theta_{\rm JK}$ are the correlated-interference factors derived from $\mathbf{L}_{\rm MD} (\omega)$ and $\mathbf{L}_{\rm JK} (\omega)$, respectively. $r_\theta$ ranges between 0, in which $\mathbf{L}_{\rm MD} (\omega)$ contains peaks with intensities that are negligible compared to those in $J(\omega)$ and $K(\omega)$, and 1, in which $\mathbf{L}_{\rm MD} (\omega)=\mathbf{L}_{\rm JK} (\omega)$. What is more, because $r_\theta$ tracks whether the MD-sampling of a system's ground state PES underestimates ($r_\theta < 1$), overestimates ($r_\theta > 1$), or exactly captures ($r_\theta = 1$) the Gaussian limit of cross correlation between nuclear-motion induced fluctuations of the energy gap and transition dipole, \textit{$r_\theta$ provides a simple metric for how Gaussian these cross correlated nuclear motions are}.

\subsection{Constructing $\mathbf{L}(\omega)$ from $J(\omega)$ and $K(\omega)$}\label{SI-sec:model-spectral-density-build-GNCT-JCP}

Here, we aim to elucidate how physically motivated features (such as low-frequency collective solvent modes and highly localized vibrations) in $J(\omega)$, $K(\omega)$, and $\mathbf{L}(\omega)$ impact the resulting absorption spectrum. We first remove unnecessary complexity in the MD-derived spectral densities so that we may focus on how their major features lead to qualitative differences in the resulting spectrum. In particular, we introduce three simplifications: 
\begin{enumerate}
    \item We adopt a simplified Gaussian limit of the cross energy gap-transition dipole spectral density where we replace the vector-valued components of the optical response with scalar versions:  $\mathbf{L} (\omega) \rightarrow L_{\rm JK} (\omega)$ and $\langle \hat{\boldsymbol{\mu}}_{ge} \rangle \rightarrow \lVert \langle \hat{\boldsymbol{\mu}}_{ge} \rangle \rVert$.  
    
    \item We take all modes to be either correlated or anticorrelated by asserting that
    \begin{equation}
        L_{\rm JK} (\omega) = \pm \sqrt{J(\omega) K(\omega)}.
    \end{equation}
    
    \item We construct approximate spectral densities inspired by the MD-based spectral densities for phenolate (Sec.~S\ref{SI-section:computational-details-GNCT-JCP}) with combinations of physically motivated models of coupling to collective solvent motions and localized vibrations such that the spectral densities can be decomposed into simple low- and high-frequency regions:
    \begin{subequations}\label{SIeq:model-sd-formulas}
    \begin{align}
        J(\omega) &= 2 \lambda_{\rm l} \omega_{J, l}\frac{ \omega}{\omega^2 + \omega^2_{J, l}} + \lambda_h \erm^{-\frac{1}{2 \sigma_J^2}(\omega - \omega_{J, h})^2}, \label{SIeq:model-egap-spectral-density-GNCT-JCP} \\
        K(\omega) &= 2 D_{ l} \omega_{K, l} \frac{\omega}{\omega^2 + \omega^2_{K, l}} + D_h \erm^{-\frac{1}{2 \sigma_K^2}(\omega - \omega_{K,h})^2} \label{SIeq:model-dipole-spectral-density-GNCT-JCP}.
    \end{align}
    \end{subequations}
\end{enumerate}
The first (low-frequency) terms in Eq.~\eqref{SIeq:model-egap-spectral-density-GNCT-JCP} and Eq.~\eqref{SIeq:model-dipole-spectral-density-GNCT-JCP} have a Debye form and account for the collective, overdamped nuclear motions, characteristic of condensed phase environments. The second (high-frequency) terms are Gaussian profiles used to mimic coupling to localized vibrations. The subscripts $l$ and $h$ denote low- and high-frequency spectral density features, respectively; $\omega_{J, x}$ and $\omega_{K, x}$ with $x \in \{l, h \}$ denote low and high characteristic frequencies (average velocities) of nuclear motions that cause fluctuations in the energy gap and transition dipole respectively; $\lambda$ is the energy gap reorganization energy; $D \propto \gamma_{\mu}$ quantifies the extent to which nuclear fluctuations cause the transition dipole to change and scales with its variance; and $\sigma$ is the standard deviation that encodes the range of high-frequency vibronic interactions which couple to the energy gap and transition dipole. 

\begin{figure*}
    \centering
\includegraphics[width=0.6\columnwidth]{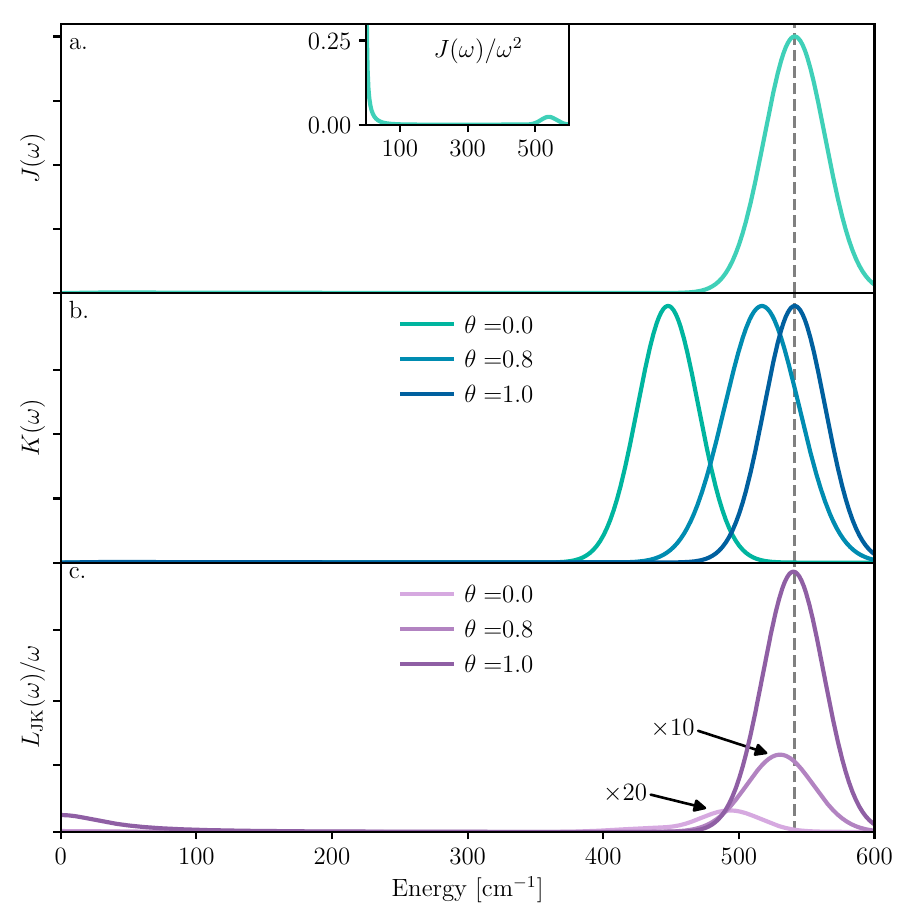}
\caption{Model spectral densities used to generate spectral interference patterns for the analysis in the main text and in Figs.~\ref{SI-fig:theta-1-limit-of-linear-spectral-interference}. and \ref{SI-fig:theta-1-limit-of-quadratic-spectral-interference} }
\label{SI-fig:model-spectral-densities-GNCT-JCP}
\end{figure*}

Figure~\ref{SI-fig:model-spectral-densities-GNCT-JCP} shows our model spectral densities inspired by features in the MD-based $J(\omega)$ and $K(\omega)$ for phenolate in cyclohexane and used to predict spectra and describe spectral interference effects in Secs.~\ref{section:tight-tdp}, \ref{section:broad-tdp} and \ref{section:beyond-physics} of the main text. Beginning with Fig.~\ref{SI-fig:model-spectral-densities-GNCT-JCP}a, $J(\omega)$ displays a negligible low-frequency region with $\lambda_l \sim 44 \wn$ and $\omega_{J,l}= 40 \wn$, reflecting the weak interactions between phenolate and cyclohexane. For simplicity in our analysis, the high-frequency profile of our model $J(\omega)$ remains centered at $\sim 500 \wn$ with an amplitude of $\lambda_h \sim 18000 \wn$, corresponding to in-plane deformations of the phenolate ring (See Sec.~\ref{SI-sec:spectral-density-normal-mode-analysis-GNCT-JCP}). 

Figure~\ref{SI-fig:model-spectral-densities-GNCT-JCP}b shows three models for $K(\omega)$ also showing negligible low-frequency profiles characteristic of the cyclohexane environment. Here, as we tune $\theta$ from $\sim 0$ to $\sim 0.8$, we shift the center of the high-frequency peak in $K(\omega)$ from $\sim 450 \wn$ to $\sim 550 \wn$ (See Figs.~\ref{fig:linear-interference-GNCT} and \ref{fig:peak_splitting_interference_GNCT}a and b). These two values for the center of the high-frequency peak correspond to out-of-plane ring deformations. Lastly, to obtain $\theta = 1$, we center the high-frequency component in $K(\omega)$ at $\sim 550 \wn$ and assign it a shape \textit{identical} to that of $J(\omega)$. We then tune $\varphi$ by scaling the initial $D_l \sim 2.57 \times 10^{-8}$ $\rm D^2/cm^{-1}$ and $D_h \sim 2.25 \times 10^-3$ $\rm D^2/cm^{-1}$ identically by 0.3 for $\varphi = -0.35$ to calculate the spectra in \ref{fig:linear-interference-GNCT} of the main text and 0, 200, and 100000 for $\varphi$ values of -1, 0, 1 for the spectra in Fig.~\ref{fig:peak_splitting_interference_GNCT}a, b, and c(d), respectively. 

Lastly, Fig.~\ref{SI-fig:model-spectral-densities-GNCT-JCP}c shows the $\omega$-effective $L_{\rm JK}(\omega)/\omega$ from $J(\omega)$ in panel a and the three curves for $K(\omega)$ in panel b. Here, one observes that as $\theta$ increases from 0 to 1, the high-frequency profile of $L_{\rm JK} (\omega)/\omega$ begins to overlap well with those of $J(\omega)$ and $K(\omega)$ as it approaches a center frequency of $\sim 550 \wn$. Additionally, because of the increasing overlap with $\theta$ approaching 1, $L_{\rm JK} (\omega)/\omega$ incrementally gains intensity across both low- and high-frequency regions with those constructed from $\theta$ values of 0.0 and 0.8 being scaled by factors of 10 and 20 to compare. In all calculated spectra, when scaling $K(\omega)$ by a factor of $a$, $L_{\rm JK}(\omega)$ is scaled by a corresponding $\sqrt{a}$.

\subsection{The value of $\theta$ tracks the relationship between spectral interference from $\langle \hat{\boldsymbol{\mu}}_{ge}\rangle \cdot \boldsymbol{\sigma_{\mathcal{A}}}(\omega)$ and $\sigma_{\mathcal{AA}}(\omega)$ to $\omega_{eg}^{\rm av}$}
\label{SI-section:symmetries-of-spectral-spectral-interference-GNCT-JCP}

In Eqs.~\eqref{SIeq:A-continuous} and \eqref{SIeq:B-continuous} of the main text, we show how our non-Condon dynamical quantities $\boldsymbol{\mathcal{A}}(t)$ 
and $\mathcal{B}(t)$ can be expressed in terms of time derivatives of the universal function derived from the Condon limit, $\Omega(\beta, \omega, t)$. Inspired by the insight our correlated interference factor $\theta$ can provide on spectral interference patterns based on how well the features of $J(\omega)$, $K(\omega)$ and $\mathbf{L}(\omega)$ overlap, we rewrite the non-Condon spectral densities in terms of a component that corresponds to $J(\omega)$ and a second that deviates, 
\begin{subequations}
\begin{align}
    \mathbf{L}(\omega) &= \boldsymbol{\eta} J(\omega) + \boldsymbol{\mathcal{L}}(\omega), \\ K(\omega) &= |\boldsymbol{\eta}|^2 J(\omega) + \mathcal{K}(\omega).
\end{align}
\end{subequations}
Here, $\boldsymbol{\eta}$ is a proportionality vector with units ${\rm D} \mdot {\rm s}$, whereas, $\boldsymbol{\mathcal{L}}(\omega)$, and $\mathcal{K}(\omega)$, are the spectral deviations of $\mathbf{L}(\omega)$ and $K(\omega)$ from the shape of $J(\omega)$. The purpose of this re-writing is to show how the spectral contributions from $\boldsymbol{\sigma}_{\mathcal{A}}(\omega)$ and $\sigma_{\mathcal{AA}}(\omega) + \sigma_{\mathcal{B}}(\omega)$ take on a simple form as $\theta$
approaches 1. In the limit where $\theta = 1$, both $\mathcal{K}(\omega)$ and $\boldsymbol{\mathcal{L}}(\omega)$ are equivalently zero and all spectral densities have the same shape. 

Before we analyze spectral interference in the highly correlated limit of the spectral densities, we need to establish how the dynamical quantities $\boldsymbol{\mathcal{A}}(t)$ and $\mathcal{B}(t)$ can be simplified when $\theta = 1$. In the limit where $K(\omega)$ and $\mathbf{L}(\omega)$ become proportional to $J(\omega)$, Eqs.~\eqref{eq:non-Condon-A-spectral-density-GNCT-JCP} and \eqref{eq:non-Condon-B-spectral-density-GNCT-JCP} from the main text simplify to,
\begin{subequations}
\begin{align}
    \boldsymbol{\mathcal{A}}_{\theta=1}(t) &= i \frac{\boldsymbol{\eta}}{\pi} \int_0^{\infty} \text{d}\omega \, \frac{J(\omega)}{\omega^2} \pdiv{\Omega(\beta, \omega, t)}{t} \nonumber \\ 
    &= i \boldsymbol{\eta} \pdiv{g_2(t)}{t}\label{SIeq:theta-1-non-Condon-A-spectral-density-GNCT-JCP}, \\
    \mathcal{B}_{\theta=1}(t) &= \frac{|\boldsymbol{\eta}|^2}{\pi} \int_0^{\infty} \text{d}\omega \, \frac{J(\omega)}{\omega^2} \pdiv{^2 \Omega(\beta, \omega, t)}{t^2} \nonumber \\ &= |\boldsymbol{\eta}|^2 \pdiv{^2 g_2(t)}{t^2}\label{SIeq:theta-1-non-Condon-B-spectral-density-GNCT-JCP}.
\end{align}
\end{subequations}
Hence, when $\theta = 1$, all dynamics that arise from symmetry-breaking fluctuations are captured by time derivatives of the lineshape function, $g_2(t)$. Now that we have established these simple relationships, we are equipped to interrogate spectral features that arise when $\theta=1$ in both the low and high variance limits of spectral interference.  

\begin{figure*}[b]
    \centering
\includegraphics[width=1\columnwidth]{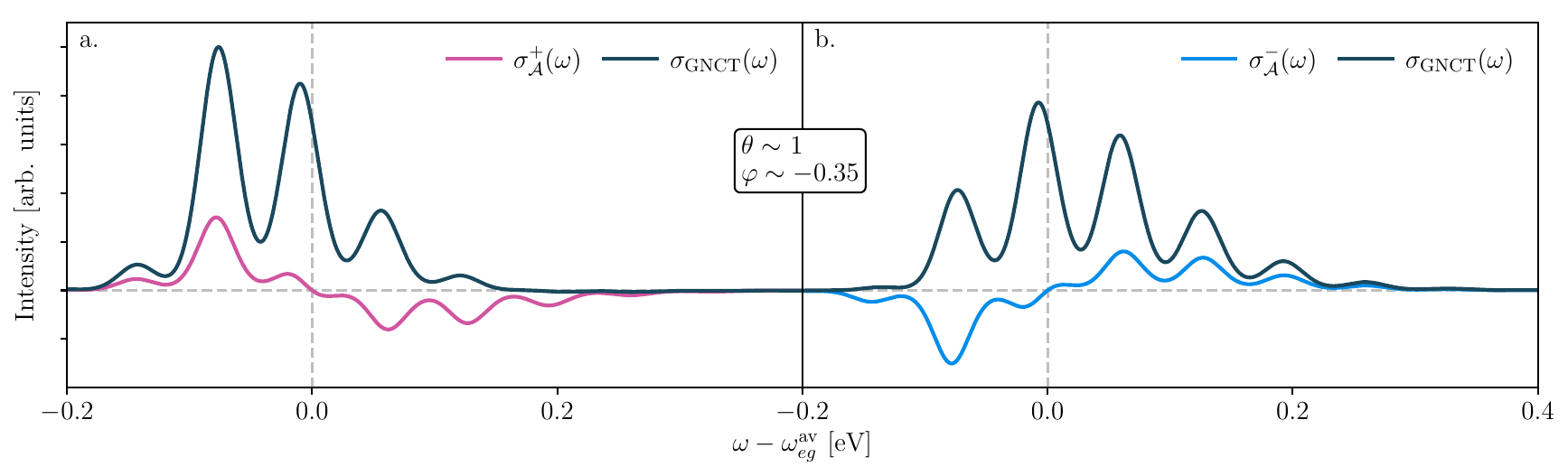}
\caption{Spectral interference in the low variance ($\varphi < 0$) and highly correlated limit ($\theta \sim 1$) causes intensity to shift either to the (a) red or (b) blue of the thermally averaged transition frequency $\omega_{eg}^{\rm av}$. The red-shifting contribution ($\sigma^{+} (\omega)$) arises from correlated peaks, $\langle \hat{\boldsymbol{\mu}}_{eg} \rangle \cdot \boldsymbol{\eta} J(\omega) > 0$, and the blue-shifting contribution ($\sigma^{-} (\omega)$) arises from anticorrelated peaks, $\langle \hat{\boldsymbol{\mu}}_{eg} \rangle \cdot \boldsymbol{\eta} < 0$.
}
\label{SI-fig:theta-1-limit-of-linear-spectral-interference}
\end{figure*}
To begin, we consider non-Condon \textit{spectral shifts in the low variance limit} where the non-Condon factor $\varphi<0$ and $\theta$ can range from 0 (i.e., the Condon limit) and the highly-correlated limit with a value of 1. Using Eq.~\eqref{SIeq:theta-1-non-Condon-A-spectral-density-GNCT-JCP} one finds that for $\theta=1$ the spectral contribution from $\boldsymbol{\sigma_{\mathcal{A}}}(\omega)$ becomes, 
\begin{equation}\label{SIeq:theta-1-non-Condon-spectral-A-term-GNCT-JCP}
\begin{split}
    \boldsymbol{\sigma}_{\mathcal{A},\theta=1}(\omega) &= \int_{-\infty}^{\infty} \text{d}t \, \erm^{i(\omega-\omega_{eg}^{\rm av})t}  \boldsymbol{\mathcal{A}}_{\theta=1}(t) \\ &= -i \boldsymbol{\eta} \int_{-\infty}^{\infty} \text{d}t \, \erm^{i(\omega-\omega_{eg}^{\rm av})t} \pdiv{}{t} \erm^{- g_2(t)} \\ &= \boldsymbol{\eta}(\omega - \omega_{eg}^{\rm av}) \sigma_\smlsub{\rm GCT}(\omega).
\end{split}
\end{equation}
Equation~\eqref{SIeq:theta-1-non-Condon-spectral-A-term-GNCT-JCP} reveals that in the highly correlated limit, non-Condon effects from $\boldsymbol{\sigma_{\mathcal{A}
}}(\omega)$ shift spectral intensity \textit{exactly about} $\omega_{eq}^{\rm av}$!

We then turn to the non-Condon \textit{spectral splitting in the high variance limit} arising as $\varphi$ and $\theta$ approach 1. Specifically, in the highly correlated limit, we can use Eqs.~\eqref{SIeq:theta-1-non-Condon-A-spectral-density-GNCT-JCP} and \eqref{SIeq:theta-1-non-Condon-B-spectral-density-GNCT-JCP} to express $\sigma_{\mathcal{AA}}(\omega) + \sigma_{\mathcal{B}}(\omega)$ as,   
\begin{equation}\label{SIeq:theta-1-non-Condon-spectral-AA-B-sum-GNCT-JCP}
\begin{split}
     \sigma_{\mathcal{AA}, \theta=1}(\omega) + \sigma_{\mathcal{B}, \theta=1}(\omega) &= \int_{-\infty}^{\infty} \text{d}t \, \erm^{i(\omega-\omega_{eg}^{\rm av})t} \left[ \mathcal{A}^2_{\theta=1} (t) + \mathcal{B}_{\theta=1}(t) \right] \erm^{-g_2(t)} \\&= |\boldsymbol{\eta}|^2 \int_{-\infty}^{\infty} \text{d}t \, \erm^{i(\omega-\omega_{eg}^{\rm av})t} \pdiv{^2}{t^2} \erm^{-g_2(t)} \\&= |\boldsymbol{\eta}|^2 (\omega- \omega_{eg}^{\rm av})^2 \sigma_\smlsub{\rm GCT} (\omega).
\end{split}
\end{equation}
This reveals that spectral interference is also related $\omega_{eg}^{\rm av}$ in this limit with peaks splitting \textit{exactly} at $\omega_{eg}^{\rm av}$. In contrast to Eq.~\eqref{SIeq:theta-1-non-Condon-spectral-A-term-GNCT-JCP}, Eq.~\eqref{SIeq:theta-1-non-Condon-spectral-AA-B-sum-GNCT-JCP} shows that the spectral interference is related to $\omega_{eg}^{\rm av}$ through the sum of two terms. 

Figure~\ref{SI-fig:theta-1-limit-of-linear-spectral-interference} shows the spectral signatures caused by non-Condon effects important in the low variance limit where the total spectrum reduces to $\sigma\smlsub{\rm GNCT}(\omega) = |\langle \hat{\boldsymbol{\mu}}_{ge} \rangle|^2 \delta(\omega) + 2 \text{Re}\{\langle \hat{\boldsymbol{\mu}}_{ge} \rangle\} \mdot \boldsymbol{\sigma_{\mathcal{A}}} (\omega) $. As in Sec.~\ref{section:tight-tdp} of the main text, here we show the total GNCT spectrum and the underlying red- and blue-shifting spectral contributions $\sigma^{+}_{\mathcal{A}} (\omega)$ (panel a) and $\sigma^{-}_{\mathcal{A}} (\omega)$ (panel b) that arise from $\langle \hat{\boldsymbol{\mu}}_{ge} \rangle \cdot \mathbf{L}(\omega)$ containing either correlated (positive) or anticorrelated (negative) peaks, respectively. When $\theta = 1$ this simplifies to $\langle \hat{\boldsymbol{\mu}}_{ge} \rangle \cdot \boldsymbol{\eta} J(\omega)$, and a point of zero-intensity emerges in both $\sigma_{\mathcal{A}}^+(\omega)$ and $\sigma_{\mathcal{A}}^-(\omega)$ \textit{exactly} at $\omega_{eg}^{\rm av}$, around which spectral intensity shifts.

\begin{figure*}[h]
    \centering
\includegraphics[width=.65\columnwidth]{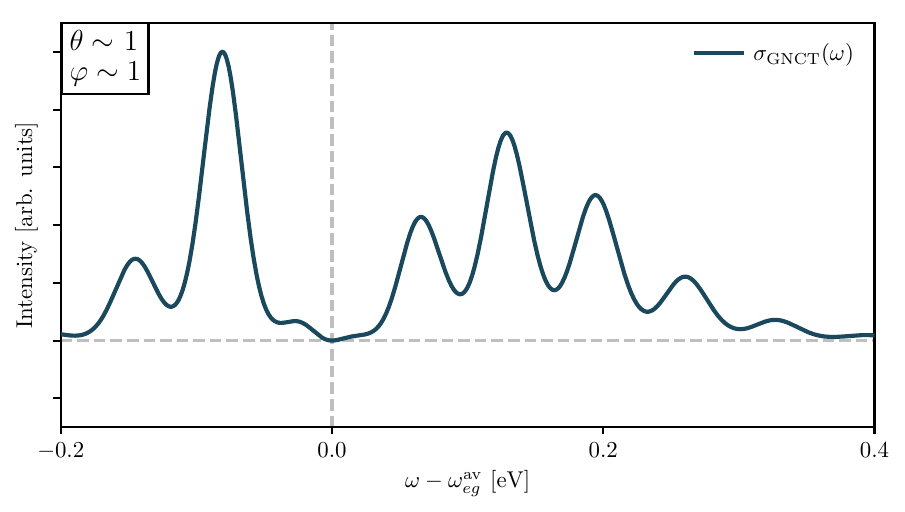}
\caption{Spectral interference in the high variance ($\varphi \sim 1$) and highly correlated limit ($\theta \sim 1$) causes the lineshape of a single electronic transition to split around the thermally averaged transition frequency $\omega_{eg}^{\rm av}$. The splitting is a consequence of the interference between $\sigma_{\mathcal{B}} (\omega)$, a purely positive contribution, and $\sigma_{\mathcal{AA}} (\omega)$, the contribution with regions of both positive and negative intensity.   
}
\label{SI-fig:theta-1-limit-of-quadratic-spectral-interference}
\end{figure*}

Figure \ref{SI-fig:theta-1-limit-of-quadratic-spectral-interference} shows the impact of spectral signatures caused by non-Condon effects in the high variance and highly correlated limit where the total spectrum becomes $\sigma\smlsub{\rm GNCT} (\omega) = \sigma_{\mathcal{AA}}(\omega) + \sigma_{\mathcal{B}}(\omega)$. Here, spectral interference dominates ($\theta = 1$), leading a single vibronic progression to split into two. With this optimal correlation, $\sigma_{\mathcal{AA}}(\omega)$ and $\sigma_{\mathcal{B}}(\omega)$ add together perfectly, leading to two distinct peaks separated by a point of zero-intensity occurring at \textit{exactly} $\omega_{eg}^{\rm av}$.

\subsection{Spectral interference effects generalized}\label{SI-sec:spectral-interference-GNCT-JCP}

We now show how representing the non-Condon spectral contributions using the discrete nuclear modes that modulate the energy gap and transition dipole elucidates how distinct spectral features control the lineshape beyond the Condon limit. To do so, we substitute Eqs.~\eqref{SIeq:K-auxilliary-spec-dens} into our expression for $\sigma_{\mathcal{B}}(\omega)$ (Eq.~\eqref{SIeq:non-Condon-B-spectrum-final-convolution-GNCT-JCP}) and \eqref{SIeq:L-auxilliary-spec-dens} into the expressions for $\boldsymbol{\sigma_{\mathcal{A}}}(\omega)$ and $\sigma_{\mathcal{AA}}(\omega)$ (Eqs.~\eqref{SIeq:non-Condon-A-spectrum-final-convolution-GNCT-JCP} and \eqref{SIeq:non-Condon-AA-spectrum-final-convolution-GNCT-JCP}) to write, 
\begin{subequations}
\begin{align}
    \boldsymbol{\sigma_{\mathcal{A}}}(\omega) &= \boldsymbol{\lambda}_{\mathbf{L}} \sigma_\smlsub{\rm GCT} (\omega) - \frac{1}{2} \sum_k \boldsymbol{\alpha}_k d_k  \frac{\sigma_\smlsub{\rm GCT} (\omega - \omega_k)}{1 - \erm^{-\beta \omega_k}} \label{SIeq:non-Condon-A-spectrum-discrete}, \\ \sigma_{\mathcal{AA}} (\omega) &= |\boldsymbol{\lambda}_{\mathbf{L}}|^2 \sigma_\smlsub{\rm GCT} (\omega) - 2 \boldsymbol{\lambda}_{\mathbf{L}} \cdot  \sum_{j} \boldsymbol{\alpha}_{j} d_j \frac{ \sigma_\smlsub{\rm GCT} (\omega - \omega_j)}{1 - \erm^{-\beta \omega_k}} + \sum_{j,k} \boldsymbol{\alpha}_{j} \mdot \boldsymbol{\alpha}_{k} d_j d_k \frac{ \sigma_\smlsub{\rm GCT} (\omega - \omega_j - \omega_k)}{(1 - \erm^{-\beta \omega_j})(1 - \erm^{-\beta \omega_k})} \label{SIeq:non-Condon-AA-spectrum-discrete}, \\ 
    \sigma_{\mathcal{B}}(\omega) &= \frac{1}{2} \sum_k \frac{|\boldsymbol{\alpha}_k|^2}{\omega_k} \frac{\sigma_\smlsub{\rm GCT}(\omega - \omega_k)}{1-\erm^{-\beta \omega_k}}\label{SIeq:non-Condon-B-spectrum-discrete}.
\end{align}
\end{subequations}
Here, $\boldsymbol{\lambda}_{\mathbf{L}}=\frac{1}{2} \sum_k \boldsymbol{\alpha}_k d_k$. Although equivalent to the expressions derived in Eqs.~\eqref{SIeq:non-Condon-A-spectrum-final-convolution-GNCT-JCP}-\eqref{SIeq:non-Condon-AA-spectrum-final-convolution-GNCT-JCP}, these discrete representations allow us to articulate, and semi-quantitatively determine, \textit{how non-Condon spectral contributions can either amplify, shift or split peaks in a spectrum}. First, a remarkable similarity between these expressions is that $\boldsymbol{\sigma_\mathcal{A}}(\omega)$, $\sigma_\mathcal{AA}(\omega)$, and $\sigma_{\mathcal{B}}(\omega)$ all contain terms in which $\sigma_{\rm GCT}(\omega)$ is shifted by each frequency, $\omega_n$, corresponding to a given nuclear motion that induces fluctuations in the transition dipole $(\sigma_\mathcal{B}(\omega))$ or correlated fluctuations between the energy gap and transition dipole $(\boldsymbol{\sigma_\mathcal{A}}(\omega)$ and $\sigma_\mathcal{AA}(\omega))$. This analysis describes the impact that $\sigma_{\mathcal{B}}(\omega)$ has on the lineshape. For further details, we refer the reader to our work that shows how this term controls the $Q$-band splitting of metal-free porphyrins \cite{Wiethorn2023-sup}. Henceforth, we restrict our analysis to the terms, $\boldsymbol{\sigma_\mathcal{A}}(\omega)$ and $\sigma_{\mathcal{AA}}(\omega)$, using Eqs.~\eqref{SIeq:non-Condon-A-spectrum-discrete} and \eqref{SIeq:non-Condon-AA-spectrum-discrete} to shed light on how these terms can manipulate linehapes.

We start with $\boldsymbol{\sigma_{\mathcal{A}}}(\omega)$. The expression for $\boldsymbol{\sigma_{\mathcal{A}}}(\omega)$ in Eq.~\eqref{SIeq:non-Condon-A-spectrum-discrete} contains a difference between two $\sigma_\smlsub{\rm GCT}(\omega)$. The first term is the GCT lineshape scaled by $\boldsymbol{\lambda}_{\mathbf{L}}$, whereas the second term is a sum over shifted GCT spectra, $\sigma_\smlsub{\rm GCT}(\omega-\omega_k)$, that are individually amplified by the set of temperature-dependent coefficients, $\{ \frac{\boldsymbol{\alpha}_k d_k}{1 - \erm^{-\beta \omega_k}} \}$. Here, Eq.~\eqref{SIeq:non-Condon-A-spectrum-discrete} indicates that each peak in a given $\mathbf{L}(\omega)/\omega$ contributes a new Condon lineshape that is shifted by $\omega_{eg}^{\rm av} + \omega_n^{\rm cm}$, where $\omega_n^{\rm cm}$ is the center-of-mass frequency for the $n$th peak in $\mathbf{L}(\omega)/\omega$. By subtracting intensity from the first term, these shifted Condon lineshapes cause a feature with negative intensity based on the relative magnitudes between the un-shifted spectrum and the subset of $\{ \frac{\boldsymbol{\alpha}_k d_k}{1 - \erm^{-\beta \omega_k}} \}$ about the frequencies they are defined, leading to spectra like those in Fig.~\ref{SI-fig:theta-1-limit-of-linear-spectral-interference}.

Lastly, we decompose $\sigma_{\mathcal{AA}}(\omega)$ into three regions that contribute unique spectral features based on the three terms shown in Eq.~\eqref{SIeq:non-Condon-AA-spectrum-discrete}. The first two terms that compose $\sigma_{\mathcal{AA}}(\omega)$ are analogous to the difference that composes $\boldsymbol{\sigma_{\mathcal{A}}}(\omega)$, which introduce two regions in the spectrum that are respectively positive or negative in intensity. Unlike $\boldsymbol{\sigma_{\mathcal{A}}}(\omega)$, the first term in $\sigma_{\mathcal{AA}}(\omega)$ is scaled by $|\boldsymbol{\lambda}_{\mathbf{L}}|^2>0$, causing the first region of $\sigma_{\mathcal{AA}}(\omega)$ to show positive intensities, and as the interference with the second term in Eq.~\eqref{SIeq:non-Condon-AA-spectrum-discrete} becomes more prominent, $\sigma_{\mathcal{AA}} (\omega)$ begins to show negative intensities. The last term in Eq.~\eqref{SIeq:non-Condon-AA-spectrum-discrete} constitutes the third region of $\sigma_{\mathcal{AA}}(\omega)$, which contributes scaled Condon lineshapes that center at frequencies, $\omega_{eg}^{\rm av} + \omega^{\rm cm}_n + \omega^{\rm cm}_m $, for a given $\mathbf{L}(\omega)/\omega$ and contribute intensity at the highest frequencies under the $\sigma_{\mathcal{AA}}(\omega)$ lineshape. This last term interferes with the second term in Eq.~\eqref{SIeq:non-Condon-AA-spectrum-discrete} to introduce a third region in $\sigma_{\mathcal{AA}} (\omega)$ with intensities that are again positive.

\end{document}